\documentclass[10pt]{article}
\pdfoutput=1

\usepackage[utf8]{inputenc}
\usepackage[T2A]{fontenc}

\input{epsf}
\usepackage{epsfig}
\usepackage{amssymb}
\usepackage{amsfonts}
\usepackage{amsbsy}
\usepackage[cmtip,all]{xy}
\usepackage{amsmath}
\usepackage{esint}
\usepackage{collectbox}
\usepackage{amscd}
\usepackage{mathrsfs}
\usepackage{amsmath,amsthm}
\usepackage{enumitem}
\usepackage{dsfont}
\usepackage{soul}
\usepackage{comment}
\usepackage{cite} 
\usepackage[most]{tcolorbox}
\usepackage{stmaryrd}
\usepackage{xcolor}
\definecolor{burgundy}{rgb}{0.5, 0.0, 0.13}
\definecolor{olive}{rgb}{0.50, 0.50, 0.0}
\usepackage[linktocpage=true,colorlinks=true,linkcolor=burgundy,citecolor=black!20!blue,urlcolor=violet]{hyperref}
\usepackage{accents}
\usepackage{xfrac}
\usepackage{bm}

\usepackage{tikz}
\usepackage{tikz-cd}
\usetikzlibrary{arrows}
\usetikzlibrary{arrows.meta}
\usetikzlibrary{positioning}
\usetikzlibrary{shapes}
\usetikzlibrary{fit}
\usetikzlibrary{decorations.pathmorphing,decorations.pathreplacing,decorations.markings}
\usetikzlibrary{calc}

\usepackage[font=footnotesize,labelfont=bf]{caption}

\theoremstyle{definition}

\DeclareMathAlphabet{\mathpzc}{OT1}{pzc}{m}{it}

\def\I{{\rm i}}

\def\p{\partial}

\def\CH {{\cal H}}

\def\CN {{\cal N}}

\def\CR {{\cal R}}

\def\CH {{\cal H}}

\def\CS {{\cal S}}

\def\IQ{\mathbb{Q}} 

\def\fD{\mathfrak{D}}

\def\fl{\mathfrak{l}}

\def\fs{\mathfrak{s}}

\def\fs{\mathfrak{s}}

\def\fu{\mathfrak{u}}

\def\fQ{\mathfrak{Q}}

\def\bphi{{\boldsymbol{\phi}}}
\def\bpsi{{\boldsymbol{\psi}}}

\def\lm{\limits}
\def\be{\begin{eqnarray}}
\def\ee{\end{eqnarray}}

\hoffset= - 1.0in

\numberwithin{equation}{section}

\tcbset{boxrule=0.6pt,colback=white!97!green}

\DeclareSymbolFont{bbsymbol}{U}{bbold}{m}{n}
\DeclareMathSymbol{\bbzero}{\mathbin}{bbsymbol}{"30}
\DeclareMathSymbol{\bbone}{\mathbin}{bbsymbol}{"31}
\DeclareMathSymbol{\bbtwo}{\mathbin}{bbsymbol}{"32}
\DeclareMathSymbol{\bbthree}{\mathbin}{bbsymbol}{"33}
\DeclareMathSymbol{\bbfour}{\mathbin}{bbsymbol}{"34}
\DeclareMathSymbol{\bbfive}{\mathbin}{bbsymbol}{"35}
\DeclareMathSymbol{\bbsix}{\mathbin}{bbsymbol}{"36}
\DeclareMathSymbol{\bbseven}{\mathbin}{bbsymbol}{"37}
\DeclareMathSymbol{\bbeight}{\mathbin}{bbsymbol}{"38}
\DeclareMathSymbol{\bbnine}{\mathbin}{bbsymbol}{"39}

\def\myblue{white!40!blue}
\def\mygreen{black!30!green}
\def\myred{black!30!red}

\newcommand\sqbox[1]{{
	\setbox0=\hbox{\mbox{$\Box$}}
	\setbox1=\hbox{\mbox{\raisebox{0.35ex}{\tiny #1}}}
	\mbox{\raisebox{-0.2ex}{\rlap{\hbox to \wd0{\hss{\box1}\hss}}\box0}}
}}

\def\MF{{\mathscr{M}}}
\def\bMF{\overline{\MF} }
\def\MOY{\Gamma}
\def\Op{{\mathscr{O}}}

\begin{document}

\hfill MIPT/TH-14/25

\hfill ITEP/TH-17/25

\hfill IITP/TH-15/25

\vskip 1.5in
\begin{center}
	
    {\bf\Large Operator lift of Reshetikhin-Turaev formalism to \\
    	Khovanov-Rozansky TQFTs}

	\vskip 0.2in
	\renewcommand{\thefootnote}{\fnsymbol{footnote}}
	{Dmitry Galakhov
		\footnote[2]{e-mail: d.galakhov.pion@gmail.com, galakhov@itep.ru},
		Elena Lanina
		\footnote[3]{e-mail: lanina.en@phystech.edu}
		 and  Alexei Morozov
		\footnote[4]{e-mail: morozov@itep.ru}}\\
	
	\vskip 0.2in
	\renewcommand{\thefootnote}{\roman{footnote}}
	{\small{
			\textit{
				MIPT, 141701, Dolgoprudny, Russia}
			\vskip 0 cm
			\textit{
				NRC ``Kurchatov Institute'', 123182, Moscow, Russia}
			\vskip 0 cm
			\textit{
				IITP RAS, 127051, Moscow, Russia}
			\vskip 0 cm
			\textit{
				ITEP, Moscow, Russia}
	}}
\end{center}

\vskip 0.2in
\baselineskip 16pt

\centerline{ABSTRACT}

\bigskip

{\footnotesize
    Topological quantum field theory (TQFT) is a powerful tool to describe homologies,
    which normally involve complexes and a variety of maps/morphisms,
    what makes a functional integration approach with a sum over a single kind of maps seemingly problematic.
    In TQFT  this problem is overcame by exploiting the  rich set of zero modes of BRST operators,
    which appear sufficient to describe complexes.
    We explain what this approach looks like for the important class of Khovanov-Rozansky (KR) cohomologies,
    which categorify the observables (Wilson lines or knot polynomials) in 3d Chern-Simons theory.
    We develop a construction of odd differential operators, associated with all link diagrams,
    including tangles with open ends.
    These operators become nilpotent only for diagram with no external legs,
    but even for open tangles one can develop a factorization formalism, which preserve
    Reidemeister/topological invariance -- the symmetry of the problem.
    This technique seems much more ``physical'' than conventional language of homological algebra
    and should have many applications to various problems beyond Chern-Simons theory.
    We also hope that this language will provide efficient algorithms, and finally allow to computerize the calculation
    of KR cohomologies -- for closed diagrams and for open tangles.

}

\bigskip

\bigskip

\tableofcontents

\bigskip

\section{Introduction}

The Khovanov-Rozansky (KR) theory \cite{KRI,KRII,Carqueville:2011zea,Dolotin:2013osa,Wu} is known to ``categorify'', refine or generalize  a theory of HOMFLY polynomial invariants \cite{Freyd:1985dx} of knots and links.
Physically HOMFLY polynomials are treated as averages of knotted Wilson loop operators in the 3d Chern-Simons (CS) theory with gauge group $SU(n)$ calculated with the help of the quantum group theory, a so-called Reshetikhin-Turaev (RT) formalism \cite{Reshetikhin:1990pr,Reshetikhin:1991tc}.
It is expected that the CS theory is a string field theory \cite{Witten:1992fb} for topological strings, or M-theory in more general terms.
The never fading interest to knot/link theory categorification in the physicist community \cite{Gukov:2004hz,Dunfield:2005si,Witten:2011xiq,Aganagic:2011sg,Aganagic:2012ne,Anokhina:2014hha, Morozov:2015sda,Gukov:2016gkn,Dasgupta:2016rhc,Arthamonov:2017oxw,Gukov:2017kmk,Anokhina:2024lzh,Anokhina:2025nti} is provided by expectations to catch an echo of the high energy M-theory in this deformations.

In this brief note we would like to initiate an attempt to translate the Khovanov-Rozansky theory from the homological algebra language to more pedestrian terms of ordinary differential operators more often used in the physicist community.
We believe these new terms would have better algorithmizability properties and would allow to calculate KR invariants of links on machines with higher efficiency.

In short, to a given link diagram $L$ (possibly a tangle diagram with open outgoing and incoming legs) we would associate a differential operator $\Op_L$ acting on functions of even (bosonic) and odd (fermionic, Grassmann) variables following the rules:
\begin{enumerate}
	\item{}
	We associate with the crossing (valence-2,2) diagrams two odd operator $\Op^\pm_v$, depending on two pairs
	of $ x$- and $\bar  x$-variables on incoming and outgoing lines respectively.
	There is also a triplet of additional odd variables
	which are associated with the vertex itself we will denote by Greek letters $\theta$, $\eta$ and $\epsilon$.
	Operators $\Op^\pm_v$ are ordinary differential operators in the odd variables one could rewrite in terms of elementary ones: we denote by $\hat \theta$ an operator of multiplication by odd variable $\theta$, and by $\hat\theta^{\dagger}$ an operator of differentiation with respect to $\theta$.

	The label $\pm$ refers to two different orientations,
	and there is also a pair of  conjugate odd parameters $\epsilon$ which allow to distinguish between
	orientations.
	All the operators $\Op_v^{\pm}$ anti-commute at different vertices and in this sense are \emph{local}.
	
	\item{}
	The squares $\Op^2 \sim  \sum_{i=1}^2 \Big(w( x_i)-w( \bar x_i)\Big)$
	with some function (superpotential) $w$.
	In the simplest case of the fundamental representation of $\fs\fl_n$ all $x$ are numbers
	and we choose $w(x) = x^{n+1}$.
	A choice of a generic polynomial would lead to so called \emph{equivariant} KR cohomology \cite{wu2012equivariant,krasner2010equivariant}.
	
	\item{}
	To consider generic link diagrams (tangles) $L$ drawn inside a disk, we have to take into account the disk boundary.
	In the 3d CS theory such a diagram would correspond to a collection of possibly open Wilson lines in a ball, and on the ball boundary we would put necessarily WZW theory to preserve topological invariance.
	Here we treat the disk boundary as a multi-valence vertex and associate to it a generic operator $\Op_{\p}(x_l,\bar x_l)$ depending only on $x_l$ and $\bar x_l$
	for all incoming and outgoing legs.
	This operator is constrained solely to the following natural conditions:
	\begin{equation}
		\left\{\Op_{\p},\Op_v^{\pm}\right\}=0,\; \forall v,\quad \Op_{\p}^2=\sum_{l} \Big(-w( x_l)+w( \bar x_l)\Big)\,,
	\end{equation}
	as for the ``boundary vertex'' outgoing legs are inflowing edges and vice versa.
	Then with arbitrary link diagram (tangle) $L$ one can associate
	\begin{equation}
		\Op_L(x_e,\hat\theta_v, \hat\eta_v,\hat\epsilon_v,\hat\theta_v^{\dagger}, \hat\eta_v^{\dagger},\hat\epsilon_v^{\dagger}|x_l,\bar x_l) = \sum_{{\rm vertices\; of\;} L} \Op^\pm_v+\Op_{\p}\,,
	\end{equation}
	which depends on $x_e$ on the internal edges of $L$ (each $x_e$ appears as incoming for one $\MF^\pm_v$ and outgoing for another),
	on triplets of $\theta_v, \eta_v,\epsilon_v$ and their conjugates at vertices, and also on $x_l$ and $\bar x_l$
	for the incoming and outgoing legs.
	For the square we have:
	\begin{equation}
		\Op_L^2 = 0\,.
	\end{equation}
	
	\item{}
	What are invariants of topological diagram moves are ``invariant wave functions'' --
	zero modes of $\Op_L$ modulo ``gauge transform'' $g_{\omega}$:
	\begin{equation}
		\Op_L\Psi_L = 0,\quad g_{\omega}(\Psi_L)=\Psi_L+\Op_L\omega\,.
	\end{equation}
	Equivalently we say that wave functions are cohomologies $H^*(\Op_L)$ of $\Op_L$.
	Since $\Op_L$ is the  differential operator w.r.t. the odd variables only,
	this is a simple equation, solved by methods of linear algebra.
	Products of $R$-matrices are obtained by an equally straightforward factorization of $\Psi_L$
	modulo $\Op_L$.
	This whole procedure can be regarded as a generalization of KR cohomology construction --
	a necessary one if we want to deal with tangles.
	
	\item{}
	For a closed diagram $\bar L$ the boundary is empty $\Op_{\p}=0$ and one can consider соhomologies of $\Op_{\bar L}$ without boundary contribution.
	They can be related to the more familiar cohomologies of the KR {\it bi}\,complex (or double complex),
	associated to the collection $\{\MOY\}$ of MOY resolutions \cite{MOY} of $L$.
	That construction is actually two-step, with first evaluation the ``vertical cohomologies''
	$H^*_\MOY$ for a given MOY resolution $\MOY$ of $L$,
	and then those of the induced ``horizontal'' complex, made from these $H^*_\MOY$ for different $\MOY$.
	The corresponding Poincare  polynomial $P_{\overline{L}}$ is known to generalize the fundamental HOMFLY polynomial, obtained from RT calculus.
\end{enumerate}

We believe this reformulation of the KR theory would make it easier to compare with other physical Morse/Floer approaches to categorified $\fs\fu_n$ link invariants originating from direct compactifications of M-theory \cite{Witten:2011zz,Gaiotto:2011nm,Gaiotto:2015aoa,Galakhov:2016cji,Galakhov:2017pod,Galakhov:2020upa,Aganagic:2020olg,Aganagic:2021ubp,Aganagic:2023amh}.
It is always easy to reformulate the problem of finding cohomologies of operator $\Op_L$ as a search for ground state wave functions of a Laplacian-like Hamiltonian \cite{Witten:1982im}:\footnote{Here operator $\Op_L^{\dagger}$ corresponds to $\Op_L$ conjugated with respect to some non-degenerate norm on polynomials of even and odd variables.
Seemingly the simplest choice would be to promote variables $x_k$ to Heisenberg operators $\hat x_k=x_k\cdot$ so that trivial polynomial $1$ corresponds to vacuum $|0\rangle$, $\langle 0|0\rangle=1$. 
In this representation $\hat x_k^{\dagger}=\p_{x_k}$.
}
\begin{equation}\label{Hamiltonian}
	\begin{aligned}
		&\CH_L=\frac{1}{2}\left(\Op_L\Op_L^{\dagger}+\Op_L^{\dagger}\Op_L\right)\,,\\
		& H^*(\Op_L):=\frac{{\rm Ker}\,\Op_L}{{\rm Im}\,\Op_L}\;\cong \; \left\{\Psi\;\big|\;\CH_L\Psi=0\right\}\,.
	\end{aligned}
\end{equation}
There is almost no chance that $\CH_L$ would be somewhat reminiscent of a Hamiltonian $H_M$ in M-theory.
Nevertheless, one might expect in fact those Hamiltonians are connected by a RG flow trajectory in the space of theories preserving zero modes.
In this case an equivalence of the two approaches would be transparent.
In the modern literature a rigorous proof of this equivalence or observing explicit discrepancies remains an open question.

We should stress these approaches have rather different spirits.
In the KR story operators $\Op_L$ are constructed from local mutually anti-commuting quantities $\Op_v^{\pm}$ corresponding to diagram vertices, whereas on the other side the Morse instanton flow could connect diagram ``colorings'' (wave functions) differing non-locally.
However, on the other side, Morse/Floer M-theory approaches are usually designed to make defect operators corresponding to link manifestly topological, so the invariance with respect to homotopies of $L$ is apparent, whereas for the KR approach this invariance is a deep theorem.

Also one could look at this reformulation of the KR theory in terms of \eqref{Hamiltonian} as a Hamiltonian evolution setup of KR foam TQFTs \cite{khovanov2004sl,mackaay2009sl,Chun:2015gda,lauda2015khovanov,queffelec2016sln,Carqueville:2017ono}.
Yet in comparison to those picturesque approaches we believe the language of differential equations could be more efficient in computational applications.

For the sake of brevity we will omit many peculiarities to be considered elsewhere.
Here we only take a glimpse on the following questions:
\begin{itemize}
	\item HOMFLY polynomials are known as Jones polynomials at $n=2$.
	Respectively the KR theory is expected to reduce directly to Khovanov's categorification of Jones polynomials \cite{Khovanov:1999qla,Dolotin:2012sw,Dolotin:2012we}.
	Could this reduction be made more manifest in terms of operators $\Op$?
	\item Decategorification.
	The ways to show directly that operators $\Op_L$ are decategorified directly to $U_q(\fs\fl_n)$ $R$-matrices and how constrained is the inverse process.
	\item  Krasner \cite{Krasner} proposed a drastic simplification of computing KR complex cohomologies for bipartite knots \cite{Anokhina:2024lbn,Anokhina:2024uso,Anokhina:2025eso}.
	Could we observe this simplification in terms of operators $\Op$?
\end{itemize}

The paper is organized as follows.
In Sec.~\ref{sec:MF} we review the structure of matrix factorization as a differential operator.
In Sec.~\ref{sec:KRdc} we give a brief review of the KR double complex structure.
In Sec.~\ref{sec:vertical} we discuss a construction of vertical morphisms and in Sec.~\ref{sec:horizontal} a construction of horizontal morphisms.
In Sec.~\ref{sec:O} we present a construction of operators $\Op_L$ and in Sec.~\ref{sec:invariant} we indicate that they define link invariants.
Finally, Sec.~\ref{sec:misc} is devoted to examples and miscellaneous questions.


\section{Matrix factorizations (MF)} \label{sec:MF}

Matrix factorization (MF) is a correction to the supercharge of 2d $\CN=(2,2)$ supersymmetric topological string theory due to a worldsheet boundary \cite{Kapustin:2002bi,Kapustin:2003ga,Kapustin:2003rc,Hori:2013ika,Herbst:2008jq,Khovanov:2004bc,Brunner:2003dc,Brunner:2004mt}.
See also App.~\ref{app:MF}.

For us matrix factorization $\MF$ is a differential operator acting on a ring $\CR$ of even, bosonic variables $x_k$ and odd, fermionic variables $\theta_i$ (by straight brackets $[*,*]$ we imply commutators, and by curly brackets $\{*,*\}$ we imply anti-commutators):
\begin{equation}
	\left[x_k,x_l\right]=0,\quad \left[x_k,\theta_i\right]=0,\quad \left\{\theta_i,\theta_j\right\}=0\,.
\end{equation}
Let us denote operators of multiplication by $\theta_i$ as $\hat\theta_i:=\theta_i\cdot$ and operators of differentiation with respect to $\theta_i$ as $\hat\theta_i^{\dagger}:=\p/\p\theta_i$.
These operators commute with $x_k$ and form the odd Heisenberg algebra:
\begin{equation}
	\left\{\hat\theta_i,\hat\theta_j\right\}=\left\{\hat\theta_i^{\dagger},\hat\theta_j^{\dagger}\right\}=0,\quad \left\{\hat\theta_i,\hat\theta_j^{\dagger}\right\}=\delta_{ij}\,.
\end{equation}

We define a MF operator $\MF$ as an arbitrary differential operator, linear in $\hat\theta_i$, $\hat\theta_i^{\dagger}$:
\begin{equation}
	\MF=\sum\lm_i A_i\hat\theta_i+B_i\hat\theta_i^{\dagger}\,,
\end{equation}
where $A_i$, $B_i$ are arbitrary polynomials in even variables $x_k$.

It is easy to derive the basic property of the MF operator:
\begin{tcolorbox}
\begin{equation}\label{MF2}
	\MF^2=\left(\sum\lm_i A_iB_i\right)\cdot\bbone\,.
\end{equation}
\end{tcolorbox}
This property of MF explains its name, as a polynomial in the r.h.s. might not factorize in a pair of polynomials,
yet it decomposes into a sum and factorizes as a square of a linear operator.
Surely, there are generalizations when $A_i$ and $B_i$ are matrices of polynomials rather than just vectors \cite{Kapustin:2002bi}, however we will not need this prescription here.

Defined this way MF operator is equivalent to the following MF \cite{KRI}:
\begin{equation}
	\MF\cong\bigotimes\lm_{i}\left(\!\!\!\!\begin{array}{c}
		\begin{tikzpicture}
			\node (A) at (0,0) {$\IQ[x_1,x_2,\ldots]$};
			\node (B) at (3,0) {$\IQ[x_1,x_2,\ldots]$};
			\draw[-stealth] ([shift={(0,0.05)}]A.east) -- ([shift={(0,0.05)}]B.west) node[pos=0.5,above] {$\scriptstyle A_i$};
			\draw[stealth-] ([shift={(0,-0.05)}]A.east) -- ([shift={(0,-0.05)}]B.west) node[pos=0.5,below] {$\scriptstyle B_i$};
		\end{tikzpicture}
	\end{array}\!\!\!\!\right)\,.
\end{equation}

\section{Khovanov-Rozansky (KR) double complex} \label{sec:KRdc}

 Construction of \cite{KRI} can be rewritten in the proposed language of differential operators in the following way.

\begin{enumerate}
	\item The starting point is a link diagram $L$ with two types of valence-4 vertices depicting resolved intersections of link strands after projecting to a plane.
	A resolved intersection vertex saves the information which link strand went above and which one went below before projection.
	If a diagram has no external edges, we call is {\it closed} and denote by $\bar L$.
	\item Substitute each intersection with a formal sum of implanted resolutions:
	\begin{equation}\label{categR}
		\begin{aligned}
			\begin{array}{c}
				\begin{tikzpicture}[scale=0.5]
					\draw[thick,-stealth] (-0.5,-0.5) -- (0.5,0.5);
					\draw[thick,-stealth] (-0.1,0.1) -- (-0.5,0.5);
					\draw[thick] (0.5,-0.5) -- (0.1,-0.1);
				\end{tikzpicture}
			\end{array}=\begin{array}{c}
				\begin{tikzpicture}[scale=0.5]
					\draw[thick,-stealth] (-0.5,-0.5) to[out=45,in=270] (-0.15,0) to[out=90,in=315] (-0.5,0.5);
					\draw[thick,-stealth] (0.5,-0.5) to[out=135,in=270] (0.15,0) to[out=90,in=225] (0.5,0.5);
				\end{tikzpicture}
			\end{array}+\epsilon\begin{array}{c}
				\begin{tikzpicture}[scale=0.5]
					\draw[thick,-stealth] (-0.5,-0.5) -- (0.5,0.5);
					\draw[thick,-stealth] (0.5,-0.5) -- (-0.5,0.5);
					\draw[fill=\myred] (0,0) circle (0.15);
				\end{tikzpicture}
			\end{array},\quad
			\begin{array}{c}
				\begin{tikzpicture}[scale=0.5, xscale=-1]
					\draw[thick,-stealth] (-0.5,-0.5) -- (0.5,0.5);
					\draw[thick,-stealth] (-0.1,0.1) -- (-0.5,0.5);
					\draw[thick] (0.5,-0.5) -- (0.1,-0.1);
				\end{tikzpicture}
			\end{array}=\begin{array}{c}
				\begin{tikzpicture}[scale=0.5]
					\draw[thick,-stealth] (-0.5,-0.5) -- (0.5,0.5);
					\draw[thick,-stealth] (0.5,-0.5) -- (-0.5,0.5);
					\draw[fill=\myred] (0,0) circle (0.15);
				\end{tikzpicture}
			\end{array}+\epsilon\begin{array}{c}
				\begin{tikzpicture}[scale=0.5]
					\draw[thick,-stealth] (-0.5,-0.5) to[out=45,in=270] (-0.15,0) to[out=90,in=315] (-0.5,0.5);
					\draw[thick,-stealth] (0.5,-0.5) to[out=135,in=270] (0.15,0) to[out=90,in=225] (0.5,0.5);
				\end{tikzpicture}
			\end{array}\,,
		\end{aligned}
	\end{equation}
	where each fermionic odd variable $\epsilon$ is unique for each intersection.
	For example,
	\begin{equation}
			\begin{array}{c}
			\begin{tikzpicture}[scale=0.6]
				\begin{scope}[shift={(0,1)}]
					\draw[thick] (0.5,-0.5) to[out=90,in=270] (-0.5,0.5);
					\draw[white, line width = 1.5mm] (-0.5,-0.5) to[out=90,in=270] (0.5,0.5);
					\draw[thick] (-0.5,-0.5) to[out=90,in=270] (0.5,0.5);
				\end{scope}
				\draw[thick,-stealth] (0.5,-0.5) to[out=90,in=270] (-0.5,0.5) -- (-0.5,0.6);
				\draw[white, line width = 1.5mm] (-0.5,-0.5) to[out=90,in=270] (0.5,0.5);
				\draw[thick,-stealth] (-0.5,-0.5) to[out=90,in=270] (0.5,0.5) -- (0.5,0.6);
				\draw[thick] (-0.5,1.5) to[out=90,in=90] (-1.2,1.5) -- (-1.2,-0.5) to[out=270,in=270] (-0.5,-0.5);
				\begin{scope}[xscale=-1]
					\draw[thick] (-0.5,1.5) to[out=90,in=90] (-1.2,1.5) -- (-1.2,-0.5) to[out=270,in=270] (-0.5,-0.5);
				\end{scope}
			\end{tikzpicture}
		\end{array}=\begin{array}{c}
				\begin{tikzpicture}[scale=0.6]
					\draw[thick] (0.5,-0.5) to[out=90,in=270] (0.3,0) to[out=90,in=270] (0.5,0.5) -- (0.5,0.6);
					\draw[thick] (-0.5,-0.5) to[out=90,in=270] (-0.3,0) to[out=90,in=270] (-0.5,0.5) -- (-0.5,0.6);
					\begin{scope}[shift={(0,1)}]
						\draw[thick] (0.5,-0.5) to[out=90,in=270] (0.3,0) to[out=90,in=270] (0.5,0.5);
						\draw[thick] (-0.5,-0.5) to[out=90,in=270] (-0.3,0) to[out=90,in=270] (-0.5,0.5);
					\end{scope}
					\draw[thick, postaction={decorate},decoration={markings, mark= at position 0.6 with {\arrow{stealth}}}] (-0.5,1.5) to[out=90,in=90] (-1.2,1.5) -- (-1.2,-0.5) to[out=270,in=270] (-0.5,-0.5);
					\begin{scope}[xscale=-1]
						\draw[thick, postaction={decorate},decoration={markings, mark= at position 0.6 with {\arrow{stealth}}}] (-0.5,1.5) to[out=90,in=90] (-1.2,1.5) -- (-1.2,-0.5) to[out=270,in=270] (-0.5,-0.5);
					\end{scope}
				\end{tikzpicture}
			\end{array}+\epsilon_1\begin{array}{c}
				\begin{tikzpicture}[scale=0.6]
					\draw[thick] (0.5,-0.5) to[out=90,in=270] (-0.5,0.5);
					\draw[thick] (-0.5,-0.5) to[out=90,in=270] (0.5,0.5);
					\begin{scope}[shift={(0,1)}]
						\draw[thick] (0.5,-0.5) to[out=90,in=270] (0.3,0) to[out=90,in=270] (0.5,0.5);
						\draw[thick] (-0.5,-0.5) to[out=90,in=270] (-0.3,0) to[out=90,in=270] (-0.5,0.5);
					\end{scope}
					\draw[thick, postaction={decorate},decoration={markings, mark= at position 0.6 with {\arrow{stealth}}}] (-0.5,1.5) to[out=90,in=90] (-1.2,1.5) -- (-1.2,-0.5) to[out=270,in=270] (-0.5,-0.5);
					\begin{scope}[xscale=-1]
						\draw[thick, postaction={decorate},decoration={markings, mark= at position 0.6 with {\arrow{stealth}}}] (-0.5,1.5) to[out=90,in=90] (-1.2,1.5) -- (-1.2,-0.5) to[out=270,in=270] (-0.5,-0.5);
					\end{scope}
					\draw[fill=\myred] (0,0) circle (0.13);
				\end{tikzpicture}
			\end{array}+\epsilon_2\begin{array}{c}
				\begin{tikzpicture}[scale=0.6]
					\draw[thick] (0.5,-0.5) to[out=90,in=270] (0.3,0) to[out=90,in=270] (0.5,0.5);
					\draw[thick] (-0.5,-0.5) to[out=90,in=270] (-0.3,0) to[out=90,in=270] (-0.5,0.5);
					\begin{scope}[shift={(0,1)}]
						\draw[thick] (0.5,-0.5) to[out=90,in=270] (-0.5,0.5);
						\draw[thick] (-0.5,-0.5) to[out=90,in=270] (0.5,0.5);
					\end{scope}
					\draw[thick, postaction={decorate},decoration={markings, mark= at position 0.6 with {\arrow{stealth}}}] (-0.5,1.5) to[out=90,in=90] (-1.2,1.5) -- (-1.2,-0.5) to[out=270,in=270] (-0.5,-0.5);
					\begin{scope}[xscale=-1]
						\draw[thick, postaction={decorate},decoration={markings, mark= at position 0.6 with {\arrow{stealth}}}] (-0.5,1.5) to[out=90,in=90] (-1.2,1.5) -- (-1.2,-0.5) to[out=270,in=270] (-0.5,-0.5);
					\end{scope}
					\draw[fill=\myred] (0,1) circle (0.13);
				\end{tikzpicture}
			\end{array}+\epsilon_1\epsilon_2\begin{array}{c}
				\begin{tikzpicture}[scale=0.6]
					\draw[thick,-stealth] (0.5,-0.5) to[out=90,in=270] (-0.5,0.5) -- (-0.5,0.6);
					\draw[thick,-stealth] (-0.5,-0.5) to[out=90,in=270] (0.5,0.5) -- (0.5,0.6);
					\begin{scope}[shift={(0,1)}]
						\draw[thick] (0.5,-0.5) to[out=90,in=270] (-0.5,0.5);
						\draw[thick] (-0.5,-0.5) to[out=90,in=270] (0.5,0.5);
					\end{scope}
					\draw[thick, postaction={decorate},decoration={markings, mark= at position 0.6 with {\arrow{stealth}}}] (-0.5,1.5) to[out=90,in=90] (-1.2,1.5) -- (-1.2,-0.5) to[out=270,in=270] (-0.5,-0.5);
					\begin{scope}[xscale=-1]
						\draw[thick, postaction={decorate},decoration={markings, mark= at position 0.6 with {\arrow{stealth}}}] (-0.5,1.5) to[out=90,in=90] (-1.2,1.5) -- (-1.2,-0.5) to[out=270,in=270] (-0.5,-0.5);
					\end{scope}
					\draw[fill=\myred] (0,0) circle (0.13) (0,1) circle (0.13);
				\end{tikzpicture}
			\end{array}
	\end{equation}

	\item Graphs consisting of oriented edges and 4-valent vertices,
resolved  as introduced above, will be called \emph{MOY diagrams} in analogy with \cite{MOY}.
	We will denote resolutions by letter $\MOY$:
\begin{equation}
L=\sum_\alpha^{2^{\#( \rm vertices)}} \MOY_\alpha\,.
\end{equation}
\item There is a basic parameter $n$ -- a number of colors in categorified $U_q(\fs\fu_n)$.

\bigskip

{\bf Here we should stress that so far rules where applicable both to the RT and the KR formalisms.
	This indicates that the HOMFLY polynomial constructed with the RT formalism is an Euler characteristic for the KR complex and its construction procedure is similar.
	Further we list points specific for the KR construction.}

\bigskip

\item	To each MOY diagram $\MOY$ we assign a ring $\CR_{\epsilon}$ of variables $x_k$, $\theta_i$ mixed with odd variables $\epsilon_\alpha$.
    After that we can associate with $\MOY$ an operator $\MF_\MOY$ of degree
    \begin{equation}
		d={\rm qdeg}\,\MF=n+1\,.
	\end{equation}
It depends on even variables $x_e$ at all edges as a polynomial and is a differential operator in terms of  $\theta_v$ associated to all vertices
(see details in Sec.~\ref{sec:vertical}).

    \item For closed diagrams $\bar\MOY$ operators $\bMF_{\bar\MOY}:=\MF_{\bar\MOY}$ are nilpotent, $\bMF_{\bar\MOY}^2=0$,
i.e. become {\it differentials}

	\item To {\it closed} link diagram $\bar L$ we assign a
KR double complex $(\CR_{\epsilon}^{*,*},\bMF,\fD)$ .
 Cohomologies are link invariants.\footnote{In principle we are able to extend this construction to an open link diagram $L$ by considering a new \emph{variable} diagram $G$ such that together with $L$ they form a closed diagram $\overline{L\cup G}$.
 The resulting double complex $\overline{L\cup G}$ depends explicitly on $G$, yet some conclusions, homotopies of the double complex, could be drawn for $G$ staying arbitrary.
 For instance, Reidemeister moves would produce double complex homotopies of this type.
 }

	\item KR complex is a double complex ($d=n+1$):
	\begin{equation}\label{KRbicompl}
		\begin{array}{c}
			\begin{tikzpicture}
				\node(A) at (0,0) {$\CR_\epsilon^{*,*}(\bar\MOY_1)$};
				\node(B) at (4,0) {$\CR_\epsilon^{*,*+1}(\bar\MOY_2)$};
				\node(C) at (8,0) {$\CR_\epsilon^{*,*+2}(\bar\MOY_3)$};
				\node(E) at (-3,0) {$\ldots$};
				\node(F) at (11,0) {$\ldots$};
				\node(A1) at (0,-1.5) {$\CR_\epsilon^{*+d,*}(\bar\MOY_1)$};
				\node(B1) at (4,-1.5) {$\CR_\epsilon^{*+d,*+1}(\bar\MOY_2)$};
				\node(C1) at (8,-1.5) {$\CR_\epsilon^{*+d,*+2}(\bar\MOY_3)$};
				\node(E1) at (-3,-1.5) {$\ldots$};
				\node(F1) at (11,-1.5) {$\ldots$};
				\node(A0) at (0,1) {$\ldots$};
				\node(B0) at (4,1) {$\ldots$};
				\node(C0) at (8,1) {$\ldots$};
				\node(A2) at (0,-2.5) {$\ldots$};
				\node(B2) at (4,-2.5) {$\ldots$};
				\node(C2) at (8,-2.5) {$\ldots$};
				\path (E) edge[->] (A) (A) edge[->] node[above] {$\scriptstyle \fD$} (B) (B) edge[->]  node[above] {$\scriptstyle \fD$}  (C) (C) edge[->] (F) (E1) edge[->] (A1) (A1) edge[->] node[above] {$\scriptstyle \fD$} (B1) (B1) edge[->]  node[above] {$\scriptstyle \fD$}  (C1) (C1) edge[->] (F1) (A0) edge[->] (A) (A) edge[->] node[right] {$\scriptstyle \bMF_{\bar\MOY_1}$} (A1) (A1) edge[->] (A2) (B0) edge[->] (B) (B) edge[->] node[right] {$\scriptstyle \bMF_{\bar\MOY_2}$} (B1) (B1) edge[->] (B2) (C0) edge[->] (C) (C) edge[->] node[right] {$\scriptstyle \bMF_{\bar\MOY_3}$} (C1) (C1) edge[->] (C2);
			\end{tikzpicture}
		\end{array}
	\end{equation}
	This double complex is \emph{equivalent} to a double complex on a ring $\CR_{\epsilon}$
with two kinds of  differentials $\bMF_{\bar\MOY}$ and $\fD$.
	\item The vertical degree $\rm qdeg$ is fixed for $x_k$ and $\theta_i$ elements of each $\CR$:
	\begin{equation}
		{\rm qdeg}\,x_k=2,\quad {\rm qdeg}\,\theta_i=1-n\mbox{ or }3-n\,.
	\end{equation}
	
	\item The horizontal degree is defined by the powers of $\epsilon$, ${\rm qdeg}\,\epsilon_i=-2$ or $0$, ${\rm tdeg}\,\epsilon_i=1$.
	\item Horizontal differential reads (see details in Sec.~\ref{sec:horizontal}):
	\begin{equation}\label{horizontal}
		\fD=\sum\lm_{\alpha\in{\rm intersections}}\chi^{(\alpha)}\; \epsilon_{\alpha}\cdot,\quad {\rm qdeg}\,\fD=0\,,
	\end{equation}
	where $\chi^{(\alpha)}$ are local MF homomorphisms:
	\begin{equation}\label{MFhomo}
		\left[\chi^{(\alpha)},\chi^{(\beta)}\right]=0,\quad \chi^{(\alpha)} \bMF_{\bar \MOY_j}=\bMF_{\bar \MOY_{j+1}}\chi^{(\alpha)}\,.
	\end{equation}
	Apparently, $\fD^2=0$.
	\item To calculate cohomologies of a double complex one descends first on vertical cohomologies using \eqref{MFhomo}
and then consider a new induced mono-complex:
	\begin{equation}\label{horizcompl}
		\begin{array}{c}
			\begin{tikzpicture}
				\node(A) at (0,0) {$H^{z,f}\left(\bMF_{\bar \MOY_1}\right)$};
				\node(B) at (4,0) {$H^{z+1,f+1}\left(\bMF_{\bar\MOY_2}\right)$};
				\node(C) at (8,0) {$H^{z+2,f+2}\left(\bMF_{\bar\MOY_3}\right)$};
				\node(E) at (-2.5,0) {$\ldots$};
				\node(F) at (10.5,0) {$\ldots$};
				\draw[->] (E) -- (A);
				\draw[->] (A) -- (B) node[pos=0.5,above] {$\scriptstyle \tilde\fD$};
				\draw[->] (B) -- (C) node[pos=0.5,above] {$\scriptstyle \tilde\fD$};
				\draw[->] (C) -- (F);
			\end{tikzpicture}
		\end{array}
	\end{equation}
	\item A Khovanov-Rozansky categorified HOMFLY polynomial is a Poincar\'e polynomial of the double complex:
	\begin{equation}
		P(q,t)=\sum\lm_{z,f} q^z\, t^f\;{\rm dim}\,H^{z,f}( \bMF,\tilde \fD)\,
	\end{equation}
	The HOMFLY polynomial corresponds to $P(q,-1)$.
	In Morse/Floer approaches (see e.g.\cite{Galakhov:2016cji}) to the Khovanov-Rozansky construction grading $z$ corresponds to the central charge of the superalgebra, and grading $f$ corresponds to the fermion number operator in the respective TQFT.
\end{enumerate}

\section{Vertical morphisms}\label{sec:vertical}

\subsection{Constructing MF from MOY}
A MOY diagram \cite{MOY} is a graph consisting of  oriented edges $\begin{array}{c}
		\begin{tikzpicture}[scale=0.5]
			\draw[thick, -stealth] (0,-0.5) -- (0,0.5);
		\end{tikzpicture}
	\end{array}$ and (2,2)-valent vertices $\begin{array}{c}
		\begin{tikzpicture}[scale=0.5]
			\draw[thick, -stealth] (-0.5,-0.5) -- (0.5,0.5);
			\draw[thick, -stealth] (0.5,-0.5) -- (-0.5,0.5);
			\draw[fill=\myred] (0,0) circle (0.15);
		\end{tikzpicture}
	\end{array}$.
Edges correspond to fundamental or anti-fundamental factors in tensor power objects in the RT formalism \cite{Reshetikhin:1990pr,Reshetikhin:1991tc}.
And the 4-valent vertices correspond to projectors to the antisymmetric representation.
We will use both equivalent notations when a 4-valent vertex is short and when it is stretched to an edge of different color:
$\begin{array}{c}
		\begin{tikzpicture}[scale=0.5,rotate=-90]
			\draw[thick, -stealth] (-0.5,-0.5) -- (0.5,0.5);
			\draw[thick, -stealth] (0.5,-0.5) -- (-0.5,0.5);
			\draw[fill=\myred] (0,0) circle (0.12);
		\end{tikzpicture}
	\end{array} = \begin{array}{c}
	\begin{tikzpicture}[scale=0.5,rotate=-90]
		\draw[thick, -stealth] (0,0) -- (0.5,0.5);
		\draw[thick, -stealth] (0,0) -- (-0.5,0.5);
		\draw[thick] (-0.5,-1) -- (0,-0.5) -- (0.5,-1);
		\draw[fill=\myred] (-0.12,-0.5) to[out=270,in=180] (0,-0.62) to[out=0,in=270] (0.12,-0.5) -- (0.12,0) to[out=90,in=0] (0,0.12) to[out=180,in=90] (-0.12,0) -- cycle;
	\end{tikzpicture}
	\end{array}$.
	
As an addition to the canonical MOY construction we would like to allow to add \emph{anywhere} on any edge simple 2-valent vertices $\begin{array}{c}
	\begin{tikzpicture}[scale=0.7]
		\draw[thick, -stealth] (-0.5,0) -- (0.5,0);
		\draw[fill=\mygreen] (0,0) circle (0.1);
	\end{tikzpicture}
\end{array}$.
These 2-valent vertices might seem redundant at the first sight, moreover MOY equivalence \eqref{MOYloc_I} we will discuss in what follows allow to eliminate or add such a type of a vertex freely.
Nevertheless, they turn out to be relevant for uniformizing ring representations of MOY diagrams to make horizontal morphisms explicitly local (see Sec.~\ref{sec:horizontal}).

We construct a ring $\CR$ as a polynomial ring from distinct \emph{even} variables assigned to \emph{edges} and \emph{odd} variables assigned to \emph{vertices}  of a MOY graph.
	See App.~\ref{app:MF}.
To any segment of an oriented edge $\begin{array}{c}
	\begin{tikzpicture}[scale=0.7]
		\draw[thick, postaction={decorate},decoration={markings, mark = at position 0.7 with {\arrow{stealth}}}] (-0.5,0) -- (0.5,0);
		\node[above] at (0,0) {$\scriptstyle x_k$};
		\draw[fill=white] (-0.5,0) circle (0.1) (0.5,0) circle (0.1);
	\end{tikzpicture}
\end{array}$ between two vertices of any type we assign the respective even variable $x_k$.

To a 2-valent vertex between two segments $x$ and $y$ we assign an odd variable $\theta$ and the respective \emph{local} MF operator:
	\begin{equation} \label{2-valentQ}
		\begin{array}{c}
			\begin{tikzpicture}
			\draw[thick, stealth-] (0,0) -- (1,0);
			\node[above] at (0,0) {$\scriptstyle x$};
			\node[above] at (0.5,0) {$\scriptstyle \theta$};
			\node[above] at (1,0) {$\scriptstyle y$};
			\draw[fill=\mygreen] (0.5,0) circle (0.07);
			\end{tikzpicture}
		\end{array}\quad\Longrightarrow\quad \MF_{x,y}=\pi_{xy}\hat\theta+(x-y)\hat\theta^{\dagger}\,,
\end{equation}
\begin{equation}\label{2vsq}
\quad \MF_{x,y}^2=x^{n+1}-y^{n+1},\quad\mbox{here\ \ }{\rm qdeg}\,\theta=-{\rm qdeg}\,\theta^\dagger= 1-n\,,
\end{equation}
From the factorization condition \eqref{MF2} we easily find polynomial $\pi_{xy}$:
\begin{equation}
\pi_{xy} = \frac{x^{n+1}-y^{n+1}}{x-y}\,.
\end{equation}

To a 4-valent vertex we assign two odd variables $\theta_1$, $\theta_2$ and the following \emph{local} MF operator:
	\begin{equation}
	\begin{aligned}\label{4-valentQ}
		&\begin{array}{c}
			\begin{tikzpicture}[scale=0.7]
				\draw[thick, -stealth] (-0.5,-0.5) -- (0.5,0.5);
				\draw[thick, -stealth] (0.5,-0.5) -- (-0.5,0.5);
				\draw[fill=\myred] (0,0) circle (0.1);
				\node[above left] at (-0.5,0.5) {$\scriptstyle x_1$};
				\node[above right] at (0.5,0.5) {$\scriptstyle x_2$};
				\node[below right] at (0.5,-0.5) {$\scriptstyle x_3$};
				\node[below left] at (-0.5,-0.5) {$\scriptstyle x_4$};
				\node[left] at (-0.1,0) {$\scriptstyle \theta_1$};
				\node[right] at (0.1,0) {$\scriptstyle \theta_2$};
			\end{tikzpicture}
		\end{array}\quad\Longrightarrow\quad \MF_{\rm 4v}\left[\begin{array}{cc}
		x_1 & x_2\\
		x_4 & x_3\\
		\end{array}\right]=u_1\hat \theta_1+(x_1+x_2-x_3-x_4)\hat\theta_1^{\dagger}+u_2\hat \theta_2+(x_1x_2-x_3x_4)\hat\theta_2^{\dagger}\,
\end{aligned}
	\end{equation}
\begin{equation}\label{4vsq}
 \MF_{\rm 4v}^2
=x_1^{n+1}+x_2^{n+1}-x_3^{n+1}-x_4^{n+1},\quad\mbox{here \ \  }{\rm qdeg}\,\theta_1= 1-n,\; {\rm qdeg}\,\theta_2= 3-n\,,
\end{equation}
Necessary expressions for $u_{1,2}$ could be easily derived following polynomial factorization procedure discussed in App.~\ref{app:Quad_quot}.
Note that neither $\frac{x_1^{n+1}+x_2^{n+1}-x_3^{n+1}-x_4^{n+1}}{x_1+x_2-x_3-x_4}$
nor   $\frac{x_1^{n+1}+x_2^{n+1}-x_3^{n+1}-x_4^{n+1}}{x_1x_2-x_3x_4}$
are polynomials, therefore MF in this case is a more sophisticated construction and involves two terms in (\ref{MF2})
-- and thus two Grassmann parameters $\theta_{1,2}$.
For concreteness, we remind explicit solutions derived in \cite{KRI}:
\begin{equation}\label{us}
	\begin{aligned}
	 &u_1(x_1,x_2,x_3,x_4)=\frac{g(x_1+x_2,x_1x_2)-g(x_3+x_4,x_1x_2)}{x_1+x_2-x_3-x_4}\,,\\
	 &u_2(x_1,x_2,x_3,x_4)=\frac{g(x_3+x_4,x_1x_2)-g(x_3+x_4,x_3x_4)}{x_1x_2-x_3x_4}\,, \\
 &g(p,q)=\left(\frac{1}{2}\left(p-\sqrt{p^2-4q}\right)\right)^{n+1}+\left(\frac{1}{2}\left(p+\sqrt{p^2-4q}\right)\right)^{n+1} \,.
 	\end{aligned}
\end{equation}

Quite often we will omit denoting fermion variables on diagrams and 2-valent vertices as well if there is no abuse of notations.
If we label an edge with two or more different even variables this implies that there are 2-valent vertices in between them.

The total MF operator is constructed as a sum of all local MF operators associated to 2-valent and 4-valent vertices:
	\begin{equation}
		\MF_\MOY=\sum\lm_{v\in{\rm vertices}}\MF_v\,.
	\end{equation}

As MF operators for different vertices depend on sets of differential operators $\hat \theta$ and $\hat\theta^{\dagger}$ assigned solely to a particular vertex, and therefore independent across vertices, we conclude that for distinct vertices $v$ and $v'$ MF operators anti-commute:
	\begin{equation}
		\left\{\MF_v,\MF_{v'}\right\}=0\,.
	\end{equation}
	Then for a generic tangle $T$ with open ends labeled by $x_k$ we find:
	\begin{equation}\label{square}
		\MF\left[\begin{array}{c}
			\begin{tikzpicture}[scale=0.8]
				\foreach \p in {0, ..., 9}
				{
					\begin{scope}[rotate = \p*36]
						\draw[thick] (0,0) -- (1,0);
					\end{scope}
				}
				\foreach \p in {2, ..., 8}
				{
					\begin{scope}[rotate = \p*36]
						\begin{scope}[rotate = 10]
							\node at (0.8,0) {.};
						\end{scope}
					\end{scope}
				}
				\begin{scope}[rotate = 10]
					\node at (0.8,0) {$\scriptstyle x_1$};
				\end{scope}
				\begin{scope}[rotate = 46]
					\node at (0.8,0) {$\scriptstyle x_2$};
				\end{scope}
				\begin{scope}[rotate = -26]
					\node at (0.8,0) {$\scriptstyle x_n$};
				\end{scope}
				\draw[fill=white] (0,0) circle (0.4);
				\node at (0,0) {$\scriptstyle \MOY$};
			\end{tikzpicture}
		\end{array}\right]^2=\sum\lm_{v\in{\rm vertices}}\MF_v^2=\sum\lm_k s_k x_k^{n+1}\,,
	\end{equation}
	where a sign $s_k$ is $+1$ if an edge $k$ is flowing outward $\MOY$ and $-1$ if it flows inward $\MOY$.
	It is simple to prove this relation inductively.
	Apparently, it is correct for basic 2-valent \eqref{2vsq} and 4-valent \eqref{4vsq} vertices.
	Now as a induction step let us consider two arbitrary diagrams $\MOY_1$ and $\MOY_2$ for which \eqref{square} is satisfied and join them into a diagram $\MOY_1\cup \MOY_2$ by edges labeled by $y_i$. Those edges that remain disconnected we will mark by variables $x_i$:
	\begin{equation}
	\begin{aligned}
		&\begin{array}{c}
		\begin{tikzpicture}[scale=0.8]
		\draw[thick] (0,0) -- (3,0) (0,0) to[out=36,in=180] (1.5,0.6) to[out=0,in=144] (3,0) (0,0) to[out=-36,in=180] (1.5,-0.6) to[out=0,in=216] (3,0);
		\node[above] at (1.5,0.6) {$\scriptstyle y_1$};
		\node[above] at (1.5,-0.1) {$\scriptstyle \cdot$};
		\node[below] at (1.5,-0.6) {$\scriptstyle y_K$};
		\foreach \p in {2, ..., 8}
		{
			\begin{scope}[rotate = \p*36]
			\draw[thick] (0,0) -- (1,0);
			\end{scope}
		}
		\foreach \p in {4, ..., 7}
		{
			\begin{scope}[rotate = \p*36]
			\begin{scope}[rotate = 10]
			\node at (0.8,0) {.};
			\end{scope}
			\end{scope}
		}
		\begin{scope}[rotate = 82]
		\node at (0.8,0) {$\scriptstyle x_1$};
		\end{scope}
		\begin{scope}[rotate = 118]
		\node at (0.8,0) {$\scriptstyle x_2$};
		\end{scope}
		\begin{scope}[rotate = -57]
		\node at (0.8,0) {$\scriptstyle x_N$};
		\end{scope}
		\draw[fill=white] (0,0) circle (0.4);
		\node at (0,0) {$\scriptstyle \MOY_1$};
		\begin{scope}[shift={(3,0)}]
			\begin{scope}[rotate=180]
			\foreach \p in {2, ..., 8}
			{
				\begin{scope}[rotate = \p*36]
				\draw[thick] (0,0) -- (1,0);
				\end{scope}
			}
			\foreach \p in {4, ..., 7}
			{
				\begin{scope}[rotate = \p*36]
				\begin{scope}[rotate = 10]
				\node at (0.8,0) {.};
				\end{scope}
				\end{scope}
			}
			\begin{scope}[rotate = 87]
			\node at (1.2,0) {$\scriptstyle x_{N+1}$};
			\end{scope}
			\begin{scope}[rotate = 123]
			\node at (1.3,0) {$\scriptstyle x_{N+2}$};
			\end{scope}
			\begin{scope}[rotate = -82]
			\node at (1.2,0) {$\scriptstyle x_{N+M}$};
			\end{scope}
			\draw[fill=white] (0,0) circle (0.4);
			\node at (0,0) {$\scriptstyle \MOY_2$};
			\end{scope}
		\end{scope}
		\end{tikzpicture}
		\end{array}\,,\\
		& \MF_{\MOY_1}^2=\sum\lm_{i=1}^Ns_ix_i^{n+1}+\sum\lm_{i=1}^K \sigma_i y_i^{n+1},\quad \MF_{\MOY_2}^2=\sum\lm_{i=N+1}^{M}s_ix_i^{n+1}-\sum\lm_{i=1}^K \sigma_i y_i^{n+1}\,,\\
		&\MF_{\MOY_1\cup\MOY_2}^2= \MF_{\MOY_1}^2+\MF_{\MOY_2}^2=\sum\lm_{i=1}^{N+M}s_ix_i^{n+1}\,.
	\end{aligned}
	\end{equation}
	The signs in front of $y$-terms for $L_1$ and $L_2$ are opposite since if for $L_1$ this is an incoming edge then for $L_2$ this is an outgoing edge and vice versa.
	And, in general, contributions of internal edges cancel  each other for the MF operator squared.

It is easy to see that for any \emph{closed} MOY diagram $\bar\Gamma$ without external edges
	\begin{equation}
		\overline{\MF}_{\bar \Gamma}^2=0\,,
	\end{equation}
	so that the vertical morphisms in \eqref{KRbicompl} are differentials indeed, and the double complex makes sense.

\subsection{MOY local equivalences}

	We would like to consider in what follows pictorial equivalences between generic MOY graphs with open legs at the boundary $T_1\cong T_2$.
	Precisely, by such an equivalence we imply that two tangles can be implanted in an \emph{arbitrary} graph $G$, so that both $T_1+G$ and $T_2+G$ are closed MOY diagrams $\bar\MOY_1$ and $\bar\MOY_2$.
	This prescription is equivalent to saying that external legs are connected to a multi-valent vertex $G$ at infinity.
	We say that $T_1\cong T_2$ when respective vertical cohomologies are isomorphic for any choice of $G$:
	\begin{equation}
		\begin{aligned}
			&\begin{array}{c}
				\begin{tikzpicture}[scale=0.8]
					\foreach \p in {0, ..., 9}
					{
						\begin{scope}[rotate = \p*36]
							\draw[thick] (0,0) -- (1,0);
						\end{scope}
					}
					\foreach \p in {2, ..., 8}
					{
						\begin{scope}[rotate = \p*36]
							\begin{scope}[rotate = 10]
								\node at (0.8,0) {.};
							\end{scope}
						\end{scope}
					}
					\begin{scope}[rotate = 10]
						\node at (0.8,0) {$\scriptstyle x_1$};
					\end{scope}
					\begin{scope}[rotate = 46]
						\node at (0.8,0) {$\scriptstyle x_2$};
					\end{scope}
					\begin{scope}[rotate = -26]
						\node at (0.8,0) {$\scriptstyle x_n$};
					\end{scope}
					\draw[fill=white] (0,0) circle (0.4);
					\node at (0,0) {$\scriptstyle T_1$};
				\end{tikzpicture}
			\end{array} \cong \begin{array}{c}
			\begin{tikzpicture}[scale=0.8]
				\foreach \p in {0, ..., 9}
				{
					\begin{scope}[rotate = \p*36]
						\draw[thick] (0,0) -- (1,0);
					\end{scope}
				}
				\foreach \p in {2, ..., 8}
				{
					\begin{scope}[rotate = \p*36]
						\begin{scope}[rotate = 10]
							\node at (0.8,0) {.};
						\end{scope}
					\end{scope}
				}
				\begin{scope}[rotate = 10]
					\node at (0.8,0) {$\scriptstyle x_1$};
				\end{scope}
				\begin{scope}[rotate = 46]
					\node at (0.8,0) {$\scriptstyle x_2$};
				\end{scope}
				\begin{scope}[rotate = -26]
					\node at (0.8,0) {$\scriptstyle x_n$};
				\end{scope}
				\draw[fill=white] (0,0) circle (0.4);
				\node at (0,0) {$\scriptstyle T_2$};
			\end{tikzpicture}
			\end{array} \quad \Longrightarrow\quad \begin{array}{c}
			\begin{tikzpicture}[scale=0.8]
				\foreach \p in {0, ..., 9}
				{
					\begin{scope}[rotate = \p*36]
						\draw[thick] (0,0) -- (1,0);
					\end{scope}
				}
				\foreach \p in {2, ..., 8}
				{
					\begin{scope}[rotate = \p*36]
						\begin{scope}[rotate = 10]
							\node at (0.8,0) {.};
						\end{scope}
					\end{scope}
				}
				\begin{scope}[rotate = 10]
					\node at (0.8,0) {$\scriptstyle x_1$};
				\end{scope}
				\begin{scope}[rotate = 46]
					\node at (0.8,0) {$\scriptstyle x_2$};
				\end{scope}
				\begin{scope}[rotate = -26]
					\node at (0.8,0) {$\scriptstyle x_n$};
				\end{scope}
				\draw[fill=white] (0,0) circle (0.4);
				\node at (0,0) {$\scriptstyle T_1$};
				\draw[fill=\myblue, even odd rule] (0,0) circle (1) (0,0) circle (1.3);
				\begin{scope}[rotate = 45]
					\node[\myblue] at (1.5,0) {$\scriptstyle G$};
				\end{scope}
			\end{tikzpicture}
			\end{array} \cong \begin{array}{c}
			\begin{tikzpicture}[scale=0.8]
				\foreach \p in {0, ..., 9}
				{
					\begin{scope}[rotate = \p*36]
						\draw[thick] (0,0) -- (1,0);
					\end{scope}
				}
				\foreach \p in {2, ..., 8}
				{
					\begin{scope}[rotate = \p*36]
						\begin{scope}[rotate = 10]
							\node at (0.8,0) {.};
						\end{scope}
					\end{scope}
				}
				\begin{scope}[rotate = 10]
					\node at (0.8,0) {$\scriptstyle x_1$};
				\end{scope}
				\begin{scope}[rotate = 46]
					\node at (0.8,0) {$\scriptstyle x_2$};
				\end{scope}
				\begin{scope}[rotate = -26]
					\node at (0.8,0) {$\scriptstyle x_n$};
				\end{scope}
				\draw[fill=white] (0,0) circle (0.4);
				\node at (0,0) {$\scriptstyle T_2$};
				\draw[fill=\myblue, even odd rule] (0,0) circle (1) (0,0) circle (1.3);
				\begin{scope}[rotate = 45]
					\node[\myblue] at (1.5,0) {$\scriptstyle G$};
				\end{scope}
			\end{tikzpicture}
			\end{array},\quad \forall G\,,\\
			&\MF_{\bar\MOY_i}=\MF_{T_i}(x_1,\ldots,x_n)+\MF_G(x_1,\ldots,x_n),\quad \left\{\MF_G,\MF_{T_i}\right\}=0,\quad -\MF_G^2=\MF_{T_1}^2=\MF_{T_2}^2\,,\\
			& H^*\left(\MF_{\bar\MOY_1}\right)\cong H^*\left(\MF_{\bar\MOY_2}\right),\quad \forall G\,.
		\end{aligned}
	\end{equation}
Naturally, we will use direct sum and tensor product signs for pictures to indicate that respective cohomologies decompose as vector spaces also for arbitrary $G$.

MF operators $\MF$ exhibit a list of local equivalences known as MOY moves.
We add to this list a relation allowing one to remove or to put a 2-valent vertex to any place in the MOY graph:
	\begin{subequations}
		\begin{align}
			&\label{MOYloc_I}\begin{array}{c}
					\begin{tikzpicture}
						\draw[thick, -stealth] (0,0) -- (1,0);
						\node[above] at (0,0) {$\scriptstyle x$};
						\node[above] at (1,0) {$\scriptstyle y$};
						\draw[fill=\mygreen] (0.5,0) circle (0.1);
					\end{tikzpicture}
				\end{array}\cong \begin{array}{c}
				\begin{tikzpicture}
					\draw[thick, -stealth] (0,0) -- (1,0);
					\node[above] at (0.5,0) {$\scriptstyle x$};
				\end{tikzpicture}
				\end{array}\cong \begin{array}{c}
				\begin{tikzpicture}
					\draw[thick, -stealth] (0,0) -- (1,0);
					\node[above] at (0.5,0) {$\scriptstyle y$};
				\end{tikzpicture}
				\end{array}\,,\\
			&\label{MOYloc_II}\begin{array}{c}
				\begin{tikzpicture}
					\draw[thick,postaction={decorate},decoration={markings, mark= at position 0.5 with {\arrow{stealth}}}] (0,0) circle (0.35);
					\node[right] at (0.35,0) {$\scriptstyle x$};
				\end{tikzpicture}
			\end{array}\cong V_{n},\quad{\rm ind}\,V_n=[n]_q\,,
		\end{align}
		\begin{align}
			&\label{MOYloc_III}\begin{array}{c}
				\begin{tikzpicture}[scale=0.7]
					\draw[thick,stealth-,postaction={decorate},decoration={markings, mark= at position 0.7 with {\arrow{stealth}}}] (0,1) -- (0,0.5) to[out=180,in=90] (-0.5,0) to[out=270,in=180] (0,-0.5) -- (0,-1);
					\draw[fill=\myred] (-0.1,0.5) to[out=90,in=180] (0,0.6) to[out=0,in=90] (0.6,0) to[out=270,in=0] (0,-0.6) to[out=180,in=270] (-0.1,-0.5) to[out=90,in=180] (0,-0.4) to[out=0,in=270] (0.4,0) to[out=90,in=0] (0,0.4) to[out=180,in=270] (-0.1,0.5);
					\node[right] at (0,1) {$\scriptstyle x$};
					\node[right] at (0,-1) {$\scriptstyle z$};
					\node[left] at (-0.5,0) {$\scriptstyle y$};
				\end{tikzpicture}
			\end{array}\cong\begin{array}{c}
				\begin{tikzpicture}
					\draw[thick, -stealth] (0,-0.5) -- (0,0.5);
					\node[right] at (0,0.5) {$\scriptstyle x$};
					\node[right] at (0,-0.5) {$\scriptstyle z$};
				\end{tikzpicture}
			\end{array}\otimes V_{n-1}, \quad {\rm ind}\,V_{n-1}=[n-1]_q\,,\\
			&\label{MOYloc_IV}\begin{array}{c}
				\begin{tikzpicture}[scale=0.7]
					\begin{scope}[shift={(0,1)}]
						\draw[thick,-stealth] (-0.5,-0.5) to[out=90,in=270] (0.5,0.5);
						\draw[thick,-stealth] (0.5,-0.5) to[out=90,in=270] (-0.5,0.5);
						\draw[fill=\myred] (0,0) circle (0.1);
					\end{scope}
					\draw[thick,-stealth] (-0.5,-0.5) to[out=90,in=270] (0.5,0.5) -- (0.5,0.6);
					\draw[thick,-stealth] (0.5,-0.5) to[out=90,in=270] (-0.5,0.5) -- (-0.5,0.6);
					\draw[fill=\myred] (0,0) circle (0.1);
					\node[left] at (-0.5,-0.5) {$\scriptstyle x_4$};
					\node[right] at (0.5,-0.5) {$\scriptstyle x_3$};
					\node[left] at (-0.5,0.5) {$\scriptstyle y$};
					\node[right] at (0.5,0.5) {$\scriptstyle z$};
					\node[left] at (-0.5,1.5) {$\scriptstyle x_1$};
					\node[right] at (0.5,1.5) {$\scriptstyle x_2$};
				\end{tikzpicture}
			\end{array}\cong \begin{array}{c}
				\begin{tikzpicture}[scale=0.7]
					\draw[thick,-stealth] (-0.5,-0.5) to[out=90,in=270] (0.5,0.5);
					\draw[thick,-stealth] (0.5,-0.5) to[out=90,in=270] (-0.5,0.5);
					\draw[fill=\myred] (0,0) circle (0.1);
					\node[left] at (-0.5,-0.5) {$\scriptstyle x_4$};
					\node[right] at (0.5,-0.5) {$\scriptstyle x_3$};
					\node[left] at (-0.5,0.5) {$\scriptstyle x_1$};
					\node[right] at (0.5,0.5) {$\scriptstyle x_2$};
				\end{tikzpicture}
			\end{array}\otimes V_2,\quad {\rm ind}\,V_2=[2]_q\,,\\
			&\label{MOYloc_V}\begin{array}{c}
				\begin{tikzpicture}[scale=0.7]
					\draw[thick,postaction={decorate},decoration={markings, mark= at position 0.7 with {\arrow{stealth}}}] (-1,-1) -- (-0.5,-0.5);
					\draw[thick,postaction={decorate},decoration={markings, mark= at position 0.7 with {\arrow{stealth}}}] (1,1) -- (0.5,0.5);
					\draw[thick,postaction={decorate},decoration={markings, mark= at position 0.7 with {\arrow{stealth}}}] (-0.5,0.5) -- (-1,1);
					\draw[thick,postaction={decorate},decoration={markings, mark= at position 0.7 with {\arrow{stealth}}}] (0.5,-0.5) -- (1,-1);
					\draw[thick,postaction={decorate},decoration={markings, mark= at position 0.7 with {\arrow{stealth}}}] (-0.5,0.5) -- (0.5,0.5);
					\draw[thick,postaction={decorate},decoration={markings, mark= at position 0.7 with {\arrow{stealth}}}] (0.5,-0.5) -- (-0.5,-0.5);
					\begin{scope}[shift={(-0.5,0)}]
						\draw[fill=\myred] (0.1,0.5) to[out=90,in=0] (0,0.6) to[out=180,in=90] (-0.1,0.5) -- (-0.1,-0.5) to[out=270,in=180] (0,-0.6) to[out=0,in=270] (0.1,-0.5) -- (0.1,0.5);
					\end{scope}
					\begin{scope}[shift={(0.5,0)}]
						\draw[fill=\myred] (0.1,0.5) to[out=90,in=0] (0,0.6) to[out=180,in=90] (-0.1,0.5) -- (-0.1,-0.5) to[out=270,in=180] (0,-0.6) to[out=0,in=270] (0.1,-0.5) -- (0.1,0.5);
					\end{scope}
					\node[left] at (-1,-1) {$\scriptstyle x_1$};
					\node[left] at (-1,1) {$\scriptstyle x_2$};
					\node[right] at (1,1) {$\scriptstyle x_3$};
					\node[right] at (1,-1) {$\scriptstyle x_4$};
					\node[above] at (0,0.5) {$\scriptstyle x_5$};
					\node[below] at (0,-0.5) {$\scriptstyle x_6$};
				\end{tikzpicture}
			\end{array}\;\cong\;\begin{array}{c}
				\begin{tikzpicture}
					\draw[thick, stealth-] (-0.5,0.5) to[out=315,in=225] (0.5,0.5);
					\draw[thick, stealth-] (0.5,-0.5) to[out=135,in=45] (-0.5,-0.5);
					\node[left] at (-0.5,-0.5) {$\scriptstyle x_1$};
					\node[left] at (-0.5,0.5) {$\scriptstyle x_2$};
					\node[right] at (0.5,0.5) {$\scriptstyle x_3$};
					\node[right] at (0.5,-0.5) {$\scriptstyle x_4$};
				\end{tikzpicture}
			\end{array} \oplus \left(\begin{array}{c}
				\begin{tikzpicture}
					\draw[thick, stealth-] (-0.5,0.5) to[out=315,in=45] (-0.5,-0.5);
					\draw[thick, stealth-] (0.5,-0.5) to[out=135,in=225] (0.5,0.5);
					\node[left] at (-0.5,-0.5) {$\scriptstyle x_1$};
					\node[left] at (-0.5,0.5) {$\scriptstyle x_2$};
					\node[right] at (0.5,0.5) {$\scriptstyle x_3$};
					\node[right] at (0.5,-0.5) {$\scriptstyle x_4$};
				\end{tikzpicture}
			\end{array} \otimes V_{n-2}\right),\quad {\rm ind}\,V_{n-2}=[n-2]_q\,,\\
			&\label{MOYloc_VI}\begin{array}{c}
				\begin{tikzpicture}[scale=0.6]
					\draw[thick,-stealth] (0,0) to[out=90,in=225] (1.5,1.5) to[out=45,in=270] (2,3);
					\draw[thick,-stealth] (2,0) to[out=90,in=315] (1.5,1.5) to[out=135,in=270] (0,3);
					\draw[thick,-stealth] (1,0) to[out=90,in=270] (0,1.5) to[out=90,in=270] (1,3);
					\node[left] at (0,0) {$\scriptstyle x_1$};
					\node[left] at (1,0) {$\scriptstyle x_2$};
					\node[right] at (2,0) {$\scriptstyle x_3$};
					\node[left] at (0,3) {$\scriptstyle x_4$};
					\node[left] at (1,3) {$\scriptstyle x_5$};
					\node[right] at (2,3) {$\scriptstyle x_6$};
					\draw[fill=\myred] (1.5,1.5) circle (0.13) (0.45,0.78) circle (0.13) (0.45,2.22) circle (0.13);
				\end{tikzpicture}
			\end{array}\oplus\begin{array}{c}
			\begin{tikzpicture}[scale=0.6]
				\draw[thick,-stealth] (0,0) -- (0,3);
				\draw[thick,-stealth] (1,0) to[out=90,in=270] (2,3);
				\draw[thick,-stealth] (2,0) to[out=90,in=270] (1,3);
				\node[left] at (0,0) {$\scriptstyle x_1$};
				\node[left] at (1,0) {$\scriptstyle x_2$};
				\node[right] at (2,0) {$\scriptstyle x_3$};
				\node[left] at (0,3) {$\scriptstyle x_4$};
				\node[left] at (1,3) {$\scriptstyle x_5$};
				\node[right] at (2,3) {$\scriptstyle x_6$};
				\draw[fill=\myred] (1.5,1.5) circle (0.13);
			\end{tikzpicture}
			\end{array}\cong \begin{array}{c}
			\begin{tikzpicture}[scale=0.6,xscale=-1]
				\draw[thick,-stealth] (0,0) to[out=90,in=225] (1.5,1.5) to[out=45,in=270] (2,3);
				\draw[thick,-stealth] (2,0) to[out=90,in=315] (1.5,1.5) to[out=135,in=270] (0,3);
				\draw[thick,-stealth] (1,0) to[out=90,in=270] (0,1.5) to[out=90,in=270] (1,3);
				\node[right] at (0,0) {$\scriptstyle x_3$};
				\node[left] at (1,0) {$\scriptstyle x_2$};
				\node[left] at (2,0) {$\scriptstyle x_1$};
				\node[right] at (0,3) {$\scriptstyle x_6$};
				\node[left] at (1,3) {$\scriptstyle x_5$};
				\node[left] at (2,3) {$\scriptstyle x_4$};
				\draw[fill=\myred] (1.5,1.5) circle (0.13) (0.45,0.78) circle (0.13) (0.45,2.22) circle (0.13);
			\end{tikzpicture}
			\end{array}\oplus\begin{array}{c}
			\begin{tikzpicture}[scale=0.6,xscale=-1]
				\draw[thick,-stealth] (0,0) -- (0,3);
				\draw[thick,-stealth] (1,0) to[out=90,in=270] (2,3);
				\draw[thick,-stealth] (2,0) to[out=90,in=270] (1,3);
				\node[right] at (0,0) {$\scriptstyle x_3$};
				\node[left] at (1,0) {$\scriptstyle x_2$};
				\node[left] at (2,0) {$\scriptstyle x_1$};
				\node[right] at (0,3) {$\scriptstyle x_6$};
				\node[left] at (1,3) {$\scriptstyle x_5$};
				\node[left] at (2,3) {$\scriptstyle x_4$};
				\draw[fill=\myred] (1.5,1.5) circle (0.13);
			\end{tikzpicture}
			\end{array}\,.
		\end{align}
	\end{subequations}

Proofs of these relations are rather straightforward and we will not review all of them here.
A proof for \eqref{MOYloc_I} is given in Sec.~\ref{sec:stray}.
A proof for \eqref{MOYloc_V} is discussed in Sec.~\ref{sec:biparti}.
We do not consider a proof for \eqref{MOYloc_III} in this note, however it follows directly from the proof of invariance with respect to the Reidemeister I move in Sec.~\ref{sec:invariant}.

The MOY moves are enough to reduce \emph{any} MOY graph to its $q$-dimension computed via RT formalism.
	This theorem is proven in \cite{KaufVog} when edges of MOY diagrams are colored with $\Box$ and $\wedge^2\Box$ only, and it is proven in \cite{Wu} for coloring by generic anti-symmetric rep $\wedge^k\Box$.\footnote{
		For the particular case of $n=3$ the proof is rather elementary.
		In this case $\wedge^2\Box\cong \bar\Box$ and MOY diagrams become simply 3-valent bipartite graphs implying that all the faces are $2m$-gons.
		For such a graph drawn on a sphere according to the Euler theorem we have $F-E+V=2$ where $F$, $E$ and $V$ are numbers of faces, edges and vertices respectively.
		For trivalent graphs $V=\frac{2}{3}E$, so $F=2+\frac{1}{3}E$.
		If all the faces of the graph where $2m$-gons then for the number of faces we would have $F=\frac{1}{m}E$.
		If there are no faces with $m<3$ then $F\leq \frac{1}{3}E$ that contradicts $F=2+\frac{1}{3}E$.
		The latter statement implies that for any MOY graph for $n=3$ we are able to find a face that is a 2-gon or a 4-gon.
		So each MOY graph for $n=3$ could be reduced to a number value via a sequence of moves \eqref{MOYloc_II}-\eqref{MOYloc_V}.
	}
	For us this theorem implies immediately:\footnote{Here we call $q$-dimension an ``index'' following the physical intuition.
	As we expect that parameter $t$ plays the role of fugacity for the fermion number, and a specification $t=-1$ refers to the supersymmetric Witten index \cite{Aganagic:2011sg,Galakhov:2016cji}.
	}
	\begin{equation}\label{MOYindex}
		{\rm ind}\,[\MOY]:=\sum\lm_{z}q^z\;{\rm dim}\,H^z\left(\MF[\MOY]\right)={\rm qdim}\,[\MOY]\,.
	\end{equation}

Relation \eqref{MOYindex} implies in turn that the Euler characteristic of the KR double complex given by a specific value $P_{\bar L}(q,-1)$ of the Poincar\'e polynomial coincides indeed with the respective HOMFLY polynomial.


\subsection{Solving MFs}

\subsubsection{Localization to quotient rings}

Let us start with just a few words about rather efficient methodology to construct cohomologies of MF-like operators.
We separate one of odd variables $\theta$ and denote the rest of even and odd variables as formal letter $V$.
Consider the following odd MF-like operator:
\begin{equation}\label{MF-like}
	M=M_0(V)+A(V)\hat \theta+B(V)\hat\theta^{\dagger}\,,
\end{equation}
Here we do not require that $M_0$ has the form of a matrix factorization operator, there is no requirement for it to be linear in other odd variables either.
However it is expected that $A$, $B$ and $M_0$ commute mutually.
It is also natural to split respective dependence on $\theta$ in wave functions $\Psi=\Psi_0(V)+\theta\Psi_1(V)$.
The action of operator $M$ on $\Psi$ reads:
\begin{equation}\label{zero-modes}
	M \Psi=\left(M_0(V)\Psi_0(V)+B(V)\Psi_1(V)\right)+\theta(-M_0(V)\Psi_1(V)+A(V)\Psi_0(V))\,.
\end{equation}
So for zero modes of $M$ we see that the wave function might have only one independent component, say, $\Psi_0$ then $\Psi_1=-B^{-1}M_0\Psi_0$.
On the other hand any $B$-generated ideal in the first component lies in the image of $M$:
\begin{equation}
	B(V)X(V)-\theta M_0(V)X(V)=M\left(\theta X(V)\right),\quad \forall X\,.
\end{equation}
So we conclude:
\begin{tcolorbox}
	\centering
	Cohomologies of MF-like operator \eqref{MF-like} localize to quotient ring $S:=\IQ[V]/\left(B(V)\IQ[V]\right)$.
\end{tcolorbox}
Actually, by localizing the $\theta^0$-component of \eqref{zero-modes} to $S$ we would derive a localized equation for zero modes $M_0\big|_{S}\Psi_0\big|_{S}=0$.
To acquire the localized operator $M_0\big|_{S}$ one should resolve constraint $B=0$ and substitute it back in $M_0$ that will \emph{decrease} the number of free variables and has a good chance to make it \emph{simpler}.
So our general strategy to define cohomologies of MF operators is to perform a chain of quotient ring localizations until the localized zero-mode equation becomes trivial to solve.

\subsubsection{Blinking 2-valent vertex} \label{sec:stray}

Relations~\eqref{MOYloc_I}--\eqref{MOYloc_VI} are conceptually very simple.
Still, their operator realization is somewhat verbose.
This is a well-known phenomenon in homological algebra, but not quite usual for physically oriented literature.
Operator language is already a considerable simplification, because operators themselves are simple and intuitive.
Still, the second step -- equivalence of their zero modes, moreover, of their universality classes -- remains
somewhat tedious and lengthy.
Clearly, one should look for a better and more concise terminology and reasoning.
From this point of view quite representative is already the most trivial relation~\eqref{MOYloc_I},
for which we need at least a 2-page derivation, despite the obvious simplicity of the operators~\eqref{MF-4.3.2}
and relative simplicity of the main claim~\eqref{dress1}.
This example is not quite exhaustive because of the lack of the second resolution.
However, as we will see in Sec.~\ref{sec:invariant}, this does not lead to serious complication --
after the material of the present subsection is consumed and thought through.

So, we would like to prove relation \eqref{MOYloc_I}.
Consider two closed MOY graphs $\bar \MOY_{1,2}$:
\begin{equation}
	\bar\MOY_1=\begin{array}{c}
		\begin{tikzpicture}[scale=0.7]
			\draw[thick, postaction={decorate},decoration={markings, mark = at position 0.25 with {\arrowreversed{stealth}}, mark = at position 0.75 with {\arrowreversed{stealth}}}] (0,0) to[out=45,in=180] (0.8,0.5) to[out=0,in=90] (1.2,0) to[out=270,in=0] (0.8,-0.5) to[out=180,in=315] (0,0);
			\draw[fill=\myblue] (0,0) circle (0.5);
			\node[white] at (0,0) {$\scriptstyle \MOY_0$};
			\node[right] at (1,0.5) {$\scriptstyle x$};
			\node[right] at (1,-0.5) {$\scriptstyle y$};
			\draw[fill=\mygreen] (1.2,0) circle (0.1);
		\end{tikzpicture}
	\end{array},\quad \bar\MOY_2=\begin{array}{c}
	\begin{tikzpicture}[scale=0.7]
		\draw[thick, postaction={decorate},decoration={markings, mark = at position 0.4 with {\arrowreversed{stealth}}}] (0,0) to[out=45,in=180] (0.8,0.5) to[out=0,in=90] (1.2,0) to[out=270,in=0] (0.8,-0.5) to[out=180,in=315] (0,0);
		\draw[fill=\myblue] (0,0) circle (0.5);
		\node[white] at (0,0) {$\scriptstyle \MOY_0$};
		\node[right] at (1.2,0) {$\scriptstyle x$};
	\end{tikzpicture}
	\end{array}\,.
\end{equation}
We assume that graphs differ by the fact that on some edge there is a 2-valent vertex with associated variable $\theta$ dividing an edge marked with variable $x$ in $\bar\MOY_2$ into two edges marked by two variables $x$ and $y$ in $\bar\MOY_1$.
Clearly -- and formally, according to \eqref{MOYloc_I} -- the insertion of the 2-valent green vertex changes nothing,
we call it ``blinking''.
However, to validate this fact, we need to perform  a whole exercise --
we would like to show that cohomologies of $\MF_{\bar\MOY_{1,2}}$ are isomorphic:
\begin{equation}\label{strayiso}
	\begin{array}{c}
		\begin{tikzpicture}
		\node(A) at (0,0) {$H^*(\MF_{\bar\MOY_1})$};
		\node(B) at (3,0) {$H^*(\MF_{\bar\MOY_2})$};
		\draw[-stealth] ([shift={(0,0.05)}]A.east) -- ([shift={(0,0.05)}]B.west) node[pos=0.5,above] {$\scriptstyle \Phi_1$};
		\draw[stealth-] ([shift={(0,-0.05)}]A.east) -- ([shift={(0,-0.05)}]B.west) node[pos=0.5,below] {$\scriptstyle \Phi_2$};
		\end{tikzpicture}
	\end{array}\,.
\end{equation}

Let us mark the set of all variables corresponding to vertices and internal edges of graph $\MOY_0$ as $V$.
Then for MF operators we have:
\begin{equation}\label{MF-4.3.2}
	\MF_{\bar\MOY_1}=\MF_{\MOY_0}(V,x,y)+\pi_{xy}\hat\theta+(x-y)\hat\theta^{\dagger},\quad \MF_{\bar\MOY_2}=\MF_{\MOY_0}(V,x,x)\,.
\end{equation}
In this terms we can explicitly decompose the first and the second space of polynomials as $\Psi_0(V,x,y)+\theta\Psi_1(V,x,y)$ and $\psi(V,x)$ respectively.
In these terms isomorphism $\Psi$ acts as:
\begin{equation}
	\begin{aligned}\label{dress1}
		\Phi_1[\Psi_0(V,x,y)+\theta\Psi_1(V,x,y)]&=\Psi_0(V,x,x),\\ \Phi_2[\psi(V,x)]&=\left(1-\theta\frac{\MF_{\MOY_0}(V,x,y)-\MF_{\MOY_0}(V,x,x)}{x-y}\right)\psi(V,x)\,.
	\end{aligned}
\end{equation}

{\bf Proof.}

Let us consider first:
\begin{equation}
	\begin{aligned}
	&\Phi_1\left[\MF_{\bar\MOY_1}\left(\Psi_0(V,x,y)+\theta\Psi_1(V,x,y)\right)\right]=\Phi\left[(\MF_{\MOY_0}(V,x,y)\Psi_0(V,x,y)+(x-y)\Psi_1(V,x,y))+\theta(\ldots)\right]=\\
	&=\MF_{\MOY_0}(V,x,x)\Psi_0(V,x,x)=\MF_{\bar \MOY_2}\Phi_1[\Psi_0(V,x,y)+\theta\Psi_1(V,x,y)]\,.
	\end{aligned}
\end{equation}
In simpler words we have shown that
\begin{equation}
	\Phi_1\circ\MF_{\bar\MOY_1}=\MF_{\bar\MOY_2}\circ\Phi_1\,,
\end{equation}
so $\Phi_1$ is a homomorphism of complexes.

Now consider:
\begin{equation}
	\begin{aligned}
	&\MF_{\bar\Gamma_1}\Phi_2[\psi(V,x)]=\left(\MF_{\MOY_0}(V,x,y)+\pi_{xy}\hat\theta+(x-y)\hat\theta^{\dagger}\right)\left(1-\theta\frac{\MF_{\MOY_0}(V,x,y)-\MF_{\MOY_0}(V,x,x)}{x-y}\right)\psi(V,x)=\\
	&=\theta\left(\MF_{\MOY_0}(V,x,y)\frac{\MF_{\MOY_0}(V,x,y)-\MF_{\MOY_0}(V,x,x)}{x-y}-\underbrace{\frac{\MF_{\MOY_0}(V,x,y)^2}{x-y}}_{=-\pi_{xy}}+\underbrace{\frac{\MF_{\MOY_0}(V,x,x)^2}{x-y}}_{=0}\right)\psi(V,x)+\\
	&+\MF_{\MOY_0}(V,x,x)\psi(V,x)=\Phi_2[\MF_{\bar\Gamma_2}\psi(V,x)]\,.
	\end{aligned}
\end{equation}
So we have shown that $\Phi_2$ is also a homomorphism:
\begin{equation}
	\MF_{\bar\Gamma_1}\circ\Phi_2=\Phi_2\circ\MF_{\bar\Gamma_2}\,.
\end{equation}

It remains to show that $\Phi_1$ is invertible on cohomologies and $\Phi_2$ is its inverse.
Apparently,
\begin{equation}
	\Phi_1[\Phi_2[\psi(V,x)]]=\psi(V,x)\,.
\end{equation}
On the other hand the fact that $\Psi\in {\rm Ker}\,\MF_{\bar\MOY_1}$ imposes on functions two constraints:
\begin{equation}
	\Psi_1(V,x,y)=-\frac{\MF_{\MOY_0}(V,x,y)}{x-y}\Psi_0(V,x,y),\quad \MF_{\MOY_0}(V,x,x)\Psi_0(V,x,x)=0\,.
\end{equation}
Then we derive:
\begin{equation}
	\begin{aligned}
	&\Psi_0(V,x,y)+\theta\Psi_1(V,x,y)=\\
	&=\left(1-\theta\frac{\MF_{\MOY_0}(V,x,y)-\MF_{\MOY_0}(V,x,x)}{x-y}\right)\Psi_0(V,x,x)+\MF_{\bar\MOY_1}\left(\frac{\Psi_0(V,x,y)-\Psi_0(V,x,x)}{x-y}\right)=\\
	&=\Phi_2\left[\Phi_1\left[\Psi_0(V,x,y)+\theta\Psi_1(V,x,y)\right]\right]+\MF_{\bar\MOY_1}(\ldots)\,.
	\end{aligned}
\end{equation}
The latter relation becomes an equality modulo ${\rm Im}\,\MF_{\bar\MOY_1}$  and indicates that $\Phi_1$ is indeed an isomorphism of cohomologies, and $\Phi_2=\Phi_1^{-1}$:
\begin{equation}
	\Phi_1\circ\Phi_2=\bbone_{H^*(\MF_{\bar\MOY_2})},\quad \Phi_2\circ\Phi_1=\bbone_{H^*(\MF_{\bar\MOY_1})}\,.
\end{equation}


\subsubsection{Splitting 4-valent vertex} \label{sec:4v_split}

A rather useful tool for calculating MF cohomologies is a splitting process for 4-valent vertices allowing one to represent cohomologies of the graph with a 4-valent vertex as \emph{constrained} cohomologies of a graph with the vertex eliminated.
Consider the following diagrams:
\begin{equation}
\bar\MOY_1=\begin{array}{c}
	\begin{tikzpicture}
		\draw[thick, postaction={decorate},decoration={markings, mark= at position 0.75 with {\arrow{stealth}}, mark= at position 0.25 with {\arrow{stealth}}}] (-0.5,-0.5) -- (0.5,0.5) node[left,pos=0.3] {$\scriptstyle x_4$} node[right,pos=0.7] {$\scriptstyle x_2$};
		\draw[thick, postaction={decorate},decoration={markings, mark= at position 0.75 with {\arrow{stealth}}, mark= at position 0.25 with {\arrow{stealth}}}] (0.5,-0.5) -- (-0.5,0.5) node[right,pos=0.3] {$\scriptstyle x_3$} node[left,pos=0.7] {$\scriptstyle x_1$};
		\draw[fill=\myred] (0,0) circle (0.08);
		\begin{scope}[rotate=45]
		\node[\myblue] at (1.1,0) {$\scriptstyle \MOY_0$};
		\draw[fill=\myblue, even odd rule] (0,0) circle (0.7)  (0,0) circle (0.9);
		\end{scope}
	\end{tikzpicture}
\end{array},\quad \bar\MOY_+=\begin{array}{c}
\begin{tikzpicture}
	\draw[thick, postaction={decorate},decoration={markings, mark= at position 0.75 with {\arrow{stealth}}, mark= at position 0.25 with {\arrow{stealth}}}] (-0.5,-0.5) to[out=45,in=315] node[left,pos=0.5,shift={(0.1,0)}] {$\scriptstyle x_1$} (-0.5,0.5);
	\draw[thick, postaction={decorate},decoration={markings, mark= at position 0.75 with {\arrow{stealth}}, mark= at position 0.25 with {\arrow{stealth}}}] (0.5,-0.5) to[out=135,in=225] node[right,pos=0.5,shift={(-0.1,0)}] {$\scriptstyle x_2$} (0.5,0.5);
	\begin{scope}[rotate=45]
		\node[\myblue] at (1.1,0) {$\scriptstyle \MOY_0$};
	\end{scope}
	\draw[fill=\myblue, even odd rule] (0,0) circle (0.7)  (0,0) circle (0.9);
\end{tikzpicture}
\end{array},\quad \bar\MOY_-=\begin{array}{c}
\begin{tikzpicture}
	\draw[dashed,thick, postaction={decorate},decoration={markings, mark= at position 0.75 with {\arrow{stealth}}, mark= at position 0.25 with {\arrow{stealth}}}] (-0.5,-0.5) -- (0.5,0.5) node[left,pos=0.3] {$\scriptstyle x_2$} node[right,pos=0.7] {$\scriptstyle x_2$};
	\draw[thick, postaction={decorate},decoration={markings, mark= at position 0.75 with {\arrow{stealth}}, mark= at position 0.25 with {\arrow{stealth}}}] (0.5,-0.5) -- (-0.5,0.5) node[right,pos=0.3] {$\scriptstyle x_1$} node[left,pos=0.7] {$\scriptstyle x_1$};
	\begin{scope}[rotate=45]
		\node[\myblue] at (1.1,0) {$\scriptstyle \MOY_0$};
	\end{scope}
	\draw[fill=\myblue, even odd rule] (0,0) circle (0.7)  (0,0) circle (0.9);
\end{tikzpicture}
\end{array}\mbox{(NO VERTEX!!!)}\,.
\end{equation}
In diagram $\bar\MOY_{\pm}$ we have identified edges of diagram $\MOY_0$ accordingly and identified respective variables.
In diagram $\bar\MOY_-$ edges are of the same type, one of them is simply marked as dashed to indicate that identification is criss-cross without any intersection or vertex.

Again let us denote the set of internal variables for subgraph $\MOY_0$ as $V$.
Respective MF operators read:
\begin{equation}
	\begin{aligned}
		&\MF_{\bar\MOY_1}=\MF_{\MOY_0}(V,x_1,x_2,x_3,x_4)+u_1\hat\theta_1+s_1\hat\theta_1^{\dagger}+u_2\hat\theta_2+s_2\hat\theta_2^{\dagger}\,,\\
		&\MF_{\bar\MOY_+}=\MF_{\MOY_0}(V,x_1,x_2,x_2,x_1),\quad \MF_{\bar\MOY_-}=\MF_{\MOY_0}(V,x_1,x_2,x_1,x_2)\,,
	\end{aligned}
\end{equation}
where $s_1=x_1+x_2-x_3-x_4$, $s_2=x_1x_2-x_3x_4$ and $u_i$ are defined by \eqref{us}.
Consider a set of two dimensional vectors of $(x,y)$-polynomial components constrained to coincide in points $x=y$:
\begin{equation}\label{xi_const}
	\Xi(x,y):=\left\{\left(\begin{array}{c}
		\xi_+(V,x,y)\\
		\xi_-(V,x,y)
	\end{array}\right)\;\bigg|\;\xi_+(V,z,z)=\xi_-(V,z,z), \;\forall z\right\}\,.
\end{equation}

Then we would like to prove that there is the following isomorphism\footnote{At the r.h.s. inside the cohomology, we indicate the space at the first place and the differential acting on this space at the second place.}:
\begin{equation}\label{4viso}
	\begin{array}{c}
		\begin{tikzpicture}
			\node(A) at (0,0) {$H^*(\MF_{\bar\MOY_1})$};
			\node(B) at (5,0) {$H^*\left(\Xi(x_1,x_2),{\rm diag}\left(\MF_{\bar\MOY_+},\MF_{\bar\MOY_-}\right)\right)$};
			\draw[-stealth] ([shift={(0,0.05)}]A.east) -- ([shift={(0,0.05)}]B.west) node[pos=0.5,above] {$\scriptstyle \Phi$};
			\draw[stealth-] ([shift={(0,-0.05)}]A.east) -- ([shift={(0,-0.05)}]B.west) node[pos=0.5,below] {$\scriptstyle \Phi^{-1}$};
		\end{tikzpicture}
	\end{array}\,.
\end{equation}

In particular, we have:
\begin{equation}\label{dress2}
\begin{aligned}
	&\Phi\left[\Psi_0(V,x_1,x_2,x_3,x_4)+\theta_1\Psi_1(V,x_1,x_2,x_3,x_4)+\theta_2\Psi_2(V,x_1,x_2,x_3,x_4)+\theta_1\theta_2\Psi_{12}(V,x_1,x_2,x_3,x_4)\right]=\\
	&=\left(\Psi_0(x_1,x_2,x_2,x_1),\Psi_0(x_1,x_2,x_1,x_2)\right)\,,\\
	&\Phi^{-1}\left[\left(\xi_+(x_1,x_2),\xi_-(x_1,x_2)\right)\right]=\left(1-\theta_1v_1\MF_{\bar\MOY_1}-\theta_2v_2\MF_{\bar\MOY_1}+\theta_1\theta_2\frac{u_2+\MF_{\bar\MOY_1}v_2\MF_{\bar\MOY_1}}{s_1}\right)\times\\
	&\times\left(\frac{1}{2}\left(1-\frac{x_3-x_4}{x_1-x_2}\right)\xi_+(x_1,x_2)+\left(1+\frac{x_3-x_4}{x_1-x_2}\right)\frac{1}{2}\xi_-(x_1,x_2)\right)\,,
\end{aligned}
\end{equation}
where linear maps $v_{1,2}$ are defined in \eqref{v12}.

The {\bf proof} of the statement that $\Phi$ is an isomorphism of cohomologies and $\Phi^{-1}$ is its inverse is totally analogous to the case of a blinking 2-valent vertex (see Sec.~\ref{sec:stray}) and exploits quadratic quotients discussed in App.~\ref{app:Quad_quot}.
We will omit it here and present its details elsewhere.


\section{Horizontal morphisms}\label{sec:horizontal}
Horizontal morphisms are fixed \emph{uniquely} up to homotopy by locality and homomorphism requirement \eqref{MFhomo}.
See also \cite{Carqueville:2011zea}.

To describe horizontal morphisms as ring morphisms it would be more natural to assign variables to MOY diagrams appearing in decomposition \eqref{categR} in a uniform way.
To do so we assign even variables to the edges of the very link diagram in question, and assign a couple of odd variables to both intersection resolutions via a single 4-valent vertex or a pair of 2-valent vertices.
Then we would like to search for horizontal morphisms in terms of a uniform variable choice, there are two types of them:
\begin{equation}\label{categRvar}
	\begin{aligned}
		& \begin{array}{c}
			\begin{tikzpicture}[scale=0.7]
				\draw[thick, -stealth] (0.5,-0.5) -- (-0.5,0.5);
				\draw[white, line width = 1.4mm] (-0.5,-0.5) -- (0.5,0.5);
				\draw[thick, -stealth] (-0.5,-0.5) -- (0.5,0.5);
				\node[above left] at (-0.5,0.5) {$\scriptstyle x_1$};
				\node[above right] at (0.5,0.5) {$\scriptstyle x_2$};
				\node[below right] at (0.5,-0.5) {$\scriptstyle x_3$};
				\node[below left] at (-0.5,-0.5) {$\scriptstyle x_4$};
			\end{tikzpicture}
		\end{array}=\begin{array}{c}
		\begin{tikzpicture}[scale=0.7]
			\draw[thick, -stealth] (-0.5,-0.5) to[out=45,in=270] (-0.2,0) to[out=90,in=315] (-0.5,0.5);
			\draw[thick, -stealth] (0.5,-0.5) to[out=135,in=270] (0.2,0) to[out=90,in=225] (0.5,0.5);
			\draw[fill=\mygreen] (-0.2,0) circle (0.08) (0.2,0) circle (0.08);
			\node[above left] at (-0.5,0.5) {$\scriptstyle x_1$};
			\node[above right] at (0.5,0.5) {$\scriptstyle x_2$};
			\node[below right] at (0.5,-0.5) {$\scriptstyle x_3$};
			\node[below left] at (-0.5,-0.5) {$\scriptstyle x_4$};
			\node[left] at (-0.2,0) {$\scriptstyle \theta_1$};
			\node[right] at (0.2,0) {$\scriptstyle \theta_2$};
		\end{tikzpicture}
		\end{array}+\epsilon^{(+)}\begin{array}{c}
			\begin{tikzpicture}[scale=0.7]
				\draw[thick, -stealth] (-0.5,-0.5) -- (0.5,0.5);
				\draw[thick, -stealth] (0.5,-0.5) -- (-0.5,0.5);
				\draw[fill=\myred] (0,0) circle (0.1);
				\node[above left] at (-0.5,0.5) {$\scriptstyle x_1$};
				\node[above right] at (0.5,0.5) {$\scriptstyle x_2$};
				\node[below right] at (0.5,-0.5) {$\scriptstyle x_3$};
				\node[below left] at (-0.5,-0.5) {$\scriptstyle x_4$};
				\node[left] at (-0.1,0) {$\scriptstyle \theta_1$};
				\node[right] at (0.1,0) {$\scriptstyle \theta_2$};
			\end{tikzpicture}
		\end{array},\quad \chi_0\left[\begin{array}{cc|c}
		x_1 & x_2 & \theta_1\\
		x_4 & x_3 & \theta_2
		\end{array}\right]:\begin{array}{c}
		\begin{tikzpicture}[scale=0.7]
			\draw[thick, -stealth] (-0.5,-0.5) to[out=45,in=270] (-0.2,0) to[out=90,in=315] (-0.5,0.5);
			\draw[thick, -stealth] (0.5,-0.5) to[out=135,in=270] (0.2,0) to[out=90,in=225] (0.5,0.5);
			\draw[fill=\mygreen] (-0.2,0) circle (0.08) (0.2,0) circle (0.08);
			\node[above left] at (-0.5,0.5) {$\scriptstyle x_1$};
			\node[above right] at (0.5,0.5) {$\scriptstyle x_2$};
			\node[below right] at (0.5,-0.5) {$\scriptstyle x_3$};
			\node[below left] at (-0.5,-0.5) {$\scriptstyle x_4$};
			\node[left] at (-0.2,0) {$\scriptstyle \theta_1$};
			\node[right] at (0.2,0) {$\scriptstyle \theta_2$};
		\end{tikzpicture}
		\end{array}\;\longrightarrow\;\begin{array}{c}
		\begin{tikzpicture}[scale=0.7]
			\draw[thick, -stealth] (-0.5,-0.5) -- (0.5,0.5);
			\draw[thick, -stealth] (0.5,-0.5) -- (-0.5,0.5);
			\draw[fill=\myred] (0,0) circle (0.1);
			\node[above left] at (-0.5,0.5) {$\scriptstyle x_1$};
			\node[above right] at (0.5,0.5) {$\scriptstyle x_2$};
			\node[below right] at (0.5,-0.5) {$\scriptstyle x_3$};
			\node[below left] at (-0.5,-0.5) {$\scriptstyle x_4$};
			\node[left] at (-0.1,0) {$\scriptstyle \theta_1$};
			\node[right] at (0.1,0) {$\scriptstyle \theta_2$};
		\end{tikzpicture}
		\end{array}\,;\\
		& \begin{array}{c}
			\begin{tikzpicture}[scale=0.7]
				\draw[thick, -stealth] (-0.5,-0.5) -- (0.5,0.5);
				\draw[white, line width = 1.4mm] (0.5,-0.5) -- (-0.5,0.5);
				\draw[thick, -stealth] (0.5,-0.5) -- (-0.5,0.5);
				\node[above left] at (-0.5,0.5) {$\scriptstyle x_1$};
				\node[above right] at (0.5,0.5) {$\scriptstyle x_2$};
				\node[below right] at (0.5,-0.5) {$\scriptstyle x_3$};
				\node[below left] at (-0.5,-0.5) {$\scriptstyle x_4$};
			\end{tikzpicture}
		\end{array}=\begin{array}{c}
		\begin{tikzpicture}[scale=0.7]
			\draw[thick, -stealth] (-0.5,-0.5) -- (0.5,0.5);
			\draw[thick, -stealth] (0.5,-0.5) -- (-0.5,0.5);
			\draw[fill=\myred] (0,0) circle (0.1);
			\node[above left] at (-0.5,0.5) {$\scriptstyle x_1$};
			\node[above right] at (0.5,0.5) {$\scriptstyle x_2$};
			\node[below right] at (0.5,-0.5) {$\scriptstyle x_3$};
			\node[below left] at (-0.5,-0.5) {$\scriptstyle x_4$};
			\node[left] at (-0.1,0) {$\scriptstyle \theta_1$};
			\node[right] at (0.1,0) {$\scriptstyle \theta_2$};
		\end{tikzpicture}
		\end{array}+\epsilon^{(-)}\begin{array}{c}
		\begin{tikzpicture}[scale=0.7]
			\draw[thick, -stealth] (-0.5,-0.5) to[out=45,in=270] (-0.2,0) to[out=90,in=315] (-0.5,0.5);
			\draw[thick, -stealth] (0.5,-0.5) to[out=135,in=270] (0.2,0) to[out=90,in=225] (0.5,0.5);
			\draw[fill=\mygreen] (-0.2,0) circle (0.08) (0.2,0) circle (0.08);
			\node[above left] at (-0.5,0.5) {$\scriptstyle x_1$};
			\node[above right] at (0.5,0.5) {$\scriptstyle x_2$};
			\node[below right] at (0.5,-0.5) {$\scriptstyle x_3$};
			\node[below left] at (-0.5,-0.5) {$\scriptstyle x_4$};
			\node[left] at (-0.2,0) {$\scriptstyle \theta_1$};
			\node[right] at (0.2,0) {$\scriptstyle \theta_2$};
		\end{tikzpicture}
		\end{array},\quad \chi_1\left[\begin{array}{cc|c}
		x_1 & x_2 & \theta_1\\
		x_4 & x_3 & \theta_2
		\end{array}\right]:\begin{array}{c}
		\begin{tikzpicture}[scale=0.7]
			\draw[thick, -stealth] (-0.5,-0.5) -- (0.5,0.5);
			\draw[thick, -stealth] (0.5,-0.5) -- (-0.5,0.5);
			\draw[fill=\myred] (0,0) circle (0.1);
			\node[above left] at (-0.5,0.5) {$\scriptstyle x_1$};
			\node[above right] at (0.5,0.5) {$\scriptstyle x_2$};
			\node[below right] at (0.5,-0.5) {$\scriptstyle x_3$};
			\node[below left] at (-0.5,-0.5) {$\scriptstyle x_4$};
			\node[left] at (-0.1,0) {$\scriptstyle \theta_1$};
			\node[right] at (0.1,0) {$\scriptstyle \theta_2$};
		\end{tikzpicture}
		\end{array}\;\longrightarrow\;\begin{array}{c}
		\begin{tikzpicture}[scale=0.7]
			\draw[thick, -stealth] (-0.5,-0.5) to[out=45,in=270] (-0.2,0) to[out=90,in=315] (-0.5,0.5);
			\draw[thick, -stealth] (0.5,-0.5) to[out=135,in=270] (0.2,0) to[out=90,in=225] (0.5,0.5);
			\draw[fill=\mygreen] (-0.2,0) circle (0.08) (0.2,0) circle (0.08);
			\node[above left] at (-0.5,0.5) {$\scriptstyle x_1$};
			\node[above right] at (0.5,0.5) {$\scriptstyle x_2$};
			\node[below right] at (0.5,-0.5) {$\scriptstyle x_3$};
			\node[below left] at (-0.5,-0.5) {$\scriptstyle x_4$};
			\node[left] at (-0.2,0) {$\scriptstyle \theta_1$};
			\node[right] at (0.2,0) {$\scriptstyle \theta_2$};
		\end{tikzpicture}
		\end{array}\,;\\
		&{\rm qdeg}\,\epsilon^{(+)}=-2,\quad {\rm tdeg}\,\epsilon^{(+)}=+1,\quad {\rm qdeg}\,\epsilon^{(-)}=0,\quad {\rm tdeg}\,\epsilon^{(-)}=+1\,.
	\end{aligned}
\end{equation}
If the uniform choice of variables is made all the rings $\CR[\Gamma]$ for various MOY diagrams $\Gamma$ are mutually isomorphic naturally.

One is able to derive expressions for morphisms $\chi_{0,1}$ explicitly based on three properties:
\begin{enumerate}
\item {\bf Parity.} Horizontal differential \eqref{horizontal} is expected to be an odd, fermionic operator, so that its nilpotency is a natural property.
Since $\epsilon$ is a fermion, it implies that morphisms $\chi_{0,1}$ are even linear operators.
\item {\bf Degree.} In Morse/Floer theory approaches to link invariant categorification homological degrees appear as eigenvalues $z$ and $f$ of commuting quantum charges in topological string theory.
These charges commute with the Hamiltonian as well.
So a differential concentrated in the Morse theory on instanton flows has also fixed degrees $z=0$ and $f=1$ as a linear operator on the Hilbert space.
Then we expect that morphisms $\chi_{0,1}$ would also have a fixed $q$-degree equal to +2 and 0 respectively.
So that in total ${\rm qdeg }\,\fD=0$.
Let us argue briefly this degree choice.
Assume that $\psi\not\in{\rm ker}\,\fD$ and consider a simple complex homotopic to zero:
$$
\left[\begin{array}{c}
	\begin{tikzpicture}
		\node (A) at (0,0) {$0$};
		\node (B) at (2,0) {$\psi$};
		\node (C) at (4,0) {$\fD\psi$};
		\node (D) at (6,0) {$0$};
		\path (A) edge[->] (B) (B) edge[->] node[above] {$\scriptstyle \fD$} (C) (C) edge[->] (D);
	\end{tikzpicture}
\end{array}\right]\;\sim \; 0\,.
$$
Its index reads $q^{{\rm qdeg}\,\psi}(1-q^{{\rm qdeg}\,\fD})$.
We expect naturally that the index is an invariant and equals to 0 for a zero complex, this reasoning imposes a constraint ${\rm qdeg }\fD=0$.
\item {\bf Locality.} This property seems to be the most important singling out the KR construction among the others.
We assume that for the uniform choice of variables morphisms $\chi_{0,1}$ are local: they depend only on variables $x_1$, $x_2$, $x_3$, $x_4$, $\theta_1$ and $\theta_2$ assigned to a particular intersection \eqref{categRvar}.
\end{enumerate}

Let us calculate morphism $\chi_0$ explicitly.
First we separate dependence on $\theta_{1,2}$ in the ring basis:
\begin{equation}
	\CR=\overset{U_0 \circlearrowleft}{\underset{{\rm parity}=+1}{\left(R\oplus R_{12}\theta_1\theta_2\right)}}\oplus\overset{U_1 \circlearrowleft}{\underset{{\rm parity}=-1}{\left(R_1\theta_1\oplus R_2\theta_2\right)}}\,,
\end{equation}
where $R$, $R_1$, $R_2$, $R_{12}$ are rings of variables $x_i$ and odd variables coming from other intersections.
We combined these subrings in pairs according to their parity.
Since $\chi_0$ preserves parity a generic $4\times 4$ linear transform $\chi_0$ on this space is split in two $2\times 2$ matrix blocks $U_0$ and $U_1$.
We can naturally represent such a map in the form of a differential operator:
\begin{equation}
\begin{aligned}
	\chi_0&=\left(U_0\right)_{11}\hat\theta_1^{\dagger}\hat\theta_1\hat\theta_2^{\dagger}\hat\theta_2+\left(U_0\right)_{22}\hat\theta_1\hat\theta_1^{\dagger}\hat\theta_2\hat\theta_2^{\dagger}+\left(U_0\right)_{21}\hat\theta_1\hat\theta_2+\left(U_0\right)_{12}\hat\theta_2^{\dagger}\hat\theta_1^{\dagger}+\\
	&+\left(U_1\right)_{11}\hat\theta_1\hat\theta_1^{\dagger}\hat\theta_2^{\dagger}\hat\theta_2 + \left(U_1\right)_{12}\hat\theta_1\hat\theta_2^{\dagger}+\left(U_1\right)_{21}\hat\theta_2\hat\theta_1^{\dagger} +\left(U_1\right)_{22} \hat\theta_1^{\dagger}\hat\theta_1\hat\theta_2\hat\theta_2^{\dagger}\,.
\end{aligned}
\end{equation}

Matrix elements of these matrices are polynomials in variables $x_{1,2,3,4}$ of fixed degree due to the $q$-degree constraint:\footnote{Here we used $q$-degree values for $\theta_i$ from \eqref{2-valentQ} and \eqref{4-valentQ}.
For a map $X:\,A\to B$ to be of degree +2 one should obtain the following degrees for matrix elements:
\begin{equation}
	{\rm qdeg}\,X_{ij}=2+{\rm qdeg}\,A_j-{\rm qdeg}\,B_i\,.
	\end{equation}}
\begin{equation}
	{\rm qdeg}\,U_0=\left(\begin{array}{cc}
		2 & 2(2-n)\\
		2(n-1)  & 0\\
	\end{array}\right),\quad {\rm qdeg}\,U_1=\left(\begin{array}{cc}
	2 & 2\\
	0  & 0\\
	\end{array}\right)\,.
\end{equation}
Let us assign respective variable homogeneous polynomials to these matrix elements:
\begin{equation}
	U_0=\left(\begin{array}{cc}
		p_1 & 0\\
		g_1  & 1\\
	\end{array}\right),\quad U_1=\left(\begin{array}{cc}
		p_2 & p_3\\
		r_1  & r_2\\
	\end{array}\right)\,,
\end{equation}
where polynomials $p_{1,2,3}$ are of degree 1, polynomial $g_1$ is of degree $n-1$, and $r_{1,2}$ are of degree 0, in other words the latter are just numbers.
Apparently, we would be able to define this differential modulo an overall numerical scale, so we fix it by choosing matrix element $(U_0)_{22}$ of degree 0 to be 1.
Respectively, matrix element $(U_0)_{12}$ has a negative degree for generic $n>2$, therefore it should be 0.

We describe local matrix factorizations for the two resolutions in the first line of \eqref{categRvar} explicitly as:
\begin{equation}\label{MFs}
	\begin{aligned}
		&\MF_{\Gamma_0}=\pi_{14}\hat\theta_1+\underbrace{\left(x_1-x_4\right)}_{x_{14}}\hat\theta_1^{\dagger}+\pi_{23}\hat\theta_2+\underbrace{\left(x_2-x_3\right)}_{x_{23}}\hat\theta_2^{\dagger}\,,\\
		&\MF_{\Gamma_1}=u_1\hat\theta_1+\underbrace{\left(x_1+x_2-x_3-x_4\right)}_{s_1}\hat\theta_1^{\dagger}+u_2\hat\theta_2+\underbrace{\left(x_1x_2-x_3x_4\right)}_{s_2}\hat\theta_2^{\dagger}\,.
	\end{aligned}
\end{equation}
For this matrix factorization operators homomorphism property descends to the following equation:
\begin{equation}
	\chi_0\MF_{\Gamma_0}=\MF_{\Gamma_1}\chi_0\,.
\end{equation}
This relation produces a collection of subconstraints:
\begin{subequations}
	\begin{align}
		&-p_2 s_1+p_1 x_{14}-r_2 s_2=0\label{sub_a}\\
		&p_3 s_1+p_1 x_{23}-r_1 s_2=0\label{sub_b}\\
		&-g_1 s_2-p_1 u_1+p_2 \pi _{14}-p_3 \pi _{23}=0\\
		&-p_3 x_{14}-p_2 x_{23}+s_2=0\\
		&r_1 x_{14}-r_2 x_{23}-s_1=0 \label{sub_e}\\
		&-g_1 x_{14}+p_2 u_2-r_2 u_1-\pi _{23}=0\label{sub_f}\\
		&-g_1 x_{23}-p_3 u_2-r_1 u_1+\pi _{14}=0
	\end{align}
\end{subequations}
Let us start with \eqref{sub_e}.
Since $s_1=x_{14}+x_{23}$ this constraint fixes uniquely $r_1=1$, $r_2=-1$.
After substituting this solution from \eqref{sub_a}, \eqref{sub_b} we re-express $p_2$ and $p_3$ in terms of $p_1$:
\begin{equation}
	p_2=\frac{s_2+p_1x_{14}}{s_1},\quad p_3=\frac{s_2-p_1x_{23}}{s_1}\,.
\end{equation}
However for these relations to make sense the pole in the denominators should be canceled.
This imposes constraints on $p_1$:
\begin{equation}
	\left(s_2+p_1x_{14}\right)\big|_{s_1=0}=\left(s_2-p_1x_{23}\right)\big|_{s_1=0}=0\,.
\end{equation}
Then we find:
\begin{equation}
	p_1\big|_{x_1=x_3+x_4-x_2}=-\frac{s_2}{x_{14}}\big|_{x_1=x_3+x_4-x_2}=\frac{s_2}{x_{23}}\big|_{x_1=x_3+x_4-x_2}=x_4-x_2\,.
\end{equation}
The most generic solution to this constraint for $p_1$ is the following:
\begin{equation}
	p_1=(x_4-x_2)+\mu\, s_1\,,
\end{equation}
And we have found a generic solution indeed since different values of $\mu$ correspond to homotopies of the double complex.

Dealing in the same way with $\chi_1$, eventually we would arrive to the following morphisms as in \cite{KRI} with free parameters $\mu$ and $\lambda$:
\begin{equation}\label{KRmorphisms}
	\begin{aligned}
		&\chi_0 =\left(\mu  \left(x_1+x_2-x_3-x_4\right)-x_2+x_4\right)\hat\theta_1^{\dagger}\hat\theta_1\hat\theta_2^{\dagger}\hat\theta_2+\hat\theta_1\hat\theta_1^{\dagger}\hat\theta_2\hat\theta_2^{\dagger}-\left((\mu-1)u_2+\frac{u_1+x_1u_2-\pi_{23}}{x_1-x_4} \right)\hat\theta_1\hat\theta_2+\\
		& +\left(x_4+\mu(x_1-x_4) \right)\hat\theta_1\hat\theta_1^{\dagger}\hat\theta_2^{\dagger}\hat\theta_2 + \left( \mu(x_2-x_3)-x_2\right)\hat\theta_1\hat\theta_2^{\dagger}-\hat\theta_2\hat\theta_1^{\dagger} + \hat\theta_1^{\dagger}\hat\theta_1\hat\theta_2\hat\theta_2^{\dagger}\,,\\
		&\chi_1=\hat\theta_1^{\dagger}\hat\theta_1\hat\theta_2^{\dagger}\hat\theta_2 +\left(\lambda\left(x_3+x_4-x_1-x_2\right)+x_1-x_3\right)\hat\theta_1\hat\theta_1^{\dagger}\hat\theta_2\hat\theta_2^{\dagger} -\left(\lambda u_2+\frac{u_1+x_1u_2-\pi_{23}}{x_4-x_1}\right)\hat\theta_1\hat\theta_2+\\
		& +\hat\theta_1\hat\theta_1^{\dagger}\hat\theta_2^{\dagger}\hat\theta_2  +\left( x_3+\lambda\left(x_2-x_3\right)\right)\hat\theta_1\hat\theta_2^{\dagger}+
		\hat\theta_2\hat\theta_1^{\dagger} +\left( x_1+\lambda\left(x_4-x_1\right)\right)\hat\theta_1^{\dagger}\hat\theta_1\hat\theta_2\hat\theta_2^{\dagger}\,.
	\end{aligned}
\end{equation}

The fact that homomorphism (commutation) relations \eqref{MFhomo} are enough under mild assumptions to derive horizontal morphisms uniquely is a strong indication that modern approaches via foam TQFT \cite{khovanov2004sl,mackaay2009sl,Chun:2015gda,lauda2015khovanov} are equivalent to the homological algebra approach of KR.
Indeed TQFT cobordisms saturating matrix elements of the horizontal morphisms correspond to evolution operators in the TQFT Hilbert space.
Evolution operators commute necessarily with supercharges we identified with MF operators in App.~\ref{app:MF}.
Then commutation (homomorphism) relations \eqref{MFhomo} constrain simultaneously both TQFT cobordism functors and  horizontal morphisms in the double complex to be compatible with supercharges (vertical morphisms) and, therefore, equivalent to each other.


\section{Operators \texorpdfstring{$\Op_L$}{OL}}\label{sec:O}

It is well known that cohomologies of a double complex could be rewritten as cohomologies of the total complex \cite{Weibel_1994}.
However in the particular case of the KR double complex the total complex could be constructed directly in terms of link diagrams $L$ omitting an additional passage to MOY diagrams.
\begin{enumerate}
	\item We assign to all edges $k$ of a link diagram independent even variables $x_k$, to each intersection $\alpha$ we assign a triplet of independent odd variables $\theta_1^{(\alpha)}$, $\theta_2^{(\alpha)}$, $\epsilon^{(\alpha)}$.
	\item To each intersection we assign a \emph{differential operator} in the following way:
	\begin{equation}\label{Ops}
		\begin{aligned}
			& \begin{array}{c}
				\begin{tikzpicture}[scale=0.7]
					\draw[thick, -stealth] (0.5,-0.5) -- (-0.5,0.5);
					\draw[white, line width = 1.4mm] (-0.5,-0.5) -- (0.5,0.5);
					\draw[thick, -stealth] (-0.5,-0.5) -- (0.5,0.5);
					\node[above left] at (-0.5,0.5) {$\scriptstyle x_1$};
					\node[above right] at (0.5,0.5) {$\scriptstyle x_2$};
					\node[below right] at (0.5,-0.5) {$\scriptstyle x_3$};
					\node[below left] at (-0.5,-0.5) {$\scriptstyle x_4$};
					\node[right] at (0.1,0) {$\scriptstyle (\theta_1,\theta_2,\epsilon)$};
				\end{tikzpicture}
			\end{array},\quad \Op_{\!\!\!\!\begin{array}{c}
				\begin{tikzpicture}[scale=0.3]
					\draw[-stealth] (0.5,-0.5) -- (-0.5,0.5);
					\draw[white, line width = 0.7mm] (-0.5,-0.5) -- (0.5,0.5);
					\draw[-stealth] (-0.5,-0.5) -- (0.5,0.5);
				\end{tikzpicture}
				\end{array}\!\!\!\!}=\MF_{\Gamma_0}\hat\epsilon^{\dagger}\hat\epsilon+\MF_{\Gamma_1}\hat\epsilon\hat\epsilon^{\dagger}+\epsilon\chi_0\,,\\
			& \begin{array}{c}
				\begin{tikzpicture}[scale=0.8]
					\draw[thick, -stealth] (-0.5,-0.5) -- (0.5,0.5);
					\draw[white, line width = 1.4mm] (0.5,-0.5) -- (-0.5,0.5);
					\draw[thick, -stealth] (0.5,-0.5) -- (-0.5,0.5);
					\node[above left] at (-0.5,0.5) {$\scriptstyle x_1$};
					\node[above right] at (0.5,0.5) {$\scriptstyle x_2$};
					\node[below right] at (0.5,-0.5) {$\scriptstyle x_3$};
					\node[below left] at (-0.5,-0.5) {$\scriptstyle x_4$};
					\node[right] at (0.1,0) {$\scriptstyle (\theta_1,\theta_2,\epsilon)$};
				\end{tikzpicture}
			\end{array},\quad \Op_{\!\!\!\!\begin{array}{c}
					\begin{tikzpicture}[scale=0.3]
						\draw[-stealth] (-0.5,-0.5) -- (0.5,0.5);
						\draw[white, line width = 0.8mm] (0.5,-0.5) -- (-0.5,0.5);
						\draw[-stealth] (0.5,-0.5) -- (-0.5,0.5);
					\end{tikzpicture}
				\end{array}\!\!\!\!}=\MF_{\Gamma_1}\hat\epsilon^{\dagger}\hat\epsilon+\MF_{\Gamma_0}\hat\epsilon\hat\epsilon^{\dagger}+\epsilon\chi_1\,,
		\end{aligned}
	\end{equation}
	where $\MF_{\Gamma_{0,1}}$ and $\chi_{0,1}$ are given by expressions \eqref{MFs} and \eqref{KRmorphisms}.
	\item The total differential operator for link subdiagram $\ell$ when we neglect external legs is defined as:
	\begin{equation}
		\Op_{\ell}=\sum\lm_{v\in{\rm vertices\; of\; }L} \Op_{v}\,.
	\end{equation}
	\item It is simple to derive that for a link diagram $\ell$ with open external edges labeled by even variables $x_k$:
	\begin{equation}
		\Op_{\ell}^2=\left(\sum\lm_k s_k x_k^{n+1}\right)\times \bbone\,,
	\end{equation}
	where sign $s_k$ equals $+1$ if edge $k$ is flowing outward $L$ and $-1$ if it flows inward $L$.
	\item To be able to discuss tangles one should take into account a contribution from the tangle boundary.
	There are two equivalent ways to deal with the  boundary $\p L$.
	One way is to consider the link diagram on a sphere and treat the boundary as another multi-valent vertex.
	The other way is to imply that the link $L$ is implanted into another bigger link $L'$ with all the legs jacked into ports of $L'$ so that the total link $L\cup L'$ is closed.
	In both cases we assign to the boundary $\p L$ a \emph{class} of operators $\Op_{\p L}$ with the following properties:
	\begin{enumerate}
		\item $\Op_{\p L}$ depends explicitly on even variables $x_k$ corresponding to external legs of $L$.
		It is allowed to depend on other even and odd variables not presented in $L$.
		\item For any internal vertex $v$: $\left\{\Op_{\p L},\Op_v\right\}=0$.
		\item $\Op_{\p L}^2=-\left(\sum\lm_k s_k x_k^{n+1}\right)\times \bbone$.
	\end{enumerate}
	Then to a link $L$ we assign the following operator:
	\begin{equation}
		\Op_{L}=\Op_{\ell}+\Op_{\p L}\,.
	\end{equation}
	It is easy to see that $\Op_{L}^2=0$.
	\item Link $L$ invariants are defined as ``invariant wave functions'' -- cohomologies $H^{*,*}(\Op_L)$.
	\item For a closed link diagram $\bar L$ $\Op_{\p\bar L}=0$, and operator $\Op_{\bar L}$ is a differential.
	Its cohomologies $H^{*,*}(\Op_{\bar L})$ are bi-graded naturally by qdeg and tdeg, they are isomorphic to the respective KR double complex cohomologies and are a \emph{link $\bar L$ invariant}.
\end{enumerate}

In what follows we will be interested in equivalence statements for tangles, for examples, we would like to say that tangle homotopy (Reidemeister moves) are tangle equivalences.
In all sorts of such statements that $L\cong L'$ with $\p L=\p L'$ we imply that respective cohomology groups are isomorphic $H^{*,*}(\Op_L)\cong H^{*,*}(\Op_{L'})$ for any choice of the boundary correction operator $\Op_{\p L}=\Op_{\p L'}$.

Now let us indicate that this structure behaves well under tangle gluing.
Consider two link diagrams $L_1$ and $L_2$ and assume that we glued them into $L_1\cup L_2$ by possibly jacking mutually some of the external legs of these diagrams.
For the glued diagram we have:
\begin{equation}
	\Op_{L_1\cup L_2}=\Op_{\ell_1}+\Op_{\ell_2}+\left[\Op_{\p(L_1\cup L_2)}\right]\,,
\end{equation}
where by square brackets $[\ldots]$ we indicate that the whole class of boundary correcting operators is considered.
However it is easy to show that $\Op_{\ell_2}+\left[\Op_{\p(L_1\cup L_2)}\right]\in\left[\Op_{\p L_1}\right]$, so that a class $\left[\Op_{L_1\cup L_2}\right]$ is a subclass in $\left[\Op_{L_1}\right]$.
Therefore any equivalence proven for sub-tangle $L_1$ would extend automatically to the glued tangle $L_1\cup L_2$.


\section{Invariants of tangles} \label{sec:invariant}

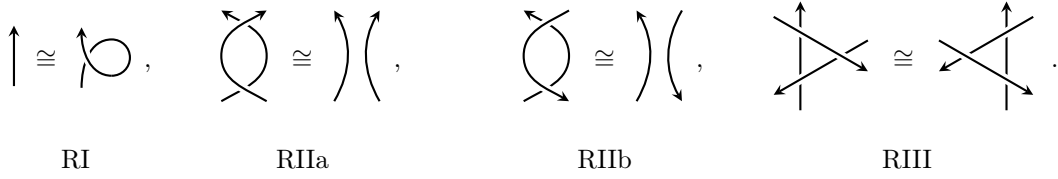
\begin{figure}[ht!]
	\centering
	\begin{tikzpicture}
		\node at (1,0) {$\begin{array}{c}
			\begin{tikzpicture}[scale=0.8]
				\draw[thick, -stealth] (0,-0.5) -- (0,0.5);
			\end{tikzpicture}
			\end{array}\cong\begin{array}{c}
			\begin{tikzpicture}[scale=0.8,xscale=-1]
				\draw[thick] (0,-0.5) to[out=90,in=0] (-0.5,0.3) to[out=180,in=90] (-0.8,0) to[out=270,in=180] (-0.5,-0.3);
				\draw[white, line width = 1.2mm] (-0.5,-0.3) to[out=0,in=270] (0,0.5);
				\draw[thick, -stealth] (-0.5,-0.3) to[out=0,in=270] (0,0.5);
			\end{tikzpicture}
			\end{array},$};
		\node at (4,0) {$\begin{array}{c}
				\begin{tikzpicture}[scale=0.6]
					\draw[thick, -stealth] (-0.5,-1) to[out=30,in=270] (0.5,0) to[out=90,in=330] (-0.5,1);
					\draw[white, line width = 1.2mm] (0.5,-1) to[out=150,in=270] (-0.5,0) to[out=90,in=210] (0.5,1);
					\draw[thick, -stealth] (0.5,-1) to[out=150,in=270] (-0.5,0) to[out=90,in=210] (0.5,1);
				\end{tikzpicture}
			\end{array}\cong \begin{array}{c}
			\begin{tikzpicture}[scale=0.6]
				\draw[thick, -stealth] (-0.5,-1) to[out=60,in=300] (-0.5,1);
				\draw[thick, -stealth] (0.5,-1) to[out=120,in=240] (0.5,1);
			\end{tikzpicture}
			\end{array},$};
		\node at (8,0) {$\begin{array}{c}
				\begin{tikzpicture}[scale=0.6]
					\draw[thick, -stealth] (-0.5,-1) to[out=30,in=270] (0.5,0) to[out=90,in=330] (-0.5,1);
					\draw[white, line width = 1.2mm] (0.5,-1) to[out=150,in=270] (-0.5,0) to[out=90,in=210] (0.5,1);
					\draw[thick, stealth-] (0.5,-1) to[out=150,in=270] (-0.5,0) to[out=90,in=210] (0.5,1);
				\end{tikzpicture}
			\end{array}\cong \begin{array}{c}
				\begin{tikzpicture}[scale=0.6]
					\draw[thick, -stealth] (-0.5,-1) to[out=60,in=300] (-0.5,1);
					\draw[thick, stealth-] (0.5,-1) to[out=120,in=240] (0.5,1);
				\end{tikzpicture}
			\end{array},$};
		\node at (12,0) {$\begin{array}{c}
				\begin{tikzpicture}[scale=0.6]
					\draw[thick,-stealth] (-0.3,-1.2) -- (-0.3,1.2);
					\begin{scope}[rotate=120]
						\draw[white,line width = 1.2mm] (-0.3,-1.2) -- (-0.3,1.2);
						\draw[thick,-stealth] (-0.3,-1.2) -- (-0.3,1.2);
					\end{scope}
					\begin{scope}[rotate=240]
						\draw[white,line width = 1.2mm] (-0.3,-1.2) -- (-0.3,1.2);
						\draw[thick,-stealth] (-0.3,-1.2) -- (-0.3,1.2);
					\end{scope}
				\end{tikzpicture}
			\end{array}\cong \begin{array}{c}
			\begin{tikzpicture}[scale=0.6]
				\draw[thick,-stealth] (0.3,-1.2) -- (0.3,1.2);
				\begin{scope}[rotate=120]
					\draw[white,line width = 1.2mm] (0.3,-1.2) -- (0.3,1.2);
					\draw[thick,-stealth] (0.3,-1.2) -- (0.3,1.2);
				\end{scope}
				\begin{scope}[rotate=240]
					\draw[white,line width = 1.2mm] (0.3,-1.2) -- (0.3,1.2);
					\draw[thick,-stealth] (0.3,-1.2) -- (0.3,1.2);
				\end{scope}
			\end{tikzpicture}
			\end{array}.$};
		\node at (1,-1.3) {RI};
		\node at (4,-1.3) {RIIa};
		\node at (8,-1.3) {RIIb};
		\node at (12,-1.3) {RIII};
	\end{tikzpicture}
	\caption{Reidemeister moves.}\label{fig:Reidemeister}
\end{figure}

Now we would like to argue that tangle wave functions -- cohomologies of $\Op_L$ -- remain invariant under homotopic moves.
Canonically this requires us to demonstrate the object is invariant with respect to basic Reidemeister moves (see Fig.~\ref{fig:Reidemeister}).
With respect to RI one expects an isomorphism of cohomologies with a degree shift in general.

Let us demonstrate explicitly the invariance of $H^{*,*}(\Op_L)$ with respect to RI here.
Remaining equivalences are proven in a similar fashion and we leave them out of the scope of the present note for the sake of brevity.

We would like to prove the following equivalence of cohomologies of $\Op$-operators for homotopically equivalent links:
\begin{equation}
	\bar L_1=\begin{array}{c}
		\begin{tikzpicture}[scale=0.7]
			\draw[thick] (0,0) to[out=270,in=180] (0.5,-0.5);
			\draw[thick,postaction={decorate},decoration={markings, mark = at position 0.65 with {\arrow{stealth}}}] (1.3,0.3) to[out=0,in=0] node[pos=0.5,right] {$\scriptstyle x_3$} (1.3,-0.3);
			\draw[thick] (1.3,-0.3) to[out=180,in=0] (0.5,0.5);
			\draw[thick] (0.5,0.5) to[out=180,in=90] (0,0);
			\draw[white,line width=1.2mm] (0.5,-0.5) to[out=0,in=180] (1.3,0.3);
			\draw[thick] (0.5,-0.5) to[out=0,in=180] (1.3,0.3);
			\draw[fill=\myblue] (0,0) circle (0.4);
			\node[white] at (0,0) {$\scriptstyle L$};
			\node[above right] at (0.5,0.5) {$\scriptstyle x_1$};
			\node[below right] at (0.5,-0.5) {$\scriptstyle x_2$};
			\node(A) [right] at (2,-0.5) {$\scriptstyle (\theta_1,\theta_2,\epsilon)$};
			\draw[-stealth] (A.west) to[out=180,in=270] (1,-0.2);
		\end{tikzpicture}
	\end{array},\quad \bar L_2=\begin{array}{c}
		\begin{tikzpicture}[scale=0.7]
			\draw[thick,postaction={decorate},decoration={markings, mark = at position 0.65 with {\arrow{stealth}}}] (0,0) to[out=270,in=180] (0.5,-0.5) to[out=0,in=0] node[pos=0.5,right] {$\scriptstyle x_1$} (0.5,0.5) to[out=180,in=90] (0,0);
			\draw[fill=\myblue] (0,0) circle (0.4);
			\node[white] at (0,0) {$\scriptstyle L$};
		\end{tikzpicture}
	\end{array}\,.
\end{equation}

We denote by $V$ all the internal variables in diagram $L$.
Here we have:
\begin{equation}
	\begin{aligned}
		&\Op_{\bar L_1}=\Op_L(V,x_1,x_2)+\MF_{\MOY_0}\hat\epsilon^{\dagger}\hat\epsilon+\MF_{\MOY_1}\hat\epsilon\hat\epsilon^{\dagger}+\hat\epsilon\chi_0\,,\\
		&\Op_{\bar L_2}=\Op_L(V,x_1,x_1)\,,
	\end{aligned}
\end{equation}
where
\begin{equation}
	\begin{aligned}
		& \MF_{\MOY_0}=\pi_{12}\hat\theta_1+\underline{(x_1-x_2)\hat\theta_{1}^{\dagger}}+x_3^n\hat\theta_2\,,\\
		& \MF_{\MOY_1}=u_1(x_1,x_3,x_3,x_2)\hat\theta_1+\underline{(x_1-x_2)\hat\theta_1^{\dagger}}+u_2(x_1,x_3,x_3,x_2)\hat\theta_2+x_3x_{12}\hat\theta_2^{\dagger}\,.
	\end{aligned}
\end{equation}

Let us separate variable $\epsilon$, then the action of $\Op_{\bar L_1}$ acts as:
\begin{equation}\label{OpL1}
	\Op_{\bar L_1}\left(\Psi_0+\epsilon \Psi_1\right)=\left(\Op_L(V,x_1,x_2)+\MF_{\MOY_0}\right)\Psi_0+\epsilon\left(-\Op_L(V,x_1,x_2)\Psi_1+\MF_{\MOY_1}\Psi_1+\chi_0\Psi_0\right)\,.
\end{equation}

Now we would like to derive cohomologies $\Phi_1$ and $\Phi_2$ of operators $\Op_L(V,x_1,x_2)+\MF_{\MOY_0}$ and $\Op_L(V,x_1,x_2)+\MF_{\MOY_1}$ respectively.
Applying \eqref{strayiso} we derive:
\begin{equation}
	\begin{aligned}
		&\Phi_1=\underbrace{\left(1-\theta_1\frac{\Op_{L}(V,x_1,x_2)-\Op_{L}(V,x_1,x_1)}{x_1-x_2}\right)\theta_2}_{S_1}\times\left(\varphi(x_3)\psi(V,x_1)\;{\rm mod}\;x_3^{n}\right)\,,\\
		&\Phi_2=\underbrace{\left(1-\theta_1\frac{\Op_{L}(V,x_1,x_2)-\Op_{L}(V,x_1,x_1)}{x_1-x_2}-x_3\theta_1\hat\theta_2^{\dagger}\right)\theta_2}_{S_2}\times\left(\varphi(x_3)\psi(V,x_1)\;{\rm mod}\;p_{n-1}(x_3,x_1)\right)\,,
	\end{aligned}
\end{equation}
where  $\varphi$ are arbitrary polynomials in one variable $x_3$, and $\psi(V,x_1)$ is cohomology of $\Op_L(V,x_1,x_1)=\Op_{\bar L_2}$, and
\begin{equation}\label{ppoly}
	p_{n-1}(x_3,x_1):=u_2(x_1,x_3,x_3,x_1)=x_3^{n-1}+\sum\lm_{i=0}^{n-2} a_i(x_1)x_3^i\,.
\end{equation}
It is easy to see that $\chi_0S_1\varphi\psi=S_2\varphi\psi$.
Let us choose bases in cohomologies:
\begin{equation}
	\begin{aligned}
		&H^*(\Op_L(V,x_1,x_2)+\MF_{\MOY_0}):\quad S_1\psi, \; S_1x_3\psi, \;  \ldots, S_1x_3^{n-2}\psi, \;  S_1p_{n-1}(x_3,x_1)\psi\,,\\
		&H^*(\Op_L(V,x_1,x_2)+\MF_{\MOY_1}):\quad S_2\psi, \;  S_2x_3\psi, \;  \ldots, \;  S_2x_3^{n-2}\psi\,,\\
	\end{aligned}
\end{equation}
So we consider the wave function in the following form:
\begin{equation}
	\Psi_{\bar L_1}(V,x_1,x_2,x_3)=\alpha S_1p_{n-1}(x_3,x_1)\psi+\sum\lm_{k=0}^{n-2}\beta_kS_1x_3^{k}\psi+\epsilon\sum\lm_{k=0}^{n-2}\gamma_kS_2x_3^{k}\psi\,,
\end{equation}
and substitute it back into \eqref{OpL1}.
This substitution constraints $\beta_k=0$ and leaves $\alpha$, $\gamma_k$ free parameters, on the other hand we find:
\begin{equation}
	\epsilon\sum\lm_{k=0}^{n-2}\gamma_kS_2x_3^{k}\psi=\Op_{\bar L_1}\sum\lm_{k=0}^{n-2}\gamma_kS_1x_3^{k}\psi\in {\rm Im}\,\Op_{\bar L_1}\,.
\end{equation}

So we conclude that the sole cohomology of $\Op_{\bar L_1}$ is isomorphic to cohomology of $\Op_{\bar L_2}$ (up to degree shifts):
\begin{equation}
	\begin{aligned}
	&\Psi_{\bar L_1}\in H^{*,*}(\Op_{\bar L_1})\;\cong\; H^{*,*}(\Op_{\bar L_2})\ni \Psi_{\bar L_2}\,,\\
	&\Psi_{\bar L_1}(V,x_1,x_2,x_3,\theta_1,\theta_2,\epsilon)=\left(1-\theta_1\frac{\Op_{L}(V,x_1,x_2)-\Op_{L}(V,x_1,x_1)}{x_1-x_2}\right)\theta_2 p_{n-1}(x_3,x_1) \times \Psi_{\bar L_2}(V,x_1)+\epsilon\cdot 0\,.
	\end{aligned}
\end{equation}


\section{Miscellaneous exercises} \label{sec:misc}

\subsection{Poincar\'e polynomial for the Hopf link}

Here we would present an exercise computation of the Poincar\'e polynomial for the Hopf link with generic $n$.
First we decompose the Hopf link into MOY diagrams:
\begin{equation}
	\begin{array}{c}
		\begin{tikzpicture}[scale=0.7]
			\begin{scope}[shift={(0,1)}]
				\draw[thick] (0.5,-0.5) to[out=90,in=270] (-0.5,0.5);
				\draw[white, line width = 1.5mm] (-0.5,-0.5) to[out=90,in=270] (0.5,0.5);
				\draw[thick] (-0.5,-0.5) to[out=90,in=270] (0.5,0.5);
			\end{scope}
			\draw[thick,-stealth] (0.5,-0.5) to[out=90,in=270] (-0.5,0.5) -- (-0.5,0.6);
			\draw[white, line width = 1.5mm] (-0.5,-0.5) to[out=90,in=270] (0.5,0.5);
			\draw[thick,-stealth] (-0.5,-0.5) to[out=90,in=270] (0.5,0.5) -- (0.5,0.6);
			\draw[thick] (-0.5,1.5) to[out=90,in=90] (-1.2,1.5) -- (-1.2,-0.5) to[out=270,in=270] (-0.5,-0.5);
			\begin{scope}[xscale=-1]
				\draw[thick] (-0.5,1.5) to[out=90,in=90] (-1.2,1.5) -- (-1.2,-0.5) to[out=270,in=270] (-0.5,-0.5);
			\end{scope}
			\node[left] at (-0.5,-0.5) {$\scriptstyle x_1$};
			\node[right] at (0.5,-0.5) {$\scriptstyle x_2$};
			\node[left] at (-0.5,0.5) {$\scriptstyle x_3$};
			\node[right] at (0.5,0.5) {$\scriptstyle x_4$};
			\node[left] at (-0.5,1.5) {$\scriptstyle x_1$};
			\node[right] at (0.5,1.5) {$\scriptstyle x_2$};
			\node[above] at (0,0) {$\scriptstyle 1$};
			\node[above] at (0,1) {$\scriptstyle 2$};
			\node[left] at (-0.3,0) {$\scriptstyle \theta_1$};
			\node[right] at (0.3,0) {$\scriptstyle \theta_2$};
			\node[left] at (-0.3,1) {$\scriptstyle \theta_3$};
			\node[right] at (0.3,1) {$\scriptstyle \theta_4$};
		\end{tikzpicture}
	\end{array}=\left[\begin{array}{c}
		\begin{tikzpicture}
			\node(A) at (0,0) {$\MOY_{00}$};
			\node(B) at (3,0.7) {$\epsilon_1\MOY_{10}$};
			\node(C) at (3,-0.7) {$\epsilon_2\MOY_{01}$};
			\node(D) at (6,0) {$\epsilon_1\epsilon_2\MOY_{11}$};
			\path (A) edge[-stealth] node[above left] {$\scriptstyle\epsilon_1\chi_1$} (B) (A) edge[-stealth] node[below left] {$\scriptstyle\epsilon_2\chi_2$} (C) (B) edge[-stealth] node[above right] {$\scriptstyle\epsilon_2\chi_2$} (D) (C) edge[-stealth] node[below right] {$\scriptstyle \epsilon_1\chi_1$} (D);
		\end{tikzpicture}
	\end{array}\right]\,,
\end{equation}
where the following MOY diagrams are used:
\begin{equation}
	\begin{aligned}
		& \MOY_{00}=\begin{array}{c}
			\begin{tikzpicture}[scale=0.5]
				\draw[thick] (0.5,-0.5) to[out=90,in=270] (0.3,0) to[out=90,in=270] (0.5,0.5) -- (0.5,0.6);
				\draw[thick] (-0.5,-0.5) to[out=90,in=270] (-0.3,0) to[out=90,in=270] (-0.5,0.5) -- (-0.5,0.6);
				\begin{scope}[shift={(0,1)}]
					\draw[thick] (0.5,-0.5) to[out=90,in=270] (0.3,0) to[out=90,in=270] (0.5,0.5);
					\draw[thick] (-0.5,-0.5) to[out=90,in=270] (-0.3,0) to[out=90,in=270] (-0.5,0.5);
				\end{scope}
				\draw[thick, postaction={decorate},decoration={markings, mark= at position 0.6 with {\arrow{stealth}}}] (-0.5,1.5) to[out=90,in=90] (-1.2,1.5) -- (-1.2,-0.5) to[out=270,in=270] (-0.5,-0.5);
				\begin{scope}[xscale=-1]
					\draw[thick, postaction={decorate},decoration={markings, mark= at position 0.6 with {\arrow{stealth}}}] (-0.5,1.5) to[out=90,in=90] (-1.2,1.5) -- (-1.2,-0.5) to[out=270,in=270] (-0.5,-0.5);
				\end{scope}
			\end{tikzpicture}
		\end{array},\quad \MOY_{10}=\begin{array}{c}
			\begin{tikzpicture}[scale=0.5]
				\draw[thick] (0.5,-0.5) to[out=90,in=270] (-0.5,0.5);
				\draw[thick] (-0.5,-0.5) to[out=90,in=270] (0.5,0.5);
				\begin{scope}[shift={(0,1)}]
					\draw[thick] (0.5,-0.5) to[out=90,in=270] (0.3,0) to[out=90,in=270] (0.5,0.5);
					\draw[thick] (-0.5,-0.5) to[out=90,in=270] (-0.3,0) to[out=90,in=270] (-0.5,0.5);
				\end{scope}
				\draw[thick, postaction={decorate},decoration={markings, mark= at position 0.6 with {\arrow{stealth}}}] (-0.5,1.5) to[out=90,in=90] (-1.2,1.5) -- (-1.2,-0.5) to[out=270,in=270] (-0.5,-0.5);
				\begin{scope}[xscale=-1]
					\draw[thick, postaction={decorate},decoration={markings, mark= at position 0.6 with {\arrow{stealth}}}] (-0.5,1.5) to[out=90,in=90] (-1.2,1.5) -- (-1.2,-0.5) to[out=270,in=270] (-0.5,-0.5);
				\end{scope}
				\draw[fill=\myred] (0,0) circle (0.13);
			\end{tikzpicture}
		\end{array},\quad\MOY_{01}=\begin{array}{c}
			\begin{tikzpicture}[scale=0.5]
				\draw[thick] (0.5,-0.5) to[out=90,in=270] (0.3,0) to[out=90,in=270] (0.5,0.5);
				\draw[thick] (-0.5,-0.5) to[out=90,in=270] (-0.3,0) to[out=90,in=270] (-0.5,0.5);
				\begin{scope}[shift={(0,1)}]
					\draw[thick] (0.5,-0.5) to[out=90,in=270] (-0.5,0.5);
					\draw[thick] (-0.5,-0.5) to[out=90,in=270] (0.5,0.5);
				\end{scope}
				\draw[thick, postaction={decorate},decoration={markings, mark= at position 0.6 with {\arrow{stealth}}}] (-0.5,1.5) to[out=90,in=90] (-1.2,1.5) -- (-1.2,-0.5) to[out=270,in=270] (-0.5,-0.5);
				\begin{scope}[xscale=-1]
					\draw[thick, postaction={decorate},decoration={markings, mark= at position 0.6 with {\arrow{stealth}}}] (-0.5,1.5) to[out=90,in=90] (-1.2,1.5) -- (-1.2,-0.5) to[out=270,in=270] (-0.5,-0.5);
				\end{scope}
				\draw[fill=\myred] (0,1) circle (0.13);
			\end{tikzpicture}
		\end{array},\quad\MOY_{11}=\begin{array}{c}
			\begin{tikzpicture}[scale=0.5]
				\draw[thick,-stealth] (0.5,-0.5) to[out=90,in=270] (-0.5,0.5) -- (-0.5,0.6);
				\draw[thick,-stealth] (-0.5,-0.5) to[out=90,in=270] (0.5,0.5) -- (0.5,0.6);
				\begin{scope}[shift={(0,1)}]
					\draw[thick] (0.5,-0.5) to[out=90,in=270] (-0.5,0.5);
					\draw[thick] (-0.5,-0.5) to[out=90,in=270] (0.5,0.5);
				\end{scope}
				\draw[thick, postaction={decorate},decoration={markings, mark= at position 0.6 with {\arrow{stealth}}}] (-0.5,1.5) to[out=90,in=90] (-1.2,1.5) -- (-1.2,-0.5) to[out=270,in=270] (-0.5,-0.5);
				\begin{scope}[xscale=-1]
					\draw[thick, postaction={decorate},decoration={markings, mark= at position 0.6 with {\arrow{stealth}}}] (-0.5,1.5) to[out=90,in=90] (-1.2,1.5) -- (-1.2,-0.5) to[out=270,in=270] (-0.5,-0.5);
				\end{scope}
				\draw[fill=\myred] (0,0) circle (0.13) (0,1) circle (0.13);
			\end{tikzpicture}
		\end{array}\,.
	\end{aligned}
\end{equation}

Now we would like to solve for zero modes of MF operators applying techniques of Sec.~\ref{sec:stray} and of Sec.~\ref{sec:4v_split}.
For the first MF operator we derive:
\begin{equation}
	\MF_{\MOY_{00}}\big|_{\substack{x_3=x_1,\;x_4=x_2\\
			\theta_3=0,\;\theta_4=0}}=x_1^n\hat\theta_1+x_2^n\hat\theta_2\,,
\end{equation}
We easily find cohomology of this differential concentrated at fermion monomial $\theta_1\theta_2$ and re-construct complete cohomological wave-functions by ``dressing'' the fermion monomial with \eqref{dress1}:
\begin{equation}
	\sum\lm_{i,j=0}^{n-1} a_{ij}x_1^ix_2^j\theta_1\theta_2\;\overset{\eqref{dress1}}{\longrightarrow}\; \Psi_{\MOY_{00}}=\sum\lm_{i,j=0}^{n-1} a_{ij}x_1^ix_2^j\Omega_{00},\quad \Omega_{00}=\theta_1\theta_2+\theta_3\theta_2+\theta_1\theta_4+\theta_3\theta_4\,.
\end{equation}

For $\MOY_{10}$ we find:
\begin{equation}
	\MF_{\MOY_{10}}\big|_{\substack{x_3=x_1,\;x_4=x_2\\
			\theta_3=0,\;\theta_4=0}}=(n+1)\left(x_1^n+x_1^{n-1}x_2+\ldots+x_2^{n}\right)\hat\theta_1+(n+1)\left(x_1^{n-1}+x_1^{n-2}x_2+\ldots+x_2^{n-1}\right)\hat\theta_2\,,
\end{equation}
Cohomologies of this differential are concentrated again in the $\theta_1\theta_2$-term and are isomorphic to $\IQ[x_1,x_2]$ modulo $x_1^n=0$, $x_2^{n-1}=0$.
So we ``dress'' these wave functions to derive the respective matrix factorization cohomology with \eqref{dress1}:
\begin{equation}
	\begin{aligned}
		&\sum\lm_{i=0}^{n-1}\sum\lm_{j=0}^{n-2}b_{ij}x_1^ix_2^j\theta_1\theta_2\quad\overset{\eqref{dress1}}{\longrightarrow}\quad\Psi_{\MOY_{10}}=\sum\lm_{i=0}^{n-1}\sum\lm_{j=0}^{n-2}b_{ij}x_1^ix_2^j\Omega_{10}\,,\\
		&\Omega_{10}=\bigg(\theta_1\theta_2+x_2\theta_1\theta_3-\theta_2\theta_3+x_3\theta_1\theta_4-\theta_2\theta_4+\left(x_3-x_2\right)\theta_3\theta_4+\\
		&+\frac{\left(u_1(x_3,x_4,x_2,x_1)+x_2u_2(x_3,x_4,x_2,x_1)\right)-\left(u_1(x_3,x_2,x_2,x_1)+x_2u_2(x_3,x_2,x_2,x_1)\right)}{x_4-x_2}\theta_1\theta_2\theta_3\theta_4\bigg)\,.
	\end{aligned}
\end{equation}

Similarly for $\MOY_{01}$ we derive cohomology concentrated at $\theta_3\theta_4$:
\begin{equation}
	\begin{aligned}
		&\sum\lm_{i=0}^{n-1}\sum\lm_{j=0}^{n-2}c_{ij}x_1^ix_2^j\theta_3\theta_4\quad\overset{\eqref{dress1}}{\longrightarrow}\quad\Psi_{\MOY_{01}}=\sum\lm_{i=0}^{n-1}\sum\lm_{j=0}^{n-2}c_{ij}x_1^ix_2^j\Omega_{01}\,,\\
		&\Omega_{01}=\bigg(\theta_3\theta_4-x_2\theta_1\theta_3+\theta_1\theta_4-x_3\theta_2\theta_3+\theta_2\theta_4+\left(x_3-x_2\right)\theta_1\theta_2+\\
		&+\frac{\left(u_1(x_1,x_2,x_4,x_3)+x_2u_2(x_1,x_2,x_4,x_3)\right)-\left(u_1(x_1,x_2,x_2,x_3)+x_2u_2(x_1,x_2,x_2,x_3)\right)}{x_2-x_4}\theta_1\theta_2\theta_3\theta_4\bigg)\,.
	\end{aligned}
\end{equation}

Eventually for $\MOY_{11}$ we find:
\begin{equation}
	\MF_{\MOY_{11}}\big|_{\substack{x_3=x_1,\;x_4=x_2\\
			\theta_3=0,\;\theta_4=0}}=\MF_{\MOY_{11}}\big|_{\substack{x_3=x_2,\;x_4=x_1\\
			\theta_3=0,\;\theta_4=0}}=\MF_{\MOY_{10}}\big|_{\substack{x_3=x_1,\;x_4=x_2\\
			\theta_3=0,\;\theta_4=0}}\,.
\end{equation}
This implies, actually, that the cohomology of this operator is representable in the form of $f(x_1,x_2)+(x_3-x_4)g(x_1,x_2)$ where both $f$ and $g$ are cohomologies of $\MF_{\MOY_{10}}$.
So in this case we obtain the following cohomological wave function:
\begin{equation}
	\begin{aligned}
		&\Psi_{\MOY_{11}}=\sum\lm_{i=0}^{n-1}\sum\lm_{j=0}^{n-2}\left(d_{ij}x_1^ix_2^j+(x_3-x_4)e_{ij}x_1^ix_2^j\right)\Omega_{11}\,,\\
		&\Omega_{11}=\left(\theta_1\theta_2-\theta_2\theta_3+\theta_1\theta_4+\theta_3\theta_4+\frac{u_1(x_1,x_2,x_4,x_3)-u_1(x_3,x_4,x_2,x_1)}{x_1x_2-x_3x_4}\theta_1\theta_2\theta_3\theta_4\right)\,.
	\end{aligned}
\end{equation}

For the action of horizontal morphisms on forms we find:
\begin{equation}
	\begin{aligned}
		&\chi_1(\Omega_{00})=\Omega_{10}\,,\\
		&\chi_2(\Omega_{00})=\Omega_{01}+\MF_{\MOY_{01}}(\theta_1\theta_2\theta_3)\,,\\
		&\chi_2(\Omega_{10})=\frac{1}{2}(x_1-x_2+x_3-x_4)\Omega_{11}+\frac{1}{2}\MF_{\MOY_{11}}(\theta_1\theta_2\theta_3)+\frac{1}{2}\MF_{\MOY_{11}}(\theta_1\theta_3\theta_4)\,,\\
		&\chi_1(\Omega_{01})=\frac{1}{2}(x_1-x_2+x_3-x_4)\Omega_{11}-\frac{1}{2}\MF_{\MOY_{11}}(\theta_1\theta_2\theta_3)+\frac{1}{2}\MF_{\MOY_{11}}(\theta_1\theta_3\theta_4)\,.
	\end{aligned}	
\end{equation}

Using these explicit expressions is easy to calculate that the cohomological wave function of $\Op_{\rm Hopf}$ reads:
\begin{equation}
	\Psi=\sum\lm_{i=0}^{n-1}a_ix_1^ix_2^{n-1}\Omega_{00}+\epsilon_1\epsilon_2\sum\lm_{i=0}^{n-1}\sum\lm_{j=0}^{n-2}b_{ij}x_1^ix_2^j\Omega_{11}\,,
\end{equation}

We calculate that the Poincar\'e polynomial for the Hopf link coincides with the known expression from the literature \cite[Table 1]{Carqueville:2011zea}:
\begin{equation}
	P_{\rm Hopf}(q,t)=q^{n+1}[n]_q+t^2q^{-1}[n]_q[n-1]_q\,.
\end{equation}

\subsection{Decategorification}

The ``golden rule'' of any refinement (categorification) procedure is that at any moment refinement could be reversed, i.e. at any stage of the categorification process we could see a simple reduction to decategorified objects.
Here we would like to argue that calculating invariant zero modes of operator $\Op_L$ decategorifies indeed to the R-matrix calculations in the RT formalism.
In the RT formalism the computation of diagram $L$ reduces to rephrasing its elements in terms of functors, simply R-matrix and cup/cap braid closure tensors, of the braided tensor category of $U_q(\fs\fu_n)$ \cite{Reshetikhin:1990pr,Reshetikhin:1991tc}.

Fortunately, explicit expressions for the R-matrix and caps/cups are known for fundamental reps $\Box$ of generic $U_q(\fs\fu_n)$.
Here we present them in the useful graphical language:
\begin{equation}\label{RT}
	\begin{aligned}
		& \begin{array}{c}
			\begin{tikzpicture}[scale=0.5]
				\draw[thick, -stealth] (0,0) to[out=90,in=180] (0.5,0.7) to[out=0,in=90] (1,0);
			\end{tikzpicture}
		\end{array}=\sum\lm_{a=1}^nq^{a-\frac{n+1}{2}}\begin{array}{c}
			\begin{tikzpicture}[scale=0.4]
				\draw[thick] (0,0) to[out=90,in=180] (0.5,0.7) to[out=0,in=90] (1,0);
				\node[below] at (0,0) {$\scriptstyle a$};
				\node[below] at (1,0) {$\scriptstyle \bar a$};
			\end{tikzpicture}
		\end{array},\quad \begin{array}{c}
			\begin{tikzpicture}[scale=0.5]
				\draw[thick, stealth-] (0,0) to[out=90,in=180] (0.5,0.7) to[out=0,in=90] (1,0);
			\end{tikzpicture}
		\end{array}=\sum\lm_{a=1}^nq^{-a+\frac{n+1}{2}}\begin{array}{c}
			\begin{tikzpicture}[scale=0.4]
				\draw[thick] (0,0) to[out=90,in=180] (0.5,0.7) to[out=0,in=90] (1,0);
				\node[below] at (0,0) {$\scriptstyle \bar a$};
				\node[below] at (1,0) {$\scriptstyle a$};
			\end{tikzpicture}
		\end{array}\,,\\
		& \begin{array}{c}
			\begin{tikzpicture}[yscale=-1,scale=0.5]
				\draw[thick, -stealth] (0,0) to[out=90,in=180] (0.5,0.7) to[out=0,in=90] (1,0);
			\end{tikzpicture}
		\end{array}=\sum\lm_{a=1}^nq^{-a+\frac{n+1}{2}}\begin{array}{c}
			\begin{tikzpicture}[yscale=-1,scale=0.4]
				\draw[thick] (0,0) to[out=90,in=180] (0.5,0.7) to[out=0,in=90] (1,0);
				\node[above] at (0,0) {$\scriptstyle a$};
				\node[above] at (1,0) {$\scriptstyle \bar a$};
			\end{tikzpicture}
		\end{array},\quad \begin{array}{c}
			\begin{tikzpicture}[yscale=-1,scale=0.5]
				\draw[thick, stealth-] (0,0) to[out=90,in=180] (0.5,0.7) to[out=0,in=90] (1,0);
			\end{tikzpicture}
		\end{array}=\sum\lm_{a=1}^nq^{a-\frac{n+1}{2}}\begin{array}{c}
			\begin{tikzpicture}[yscale=-1,scale=0.4]
				\draw[thick] (0,0) to[out=90,in=180] (0.5,0.7) to[out=0,in=90] (1,0);
				\node[above] at (0,0) {$\scriptstyle \bar a$};
				\node[above] at (1,0) {$\scriptstyle a$};
			\end{tikzpicture}
		\end{array}\,,\\
		&\begin{array}{c}
			\begin{tikzpicture}[scale=0.5]
				\draw[thick,-stealth] (0.5,-0.5) -- (-0.5,0.5);
				\draw[white, line width = 1.2mm] (-0.5,-0.5) -- (0.5,0.5);
				\draw[thick,-stealth] (-0.5,-0.5) -- (0.5,0.5);
			\end{tikzpicture}
		\end{array}=\sum\lm_{a,b}\begin{array}{c}
			\begin{tikzpicture}[scale=0.4]
				\draw[thick] (0.5,-0.5) -- (-0.5,0.5);
				\draw[thick] (-0.5,-0.5) -- (0.5,0.5);
				\node[below left] at (-0.4,-0.4) {$\scriptstyle \color{burgundy} \bm{a}$};
				\node[below right] at (0.4,-0.4) {$\scriptstyle \color{\myblue} \bm{b}$};
				\node[above left] at (-0.4,0.4) {$\scriptstyle \color{burgundy} \bm{a}$};
				\node[above right] at (0.4,0.4) {$\scriptstyle \color{\myblue} \bm{b}$};
			\end{tikzpicture}
		\end{array}-q^{-1}\sum\lm_{\color{black!60!green} \bm{a\neq b}}\left(\begin{array}{c}
			\begin{tikzpicture}[scale=0.4]
				\draw[thick] (0.5,-0.5) -- (-0.5,0.5);
				\draw[thick] (-0.5,-0.5) -- (0.5,0.5);
				\node[below left] at (-0.4,-0.4) {$\scriptstyle \color{\myblue} \bm{b}$};
				\node[below right] at (0.4,-0.4) {$\scriptstyle \color{burgundy} \bm{a}$};
				\node[above left] at (-0.4,0.4) {$\scriptstyle \color{burgundy} \bm{a}$};
				\node[above right] at (0.4,0.4) {$\scriptstyle \color{\myblue} \bm{b}$};
			\end{tikzpicture}
		\end{array}+q^{{\rm sign}(a-b)}\begin{array}{c}
			\begin{tikzpicture}[scale=0.4]
				\draw[thick] (0.5,-0.5) -- (-0.5,0.5);
				\draw[thick] (-0.5,-0.5) -- (0.5,0.5);
				\node[below left] at (-0.4,-0.4) {$\scriptstyle \color{burgundy} \bm{a}$};
				\node[below right] at (0.4,-0.4) {$\scriptstyle \color{\myblue} \bm{b}$};
				\node[above left] at (-0.4,0.4) {$\scriptstyle \color{burgundy} \bm{a}$};
				\node[above right] at (0.4,0.4) {$\scriptstyle \color{\myblue} \bm{b}$};
			\end{tikzpicture}
		\end{array}\right)\,,\\
		&\begin{array}{c}
			\begin{tikzpicture}[xscale=-1,scale=0.5]
				\draw[thick,-stealth] (0.5,-0.5) -- (-0.5,0.5);
				\draw[white, line width = 1.2mm] (-0.5,-0.5) -- (0.5,0.5);
				\draw[thick,-stealth] (-0.5,-0.5) -- (0.5,0.5);
			\end{tikzpicture}
		\end{array}=\sum\lm_{a\neq b}\left(\begin{array}{c}
			\begin{tikzpicture}[scale=0.4]
				\draw[thick] (0.5,-0.5) -- (-0.5,0.5);
				\draw[thick] (-0.5,-0.5) -- (0.5,0.5);
				\node[below left] at (-0.4,-0.4) {$\scriptstyle a$};
				\node[below right] at (0.4,-0.4) {$\scriptstyle b$};
				\node[above left] at (-0.4,0.4) {$\scriptstyle b$};
				\node[above right] at (0.4,0.4) {$\scriptstyle a$};
			\end{tikzpicture}
		\end{array}+q^{{\rm sign}(a-b)}\begin{array}{c}
			\begin{tikzpicture}[scale=0.4]
				\draw[thick] (0.5,-0.5) -- (-0.5,0.5);
				\draw[thick] (-0.5,-0.5) -- (0.5,0.5);
				\node[below left] at (-0.4,-0.4) {$\scriptstyle a$};
				\node[below right] at (0.4,-0.4) {$\scriptstyle b$};
				\node[above left] at (-0.4,0.4) {$\scriptstyle a$};
				\node[above right] at (0.4,0.4) {$\scriptstyle b$};
			\end{tikzpicture}
		\end{array}\right)-q^{-1}\sum\lm_{a,b}\begin{array}{c}
			\begin{tikzpicture}[scale=0.4]
				\draw[thick] (0.5,-0.5) -- (-0.5,0.5);
				\draw[thick] (-0.5,-0.5) -- (0.5,0.5);
				\node[below left] at (-0.4,-0.4) {$\scriptstyle a$};
				\node[below right] at (0.4,-0.4) {$\scriptstyle b$};
				\node[above left] at (-0.4,0.4) {$\scriptstyle a$};
				\node[above right] at (0.4,0.4) {$\scriptstyle b$};
			\end{tikzpicture}
		\end{array}\,.
	\end{aligned}
\end{equation}
Here in the r.h.s. indices $a$ correspond to weights of vectors of $\Box$, and barred indices $\bar a$ correspond to $\bar\Box$.
These diagrams represent propagation of weights across the strands of a tangle, so the R-matrices correspond to permutations of weights and cups/caps to conjugations.
To derive a closed link invariant using this graphical language elements of the tangle diagram are substituted by alternative expressions from the r.h.s., so the diagram is expanded in a formal sum of graphs.
In the graphs weight indices on all edges should coincide, then a graph contribution is equal to 1, otherwise it is zero.

Let us \emph{stress} here that naturally there are no limits in summations for R-matrices in \eqref{RT}.
These expressions are universal and $n$-independent.
We might assume that the $n$-dependence is concentrated in closing caps and cups.

Comparing diagrammatic expansion for the unknot with respective cohomologies $\theta$, $x\theta$, $x^2\theta$, $\ldots$ one observes that it is natural to identify weight labels $a$ on graphs with ``Fourier modes'' $x^a$ of wave functions in cohomologies of $\Op_L$.

Let us consider a single intersection as an elementary tangle, then we could construct an operator for it (where labeling matches \eqref{Ops}):
\begin{equation}
	\Op_0=\Op_{\!\!\!\!\begin{array}{c}
			\begin{tikzpicture}[scale=0.3]
				\draw[-stealth] (0.5,-0.5) -- (-0.5,0.5);
				\draw[white, line width = 0.7mm] (-0.5,-0.5) -- (0.5,0.5);
				\draw[-stealth] (-0.5,-0.5) -- (0.5,0.5);
			\end{tikzpicture}
		\end{array}\!\!\!\!}+\Op_{\p}\left[\begin{array}{cc}
		x_1 & x_2\\
		x_4 & x_3
	\end{array}\right]\,.
\end{equation}
For matrix factorization parts we apply simplifying relations of Sec.~\ref{sec:stray} and Sec.~\ref{sec:4v_split}, then for cohomologies the following ansatz is applicable:
\begin{equation}\label{categRPsi}
	\Psi=\underset{\substack{\rotatebox[origin=c]{270}{$\in$}\\ \scalebox{0.7}{$H^*\left(\Op_{\p}\left[\begin{array}{cc}
					{\color{burgundy} \bm{x_1}} & {\color{\myblue} \bm{x_2}}\\
					{\color{burgundy} \bm{x_1}} & {\color{\myblue} \bm{x_2}}\\
				\end{array}\right]\right)$}}}{\psi(x_1,x_2)}+\epsilon\Bigg(\frac{1}{2}\left(1+\frac{x_3-x_4}{x_1-x_2}\right)\underset{\substack{\rotatebox[origin=c]{270}{$\in$}\\ \scalebox{0.7}{$H^*\left(\Op_{\p}\left[\begin{array}{cc}
					{\color{burgundy} \bm{x_1}} & {\color{\myblue} \bm{x_2}}\\
					{\color{\myblue} \bm{x_2}} & {\color{burgundy} \bm{x_1}}\\
				\end{array}\right]\right)$}}}{\xi_-(x_1,x_2)}+\frac{1}{2}\left(1-\frac{x_3-x_4}{x_1-x_2}\right)\underset{\substack{\rotatebox[origin=c]{270}{$\in$}\\ \scalebox{0.7}{$H^*\left(\Op_{\p}\left[\begin{array}{cc}
					{\color{burgundy} \bm{x_1}} & {\color{\myblue} \bm{x_2}}\\
					{\color{burgundy} \bm{x_1}} & {\color{\myblue} \bm{x_2}}\\
				\end{array}\right]\right)$}}}{\xi_+(x_1,x_2)}\Bigg)+\ldots\,,\\
\end{equation}
where dotted terms contain non-zero powers of $\theta_{1,2}$ necessarily and are uniquely defined by functions $\psi$, $\xi_{\pm}$.
Now we see clearly that the pattern of \eqref{categRPsi} matches the pattern of weight distributions in the R-matrix \eqref{RT}.
Diagrams in the r.h.s. of the R-matrix expression define weight matchings: straight (no permutation, same as $x_k$ for $\psi$), criss-cross (permutation $\sigma_{12}$, same as $x_k$ for $\xi_-$), straight (same as $x_k$ for $\xi_+$), marked additionally by the respective colors.
In addition functions $\xi_{\pm}$ are constrained $\xi_+(x,x)=\xi_-(x,x)$, that usually (yet not always) leads to a simple solution $\xi_{\pm}(x_1,x_2)=(x_1-x_2)g_{\pm}(x_1,x_2)$ for unconstrained $g_{\pm}$.
However this solution implies that wave functions $\xi_{\pm}$ in matching points $x_1=x_2$ \emph{vanish} that corresponds to constraint $a\neq b$ in the second term of the R-matrix \eqref{RT}, also we note that the coefficient $(-q^{-1})$ coincides exactly with the index of $\epsilon$.

Pattern matching is a sole observation, therefore it is not as strong as the theorem of MOY moves \eqref{MOYloc_I}-\eqref{MOYloc_V}.
Nevertheless we expect this observation to be a straightforward indication that zero modes of operator $\Op_L$ are a categorification of the usual RT formalism indeed.

\subsection{Reduction to Khovanov complex at \texorpdfstring{$n=2$}{}}

Here we would like to discuss a reduction of Khovanov-Rozansky double complex to a simple Khovanov mono-complex \cite{Khovanov:1999qla} categorifying Jones polynomials at $n=2$.

First, we would like to prove that the MOY edge is actually irrelevant and is equivalent to a pair of a cap and a cup:
\begin{equation}\label{coh-isom-n=2}
	\begin{array}{c}
		\begin{tikzpicture}
			\node(A) at (0,0) {$H^*(\MF_{\bar\MOY_1})$};
			\node(B) at (3,0) {$H^*(\MF_{\bar\MOY_2})$};
			\draw[-stealth] ([shift={(0,0.05)}]A.east) -- ([shift={(0,0.05)}]B.west) node[pos=0.5,above] {$\scriptstyle \Phi$};
			\draw[stealth-] ([shift={(0,-0.05)}]A.east) -- ([shift={(0,-0.05)}]B.west) node[pos=0.5,below] {$\scriptstyle \Phi^{-1}$};
		\end{tikzpicture}
	\end{array}\,,
\end{equation}
where $\bar\MOY_1$ and $\bar\MOY_2$ are given by
\begin{equation}
\bar\MOY_1=\begin{array}{c}
	\begin{tikzpicture}[scale=1.2]
		\draw[thick, postaction={decorate},decoration={markings, mark= at position 0.75 with {\arrow{stealth}}, mark= at position 0.25 with {\arrow{stealth}}}] (-0.5,-0.5) -- (0.5,0.5) node[left,pos=0.3] {$\scriptstyle x_4$} node[right,pos=0.7] {$\scriptstyle x_2$};
		\draw[thick, postaction={decorate},decoration={markings, mark= at position 0.75 with {\arrow{stealth}}, mark= at position 0.25 with {\arrow{stealth}}}] (0.5,-0.5) -- (-0.5,0.5) node[right,pos=0.3] {$\scriptstyle x_3$} node[left,pos=0.7] {$\scriptstyle x_1$};
		\draw[fill=\myred] (0,0) circle (0.08);
		\begin{scope}[rotate=45]
		\node[\myblue] at (1.1,0) {$\scriptstyle G$};
		\draw[fill=\myblue, even odd rule] (0,0) circle (0.7)  (0,0) circle (0.9);
		\end{scope}
	\end{tikzpicture}
\end{array},\quad \bar\MOY_2=\begin{array}{c}
		\begin{tikzpicture}[scale=0.7]
			\draw[thick] (-1,-0.7) to[out=60,in=120] (1,-0.7);
			\draw[thick] (1,0.7) to[out=240,in=300] (-1,0.7);
			\draw[fill=\myblue, even odd rule] (0,0) circle (1.2) (0,0) circle (1.5);
			\node[right] at (-1,0.7) {$\scriptstyle x_1$};
			\node[left] at (1,-0.7) {$\scriptstyle x_3$};
			\node[left] at (1,0.7) {$\scriptstyle x_2$};
			\node[right] at (-1,-0.7) {$\scriptstyle x_4$};
			\begin{scope}[rotate=45]
				\node[\myblue] at (1.8,0) {$\scriptstyle G$};
			\end{scope}
		\end{tikzpicture}
	\end{array}\,.
\end{equation}
In the $n=2$ case, we have:
\begin{equation}
	\begin{aligned}
		& u_1=x_1^2-x_2 x_1+x_3 x_1+x_4 x_1+x_2^2+x_3^2+x_4^2+x_2 x_3+x_2 x_4+2 x_3 x_4\,,\\
		& u_2=-3 x_3-3 x_4\,.
	\end{aligned}
\end{equation}
We again denote the set of internal variables for a subgraph $G$ as $V$. Respective MF differentials read:
\begin{equation}
\begin{aligned}
    \MF_{\bar\MOY_1}&=\MF_G(V,x_1,x_2,x_3,x_4)+u_1\hat\theta_1-3(x_3+x_4)\hat\theta_2+(x_1+x_2-x_3-x_4)\hat\theta_1^{\dagger}+(x_1x_2-x_3x_4)\hat\theta_2^{\dagger}\,, \\
    \MF_{\bar\MOY_2}&=\MF_G(V,x_1,x_2,x_3,x_4)+\bar\pi_{12}\hat\xi_1+\bar\pi_{34}\hat\xi_2+\left(x_1+x_2\right)\hat\xi_1^{\dagger}-\left(x_3+x_4\right)\hat\xi_2^{\dagger}\,,
\end{aligned}
\end{equation}
where
\begin{equation}
	\bar\pi_{ij}:=\frac{x_i^3+x_j^3}{x_i+x_j}=x_i^2-x_i x_j+x_j^2\,.
\end{equation}
We  note that terms in $\MF_{\bar\MOY_1}$ in front of $\hat\theta_2$ and $\hat\theta_1^{\dagger}$ are linear in $x_i$, so it would be natural to have them as new fermionic variables. Moreover, it would be nice to make them ``orthogonal'' in $x$'s. So, we perform the following canonical change of fermionic operators:
\begin{equation}
	\hat\xi_1=\hat\theta_1-\frac{1}{3}\hat\theta_2^{\dagger},\quad \hat\xi_2=\frac{1}{3}\hat\theta_2^{\dagger},\quad \hat\xi_1^{\dagger}=\hat\theta_1^{\dagger},\quad \hat\xi_2^{\dagger}=\hat\theta_1^{\dagger}+3\hat\theta_2\,,
\end{equation}
so that
\begin{equation}
	\left\{\hat\xi_i,\hat\xi_j^\dagger\right\}=\delta_{ij},\quad \left\{\hat\xi_i,\hat\xi_j\right\}=\left\{\hat\xi_i^\dagger,\hat\xi_j^\dagger\right\}=0\,.
\end{equation}
Eventually, we arrive at the following matrix factorization operator:
\begin{equation}
	\MF_{\bar\MOY_1}=\MF_G(V,x_1,x_2,x_3,x_4)+\left(\bar\pi_{12}+(x_3+x_4)s\right)\hat\xi_1+\left(\bar\pi_{34}+(x_1+x_2)s\right)\hat\xi_2+\left(x_1+x_2\right)\hat\xi_1^{\dagger}-\left(x_3+x_4\right)\hat\xi_2^{\dagger}\,,
\end{equation}
where
\begin{equation}
	s:=x_1+x_2+x_3+x_4\,.
\end{equation}
It is easy to observe that:
\begin{equation}\label{DD'}
	\MF_{\bar\MOY_1}=\MF_{\bar\MOY_2}+\left[\MF_{\bar\MOY_2},s \hat\xi_1\hat\xi_2\right]=e^{-s\hat \xi_1\hat \xi_2}\MF_{\bar\MOY_2} e^{s\hat \xi_1\hat \xi_2}\,.
\end{equation}
Relation \eqref{DD'} implies that map $e^{s\hat\xi_1\hat\xi_2}$ provides the isomorphism of the cohomologies~\eqref{coh-isom-n=2}.

This isomorphism implies that all the MOY diagrams are decomposed in collections of disjoint cycles with only 2-valent vertices.
It is trivial to compute respective cohomologies to be a 2d vector space:
\begin{equation}
	H^*\left(\!\!\!\!\scalebox{0.8}{$\begin{array}{c}
		\begin{tikzpicture}
			\draw[thick] (0,0) circle (0.4);
			\draw[fill=\mygreen] (0.4,0) circle (0.06);
			\node at (0.7,0) {$\scriptstyle\theta_1$};
			\begin{scope}[rotate=30]
				\node at (0.6,0) {$\scriptstyle x_1$};
			\end{scope}
			\begin{scope}[rotate=60]
				\draw[fill=\mygreen] (0.4,0) circle (0.06);
				\node at (0.7,0) {$\scriptstyle\theta_2$};
				\begin{scope}[rotate=30]
					\node at (0.6,0) {$\scriptstyle x_2$};
				\end{scope}
			\end{scope}
			\begin{scope}[rotate=120]
				\draw[fill=\mygreen] (0.4,0) circle (0.06);
				\node at (0.7,0) {$\scriptstyle\theta_3$};
				\begin{scope}[rotate=30]
					\node at (0.6,0) {$\scriptstyle x_3$};
				\end{scope}
			\end{scope}
			\begin{scope}[rotate=180]
				\draw[fill=\mygreen] (0.4,0) circle (0.06);
				\node at (0.7,0) {$\scriptstyle\theta_4$};
				\begin{scope}[rotate=30]
					\node at (0.6,0) {$\scriptstyle x_4$};
				\end{scope}
			\end{scope}
			\begin{scope}[rotate=240]
				\draw[fill=\mygreen] (0.4,0) circle (0.06);
				\node at (0.7,0) {$\scriptstyle\theta_5$};
				\begin{scope}[rotate=30]
					\node at (0.6,0) {$\scriptstyle x_5$};
				\end{scope}
			\end{scope}
			\begin{scope}[rotate=300]
				\draw[fill=\mygreen] (0.4,0) circle (0.06);
				\node at (0.7,0) {$\scriptstyle\cdot$};
				\begin{scope}[rotate=30]
					\node at (0.6,0) {$\scriptstyle \cdot$};
				\end{scope}
			\end{scope}
		\end{tikzpicture}
	\end{array}$}\!\!\!\!\right)\;\overset{\eqref{MOYloc_I}}{\cong}\;H^*\left(\!\!\!\!\begin{array}{c}
	\begin{tikzpicture}
		\draw[thick] (0,0) circle (0.4);
		\draw[fill=\mygreen] (0.4,0) circle (0.06);
		\node at (0.6,0) {$\scriptstyle\theta$};
		\node at (-0.55,0) {$\scriptstyle x$};
	\end{tikzpicture}
	\end{array}\!\!\!\!\right)\cong H^*\left(x^2\hat\theta\right)={\rm Span}\{\theta,x\theta\}\cong\IQ[x]/\langle x^2\rangle\,.
\end{equation}

\begin{figure}[ht!]
	\centering
	\begin{tikzpicture}
		\node at (0,0) {$\chi_{0,1}'\to\left\{\begin{array}{c}
				\begin{tikzpicture}[scale=0.4]
					\draw[fill=white!80!blue,draw=none] (0,0.5) to[out=0,in=180] (4,1.5) -- (4,-1.5) to[out=180,in=0] (0,-0.5);
					\draw[thick] (0,0.5) to[out=0,in=180] (4,1.5) (4,-1.5) to[out=180,in=0] (0,-0.5);
					\begin{scope}[yscale=0.25]
						\draw[thick, fill=white] (4.01,2) to[out=180,in=90] (2.5,0) to[out=270,in=180] (4.01,-2);
					\end{scope}
					\begin{scope}[xscale=0.3]
						\draw[thick, fill=white](0,0) circle (0.5);
					\end{scope}
					\begin{scope}[shift={(4,1)}]
						\begin{scope}[xscale=0.3]
							\draw[draw=none,fill=white!80!blue](0,0) circle (0.5);
							\draw[thick, dashed] ([shift={(90:0.5)}]0,0) arc (90:270:0.5);
							\draw[thick] ([shift={(-90:0.5)}]0,0) arc (-90:90:0.5);
						\end{scope}
					\end{scope}
					\begin{scope}[shift={(4,-1)}]
						\begin{scope}[xscale=0.3]
							\draw[draw=none,fill=white!80!blue](0,0) circle (0.5);
							\draw[thick, dashed] ([shift={(90:0.5)}]0,0) arc (90:270:0.5);
							\draw[thick] ([shift={(-90:0.5)}]0,0) arc (-90:90:0.5);
						\end{scope}
					\end{scope}
				\end{tikzpicture}
			\end{array},\quad \begin{array}{c}
			\begin{tikzpicture}[scale=0.4,xscale=-1]
				\draw[fill=white!80!blue,draw=none] (0,0.5) to[out=0,in=180] (4,1.5) -- (4,-1.5) to[out=180,in=0] (0,-0.5);
				\draw[thick] (0,0.5) to[out=0,in=180] (4,1.5) (4,-1.5) to[out=180,in=0] (0,-0.5);
				\begin{scope}[yscale=0.25]
					\draw[thick, fill=white] (4.01,2) to[out=180,in=90] (2.5,0) to[out=270,in=180] (4.01,-2);
				\end{scope}
				\begin{scope}[xscale=-0.3]
					\draw[draw=none,fill=white!80!blue](0,0) circle (0.5);
					\draw[thick, dashed] ([shift={(90:0.5)}]0,0) arc (90:270:0.5);
					\draw[thick] ([shift={(-90:0.5)}]0,0) arc (-90:90:0.5);
				\end{scope}
				\begin{scope}[shift={(4,1)}]
					\begin{scope}[xscale=0.3]
						\draw[thick, fill=white](0,0) circle (0.5);
					\end{scope}
				\end{scope}
				\begin{scope}[shift={(4,-1)}]
					\begin{scope}[xscale=0.3]
						\draw[thick, fill=white](0,0) circle (0.5);
					\end{scope}
				\end{scope}
			\end{tikzpicture}
			\end{array}\right\}$};
		\node (A) at (8,0.5) {$\begin{array}{c}
				\begin{tikzpicture}[scale=0.5]
					\draw[thick, fill=white!80!blue] (0,0.5) to[out=0,in=90] (1,0) to[out=270,in=0] (0,-0.5);
					\begin{scope}[xscale=0.3]
						\draw[thick, fill=white](0,0) circle (0.5);
					\end{scope}
				\end{tikzpicture}
			\end{array}$};
		\node[left] at ([shift={(0.3,0)}]A.west) {$\langle\theta|\to $};
		\node(A) at (8,-0.5) {$\begin{array}{c}
				\begin{tikzpicture}[scale=0.5]
					\draw[thick, fill=white!80!blue] (0,0.5) to[out=0,in=90] (1,0) to[out=270,in=0] (0,-0.5);
					\begin{scope}[xscale=0.3]
						\draw[thick, fill=white](0,0) circle (0.5);
					\end{scope}
					\draw[fill=black] (0.6,0) circle (0.1);
				\end{tikzpicture}
			\end{array}$};
		\node[left] at ([shift={(0.3,0)}]A.west) {$\langle x\theta|\to $};
		\node (A) at (5.5,0.5) {$\begin{array}{c}
				\begin{tikzpicture}[scale=0.5,xscale=-1]
					\draw[thick, fill=white!80!blue] (0,0.5) to[out=0,in=90] (1,0) to[out=270,in=0] (0,-0.5);
					\begin{scope}[xscale=-0.3]
						\draw[draw=none,fill=white!80!blue](0,0) circle (0.5);
						\draw[thick, dashed] ([shift={(90:0.5)}]0,0) arc (90:270:0.5);
						\draw[thick] ([shift={(-90:0.5)}]0,0) arc (-90:90:0.5);
					\end{scope}
					\draw[fill=black] (0.6,0) circle (0.1);
				\end{tikzpicture}
			\end{array}$};
		\node[left] at ([shift={(0.3,0)}]A.west) {$|\theta\rangle\to $};
		\node(A) at (5.5,-0.5) {$\begin{array}{c}
				\begin{tikzpicture}[scale=0.5,xscale=-1]
					\draw[thick, fill=white!80!blue] (0,0.5) to[out=0,in=90] (1,0) to[out=270,in=0] (0,-0.5);
					\begin{scope}[xscale=-0.3]
						\draw[draw=none,fill=white!80!blue](0,0) circle (0.5);
						\draw[thick, dashed] ([shift={(90:0.5)}]0,0) arc (90:270:0.5);
						\draw[thick] ([shift={(-90:0.5)}]0,0) arc (-90:90:0.5);
					\end{scope}
				\end{tikzpicture}
			\end{array}$};
		\node[left] at ([shift={(0.3,0)}]A.west) {$|x\theta\rangle\to $};
		\node(A) at (10,0) {$\begin{array}{c}
				\begin{tikzpicture}
					\draw[thick, fill=white!80!blue] (0,0) circle (0.5);
					\begin{scope}[yscale=0.3]
						\draw[thick, dashed] ([shift={(0:0.5)}]0,0) arc (0:180:0.5);
						\draw[thick] ([shift={(180:0.5)}]0,0) arc (180:360:0.5);
					\end{scope}
					\draw[fill=black] (0,0.3) circle (0.05);
				\end{tikzpicture}
			\end{array}$};
		\node[right] at (A.east) {$=\delta_{1,m}$};
		\node[above] at ([shift={(0,-0.2)}]A.north) {\scriptsize $m$ marked points};
	\end{tikzpicture}
	\caption{Rules of TQFT for $n=2$.}\label{fig:n=2TQFT}
\end{figure}
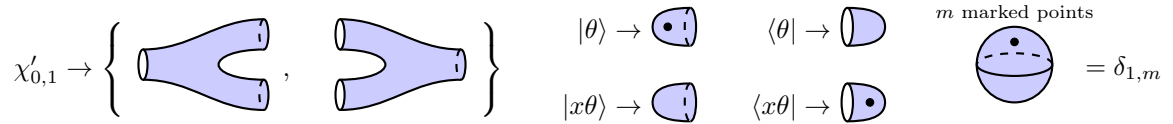

Second, due to the homomorphism requirement~\eqref{MFhomo}, the new horizontal morphism
\begin{equation}\label{chi'-map}
    \begin{array}{c}
			\begin{tikzpicture}[scale=0.7]
				\draw[thick, -stealth] (0.5,-0.5) -- (-0.5,0.5);
				\draw[white, line width = 1.4mm] (-0.5,-0.5) -- (0.5,0.5);
				\draw[thick, -stealth] (-0.5,-0.5) -- (0.5,0.5);
				\node[above left] at (-0.5,0.5) {$\scriptstyle x_1$};
				\node[above right] at (0.5,0.5) {$\scriptstyle x_2$};
				\node[below right] at (0.5,-0.5) {$\scriptstyle x_3$};
				\node[below left] at (-0.5,-0.5) {$\scriptstyle x_4$};
			\end{tikzpicture}
		\end{array}=\begin{array}{c}
		\begin{tikzpicture}[scale=0.7]
			\draw[thick, -stealth] (-0.5,-0.5) to[out=45,in=270] (-0.2,0) to[out=90,in=315] (-0.5,0.5);
			\draw[thick, -stealth] (0.5,-0.5) to[out=135,in=270] (0.2,0) to[out=90,in=225] (0.5,0.5);
			\draw[fill=\mygreen] (-0.2,0) circle (0.08) (0.2,0) circle (0.08);
			\node[above left] at (-0.5,0.5) {$\scriptstyle x_1$};
			\node[above right] at (0.5,0.5) {$\scriptstyle x_2$};
			\node[below right] at (0.5,-0.5) {$\scriptstyle x_3$};
			\node[below left] at (-0.5,-0.5) {$\scriptstyle x_4$};
			\node[left] at (-0.2,0) {$\scriptstyle \xi_1$};
			\node[right] at (0.2,0) {$\scriptstyle \xi_2$};
		\end{tikzpicture}
		\end{array}+\epsilon\begin{array}{c}
			\begin{tikzpicture}[scale=0.7]
			\draw[thick] (-0.5,-0.5) to[out=45,in=135] (0.5,-0.5);
			\draw[thick] (-0.5,0.5) to[out=315,in=225] (0.5,0.5);
            \draw[fill=\mygreen] (0,-0.3) circle (0.08) (0,0.3) circle (0.08);
			\node[above left] at (-0.5,0.5) {$\scriptstyle x_1$};
			\node[above right] at (0.5,0.5) {$\scriptstyle x_2$};
			\node[below right] at (0.5,-0.5) {$\scriptstyle x_3$};
			\node[below left] at (-0.5,-0.5) {$\scriptstyle x_4$};
            \node[left] at (0.3,0.6) {$\scriptstyle \xi_1$};
			\node[right] at (-0.3,-0.6) {$\scriptstyle \xi_2$};
		\end{tikzpicture}
		\end{array},\quad \chi'_0\left[\begin{array}{cc|c}
		x_1 & x_2 & \xi_1\\
		x_4 & x_3 & \xi_2
		\end{array}\right]:\begin{array}{c}
		\begin{tikzpicture}[scale=0.7]
			\draw[thick, -stealth] (-0.5,-0.5) to[out=45,in=270] (-0.2,0) to[out=90,in=315] (-0.5,0.5);
			\draw[thick, -stealth] (0.5,-0.5) to[out=135,in=270] (0.2,0) to[out=90,in=225] (0.5,0.5);
			\draw[fill=\mygreen] (-0.2,0) circle (0.08) (0.2,0) circle (0.08);
			\node[above left] at (-0.5,0.5) {$\scriptstyle x_1$};
			\node[above right] at (0.5,0.5) {$\scriptstyle x_2$};
			\node[below right] at (0.5,-0.5) {$\scriptstyle x_3$};
			\node[below left] at (-0.5,-0.5) {$\scriptstyle x_4$};
			\node[left] at (-0.2,0) {$\scriptstyle \xi_1$};
			\node[right] at (0.2,0) {$\scriptstyle \xi_2$};
		\end{tikzpicture}
		\end{array}\;\longrightarrow\;\begin{array}{c}
		\begin{tikzpicture}[scale=0.7]
			\draw[thick] (-0.5,-0.5) to[out=45,in=135] (0.5,-0.5);
			\draw[thick] (-0.5,0.5) to[out=315,in=225] (0.5,0.5);
            \draw[fill=\mygreen] (0,-0.3) circle (0.08) (0,0.3) circle (0.08);
			\node[above left] at (-0.5,0.5) {$\scriptstyle x_1$};
			\node[above right] at (0.5,0.5) {$\scriptstyle x_2$};
			\node[below right] at (0.5,-0.5) {$\scriptstyle x_3$};
			\node[below left] at (-0.5,-0.5) {$\scriptstyle x_4$};
            \node[left] at (0.3,0.6) {$\scriptstyle \xi_1$};
			\node[right] at (-0.3,-0.6) {$\scriptstyle \xi_2$};
		\end{tikzpicture}
		\end{array}\,.
\end{equation}
can be obtained from $\chi_0$ given by~\eqref{KRmorphisms} with $\mu=1$ and $\lambda=0$ fixed for simplicity:
\begin{equation}\label{chi'}
    \chi'_0 = e^{s\hat\xi_1\hat\xi_2} \chi_0=\chi_0 - \frac{1}{3}s\, \hat\xi_1 \hat\xi_2 \hat\xi_2^\dagger \hat\xi_1^\dagger + s\, \hat\xi_1 \hat\xi_2\,.
\end{equation}
When the free ends of tangles in~\eqref{chi'-map} are closed, this horizontal morphism would reduce to
\begin{equation}
    \mathbb{Q}[x_1,x_3]/\langle x_1^2,\,x_3^2,\,x_3=-x_1 \rangle \quad \overset{x_1 - x_3}{\longrightarrow} \quad \mathbb{Q}[x_1]/\langle x_1^2 \rangle \otimes \mathbb{Q}[x_3]/\langle x_3^2 \rangle
\end{equation}
or
\begin{equation}
    \mathbb{Q}[x_1]/\langle x_1^2 \rangle \otimes \mathbb{Q}[x_2]/\langle x_2^2 \rangle \quad \overset{1}{\longrightarrow}\quad \mathbb{Q}[x_1,x_2]/\langle x_1^2,\,x_2^2,\,x_2=-x_1 \rangle
\end{equation}
depending on the choice of the closure. 
These differentials reproduce the action of Khovanov multiplication and co-product morphisms~\cite{Khovanov:1999qla}.
Matrix elements of horizontal morphisms then could be calculated in the language of Bar-Natan's TQFT \cite{bar2005khovanov}.
Maps $\chi_{0,1}'$ are identified with ``pair-of-pants'' cobordisms.
The complete set of rules for this TQFT is depicted in Fig.~\ref{fig:n=2TQFT}, and it turns out to be algorithmic enough to boost computations of Khovanov polynomials with the help of computers \cite{bar2007fast}.
A generalization of Bar-Natan's TQFT to foam TQFT \cite{khovanov2004sl,mackaay2009sl,Chun:2015gda,lauda2015khovanov,queffelec2016sln,Carqueville:2017ono} is much more involved and much less algorithmic due to allowed surface junctions (see App.~\ref{app:MF}), nevertheless we hope a little bit more ``old fashioned'' approach with wave functions and differential equations might prove useful on this route.

In general, in a knot complex, we have full smoothings glued from several $\!\!\begin{array}{c}
	\begin{tikzpicture}[scale=0.25]
		\draw[thick] (-0.5,-0.5) to[out=45,in=270] (-0.2,0) to[out=90,in=315] (-0.5,0.5);
		\draw[thick] (0.5,-0.5) to[out=135,in=270] (0.2,0) to[out=90,in=225] (0.5,0.5);
	\end{tikzpicture}
\end{array}\!\!$ and $\!\!\begin{array}{c}
\begin{tikzpicture}[scale=0.25,rotate=90]
	\draw[thick] (-0.5,-0.5) to[out=45,in=270] (-0.2,0) to[out=90,in=315] (-0.5,0.5);
	\draw[thick] (0.5,-0.5) to[out=135,in=270] (0.2,0) to[out=90,in=225] (0.5,0.5);
\end{tikzpicture}
\end{array}\!\!$ resolutions. The cut-and-join operator $\chi'_0$ acts inside local regions of these full smoothings, when one $\!\!\begin{array}{c}
\begin{tikzpicture}[scale=0.25]
	\draw[thick] (-0.5,-0.5) to[out=45,in=270] (-0.2,0) to[out=90,in=315] (-0.5,0.5);
	\draw[thick] (0.5,-0.5) to[out=135,in=270] (0.2,0) to[out=90,in=225] (0.5,0.5);
\end{tikzpicture}
\end{array}\!\!$ resolution transforms to  $\!\!\begin{array}{c}
\begin{tikzpicture}[scale=0.25,rotate=90]
	\draw[thick] (-0.5,-0.5) to[out=45,in=270] (-0.2,0) to[out=90,in=315] (-0.5,0.5);
	\draw[thick] (0.5,-0.5) to[out=135,in=270] (0.2,0) to[out=90,in=225] (0.5,0.5);
\end{tikzpicture}
\end{array}\!\!$ resolution, and depends locally only on two respective fermionic variables, see~\eqref{chi'-map}.

Our goal is to get rid of odd variables, as proposed in Sec.~\ref{sec:stray}, in order to reduce the problem to the already analyzed case. The isomorphisms $\Phi_1$ commute with $\chi'_0$ because we get rid of ``far'' variables and act on Grassmann variables not coincident with $\chi'_0$ fermions. Proceeding this way, we transfer to new simple vertical morphisms, whose cohomologies are tensor products of 2-dimensional spaces, and to horizontal morphisms~\eqref{chi'} giving rise to Khovanov operators.

It is important to note that this reduction breaks the locality of the KR double complex. Before the reduction, all the morphisms act on local 4-valent vertices, whereas the reduction forces the morphisms to act on a whole cycle or a pair of cycles. The horizontal morphism could switch ``spin'' -- a choice of a single cycle wave function ``$1$'' or ``$x$'' -- flowing in the cycle, and switching spins in one isolated vertex might affect rather distant parts of the link diagram. We believe this non-locality is a result of entanglement \cite{Horodecki:2009zz,Melnikov:2018zfn}: despite we start with a system of mutually ``non-interacting'' operators $\Op_v^{\pm}$ corresponding to local vertices, the true invariant wave function is not a pure state, rather it is highly entangled and therefore non-local at all.


\subsection{Bipartite vertex reduction}\label{sec:biparti}

Certain simplification of KR complexes was performed by Krasner \cite{Krasner} for bipartite knots and links \cite{Anokhina:2024lbn,Anokhina:2024uso,Anokhina:2025eso}.
For bipartite links strands could be bent in a specific way to form ``locks'' of anti-parallel strands with two intersections -- bipartite vertices.
Applying the RT formalism it is possible to show that such a bipartite lock can be decomposed over untangled intersection resolutions, that allows one to decompose a bipartite link into a sum over  diagrams of non-intersecting cycles.
As it was shown in \cite{Krasner} this RT formalism simplification is lifted to the KR complex via the Gauss elimination algorithm.
Here we would like to discuss prospects of this simplification from the point of view of operators $\Op_L$.

We consider a bipartite vertex as a sole tangle.
We add the boundary correction operator as an external closing link denoted by $G$:
\begin{equation}\label{biparti}
	\begin{array}{c}
		\begin{tikzpicture}[scale=0.7]
			\draw[thick, postaction={decorate},decoration={markings, mark= at position 0.9 with {\arrow{stealth}}}] (0,0.4) to[out=0,in=120] (1,-0.7);
			\draw[thick, postaction={decorate},decoration={markings, mark= at position 0.9 with {\arrow{stealth}}}] (0,-0.4) to[out=180,in=300] (-1,0.7);
			\draw[white, line width = 1.2mm] (-1,-0.7) to[out=60,in=180] (0,0.4);
			\draw[thick, -stealth] (-1,-0.7) to[out=60,in=180] (0,0.4) -- (0.1,0.4);
			\draw[white, line width = 1.2mm] (1,0.7) to[out=240,in=0] (0,-0.4);
			\draw[thick, -stealth] (1,0.7) to[out=240,in=0] (0,-0.4) -- (-0.1,-0.4);
			\draw[fill=\myblue, even odd rule] (0,0) circle (1.2) (0,0) circle (1.5);
			\node[right] at (-1,0.7) {$\scriptstyle x_1$};
			\node[left] at (1,-0.7) {$\scriptstyle x_2$};
			\node[left] at (1,0.7) {$\scriptstyle x_3$};
			\node[right] at (-1,-0.7) {$\scriptstyle x_4$};
			\node[above] at (0,0.4) {$\scriptstyle x_5$};
			\node[below] at (0,-0.4) {$\scriptstyle x_6$};
			\begin{scope}[rotate=45]
				\node[\myblue] at (1.8,0) {$\scriptstyle G$};
			\end{scope}
		\end{tikzpicture}
	\end{array}=\underset{\MOY_{00}}{\begin{array}{c}
			\begin{tikzpicture}[scale=0.7]
				\draw[thick, postaction={decorate},decoration={markings, mark= at position 0.9 with {\arrow{stealth}}}] (-1,-0.7) to[out=60,in=300] (-1,0.7);
				\draw[thick, postaction={decorate},decoration={markings, mark= at position 0.9 with {\arrow{stealth}}}] (1,0.7) to[out=240,in=120] (1,-0.7);
				\draw[thick, postaction={decorate},decoration={markings, mark= at position 0.8 with {\arrow{stealth}}}] (0,0.4) to[out=0,in=90] (0.5,0) to[out=270,in=0] (0,-0.4);
				\draw[thick, postaction={decorate},decoration={markings, mark= at position 0.8 with {\arrow{stealth}}}] (0,-0.4) to[out=180,in=270] (-0.5,0) to[out=90,in=180] (0,0.4);
				\draw[fill=\myblue, even odd rule] (0,0) circle (1.2) (0,0) circle (1.5);
				\node[right] at (-1,0.7) {$\scriptstyle x_1$};
				\node[left] at (1,-0.7) {$\scriptstyle x_2$};
				\node[left] at (1,0.7) {$\scriptstyle x_3$};
				\node[right] at (-1,-0.7) {$\scriptstyle x_4$};
				\node[above] at (0,0.4) {$\scriptstyle x_5$};
				\node[below] at (0,-0.4) {$\scriptstyle x_6$};
				\begin{scope}[rotate=45]
					\node[\myblue] at (1.8,0) {$\scriptstyle G$};
				\end{scope}
			\end{tikzpicture}
	\end{array}}\oplus \epsilon_1 \underset{\MOY_{10}}{\begin{array}{c}
			\begin{tikzpicture}[scale=0.7]
				\draw[thick, postaction={decorate},decoration={markings, mark= at position 0.9 with {\arrow{stealth}}}] (0,-0.4) to[out=180,in=300] (-1,0.7);
				\draw[thick,-stealth] (-1,-0.7) to[out=60,in=180] (0,0.4) -- (0.1,0.4);
				\draw[thick] (0,0.4) to[out=0,in=90] (0.5,0) to[out=270,in=0] (0,-0.4);
				\draw[thick, postaction={decorate},decoration={markings, mark= at position 0.9 with {\arrow{stealth}}}] (1,0.7) to[out=240,in=120] (1,-0.7);
				\draw[fill=\myred] (-0.63,0) circle (0.1);
				\draw[fill=\myblue, even odd rule] (0,0) circle (1.2) (0,0) circle (1.5);
				\node[right] at (-1,0.7) {$\scriptstyle x_1$};
				\node[left] at (1,-0.7) {$\scriptstyle x_2$};
				\node[left] at (1,0.7) {$\scriptstyle x_3$};
				\node[right] at (-1,-0.7) {$\scriptstyle x_4$};
				\node[above] at (0,0.4) {$\scriptstyle x_5$};
				\node[below] at (0,-0.4) {$\scriptstyle x_6$};
				\begin{scope}[rotate=45]
					\node[\myblue] at (1.8,0) {$\scriptstyle G$};
				\end{scope}
			\end{tikzpicture}
	\end{array}}\oplus \epsilon_2 \underset{\MOY_{01}}{\begin{array}{c}
			\begin{tikzpicture}[scale=0.7]
				\draw[thick, postaction={decorate},decoration={markings, mark= at position 0.9 with {\arrow{stealth}}}] (0,0.4) to[out=0,in=120] (1,-0.7);
				\draw[thick, -stealth] (1,0.7) to[out=240,in=0] (0,-0.4) -- (-0.1,-0.4);
				\draw[thick] (0,-0.4) to[out=180,in=270] (-0.5,0) to[out=90,in=180] (0,0.4);
				\draw[thick, postaction={decorate},decoration={markings, mark= at position 0.9 with {\arrow{stealth}}}] (-1,-0.7) to[out=60,in=300] (-1,0.7);
				\draw[fill=\myred](0.63,0) circle (0.1);
				\draw[fill=\myblue, even odd rule] (0,0) circle (1.2) (0,0) circle (1.5);
				\node[right] at (-1,0.7) {$\scriptstyle x_1$};
				\node[left] at (1,-0.7) {$\scriptstyle x_2$};
				\node[left] at (1,0.7) {$\scriptstyle x_3$};
				\node[right] at (-1,-0.7) {$\scriptstyle x_4$};
				\node[above] at (0,0.4) {$\scriptstyle x_5$};
				\node[below] at (0,-0.4) {$\scriptstyle x_6$};
				\begin{scope}[rotate=45]
					\node[\myblue] at (1.8,0) {$\scriptstyle G$};
				\end{scope}
			\end{tikzpicture}
	\end{array}}\oplus\epsilon_1\epsilon_2 \underset{\MOY_{11}}{\begin{array}{c}
			\begin{tikzpicture}[scale=0.7]
				\draw[thick, postaction={decorate},decoration={markings, mark= at position 0.9 with {\arrow{stealth}}}] (0,0.4) to[out=0,in=120] (1,-0.7);
				\draw[thick, postaction={decorate},decoration={markings, mark= at position 0.9 with {\arrow{stealth}}}] (0,-0.4) to[out=180,in=300] (-1,0.7);
				\draw[thick, -stealth] (-1,-0.7) to[out=60,in=180] (0,0.4) -- (0.1,0.4);
				\draw[thick, -stealth] (1,0.7) to[out=240,in=0] (0,-0.4) -- (-0.1,-0.4);
				\draw[fill=\myred] (-0.63,0) circle (0.1) (0.63,0) circle (0.1);
				\draw[fill=\myblue, even odd rule] (0,0) circle (1.2) (0,0) circle (1.5);
				\node[right] at (-1,0.7) {$\scriptstyle x_1$};
				\node[left] at (1,-0.7) {$\scriptstyle x_2$};
				\node[left] at (1,0.7) {$\scriptstyle x_3$};
				\node[right] at (-1,-0.7) {$\scriptstyle x_4$};
				\node[above] at (0,0.4) {$\scriptstyle x_5$};
				\node[below] at (0,-0.4) {$\scriptstyle x_6$};
				\begin{scope}[rotate=45]
					\node[\myblue] at (1.8,0) {$\scriptstyle G$};
				\end{scope}
			\end{tikzpicture}
	\end{array}}\,.
\end{equation}
For MF operators, we have:
\begin{equation}
	\begin{aligned}
		& \MF_{\MOY_{00}}=\MF_G(x_1,x_2,x_3,x_4)+\pi_{14}\hat\theta_1+x_{14}\hat\theta_1^{\dagger}+\pi_{56}\hat\theta_2+x_{56}\hat\theta_2^{\dagger}+\pi_{23}\hat\theta_3+x_{23}\hat\theta_3^{\dagger}+\pi_{56}\hat\theta_4+x_{65}\hat\theta_4^{\dagger}\,,\\
		& \MF_{\MOY_{10}}=\MF_G(x_1,x_2,x_3,x_4)+u_1\left[\begin{array}{cc}
			x_1 & x_5\\
			x_4 & x_6\\
		\end{array}\right]\hat\theta_1+s_1\left[\begin{array}{cc}
			x_1 & x_5\\
			x_4 & x_6\\
		\end{array}\right]\hat\theta_1^{\dagger}+u_2\left[\begin{array}{cc}
			x_1 & x_5\\
			x_4 & x_6\\
		\end{array}\right]\hat\theta_2+s_2\left[\begin{array}{cc}
			x_1 & x_5\\
			x_4 & x_6\\
		\end{array}\right]\hat\theta_2^{\dagger}+\\
		&+\pi_{23}\hat\theta_3+x_{23}\hat\theta_3^{\dagger}+\pi_{56}\hat\theta_4+x_{65}\hat\theta_4^{\dagger}\,,\\
		& \MF_{\MOY_{01}}=\MF_G(x_1,x_2,x_3,x_4)+\pi_{14}\hat\theta_1+x_{14}\hat\theta_1^{\dagger}+\pi_{56}\hat\theta_2+x_{56}\hat\theta_2^{\dagger}+\\
		&+u_1\left[\begin{array}{cc}
			x_2 & x_6\\
			x_3 & x_5\\
		\end{array}\right]\hat\theta_3+s_1\left[\begin{array}{cc}
			x_2 & x_6\\
			x_3 & x_5\\
		\end{array}\right]\hat\theta_3^{\dagger}+u_2\left[\begin{array}{cc}
			x_2 & x_6\\
			x_3 & x_5\\
		\end{array}\right]\hat\theta_4+s_2\left[\begin{array}{cc}
			x_2 & x_6\\
			x_3 & x_5\\
		\end{array}\right]\hat\theta_4^{\dagger}\,,\\
		& \MF_{\MOY_{11}}=\MF_G(x_1,x_2,x_3,x_4)+u_1\left[\begin{array}{cc}
			x_1 & x_5\\
			x_4 & x_6\\
		\end{array}\right]\hat\theta_1+s_1\left[\begin{array}{cc}
			x_1 & x_5\\
			x_4 & x_6\\
		\end{array}\right]\hat\theta_1^{\dagger}+u_2\left[\begin{array}{cc}
			x_1 & x_5\\
			x_4 & x_6\\
		\end{array}\right]\hat\theta_2+s_2\left[\begin{array}{cc}
			x_1 & x_5\\
			x_4 & x_6\\
		\end{array}\right]\hat\theta_2^{\dagger}+\\
		&+u_1\left[\begin{array}{cc}
			x_2 & x_6\\
			x_3 & x_5\\
		\end{array}\right]\hat\theta_3+s_1\left[\begin{array}{cc}
			x_2 & x_6\\
			x_3 & x_5\\
		\end{array}\right]\hat\theta_3^{\dagger}+u_2\left[\begin{array}{cc}
			x_2 & x_6\\
			x_3 & x_5\\
		\end{array}\right]\hat\theta_4+s_2\left[\begin{array}{cc}
			x_2 & x_6\\
			x_3 & x_5\\
		\end{array}\right]\hat\theta_4^{\dagger}\,,\\
	\end{aligned}
\end{equation}
where we denoted
\begin{equation}
	s_1\left[\begin{array}{cc}
		x_2 & x_6\\
		x_3 & x_5\\
	\end{array}\right]:=x_2 + x_6 - x_3 - x_5\,,\quad s_2\left[\begin{array}{cc}
	x_2 & x_6\\
	x_3 & x_5\\
	\end{array}\right]:=x_2 x_6 - x_3 x_5\,.
\end{equation}

Horizontal differential in this system reads:
\begin{equation}
	\fD=\fD_G+\hat\epsilon_1\chi_0\left[\begin{array}{cc}
		x_1 & x_5\\
		x_4 & x_6\\
	\end{array}\right]+\hat\epsilon_2\chi_0\left[\begin{array}{cc}
		x_2 & x_6\\
		x_3 & x_5\\
	\end{array}\right]\,.
\end{equation}

Also we would like to introduce the following notations:
\begin{equation}
	\MF_+:=\MF_G(x_1,x_2,x_2,x_1)=\begin{array}{c}
		\begin{tikzpicture}[scale=0.7]
			\draw[thick, postaction={decorate},decoration={markings, mark= at position 0.9 with {\arrow{stealth}}}] (-1,-0.7) to[out=60,in=300] (-1,0.7);
			\draw[thick, postaction={decorate},decoration={markings, mark= at position 0.9 with {\arrow{stealth}}}] (1,0.7) to[out=240,in=120] (1,-0.7);
			\draw[fill=\myblue, even odd rule] (0,0) circle (1.2) (0,0) circle (1.5);
			\node[right] at (-1,0.7) {$\scriptstyle x_1$};
			\node[left] at (1,-0.7) {$\scriptstyle x_2$};
			\node[left] at (1,0.7) {$\scriptstyle x_2$};
			\node[right] at (-1,-0.7) {$\scriptstyle x_1$};
			\begin{scope}[rotate=45]
				\node[\myblue] at (1.8,0) {$\scriptstyle G$};
			\end{scope}
		\end{tikzpicture}
	\end{array},\quad \MF_-:=\MF_G(x_1,x_2,x_1,x_2)=\begin{array}{c}
		\begin{tikzpicture}[scale=0.7]
			\draw[thick, postaction={decorate},decoration={markings, mark= at position 0.9 with {\arrow{stealth}}}] (-1,-0.7) to[out=60,in=120] (1,-0.7);
			\draw[thick, postaction={decorate},decoration={markings, mark= at position 0.9 with {\arrow{stealth}}}] (1,0.7) to[out=240,in=300] (-1,0.7);
			\draw[fill=\myblue, even odd rule] (0,0) circle (1.2) (0,0) circle (1.5);
			\node[right] at (-1,0.7) {$\scriptstyle x_1$};
			\node[left] at (1,-0.7) {$\scriptstyle x_2$};
			\node[left] at (1,0.7) {$\scriptstyle x_1$};
			\node[right] at (-1,-0.7) {$\scriptstyle x_2$};
			\begin{scope}[rotate=45]
				\node[\myblue] at (1.8,0) {$\scriptstyle G$};
			\end{scope}
		\end{tikzpicture}
	\end{array}\,.
\end{equation}

All the MOY diagrams in the r.h.s. of \eqref{biparti} could be simplified by MOY moves \eqref{MOYloc_I}-\eqref{MOYloc_V}.
However to define the action of horizontal morphisms on these graphs it is required to calculate wave functions explicitly.
Those wave functions could be calculated using methods discussed in Sec.~\ref{sec:stray} and in Sec.~\ref{sec:4v_split}.
In particular for $\MOY_{11}$ one has to apply MOY move \eqref{MOYloc_V}.
Let us discuss this move in detail.

At the first step we split the left 4-valent vertex of $\MOY_{11}$ using techniques of Sec.~\ref{sec:4v_split}.
First consider root $x_4=x_1$, $x_6=x_5$:
\begin{equation}
	\begin{aligned}
		&\MF_{\MOY_{11}}\big|_{\substack{x_4=x_1\\
				x_6=x_5}}=\MF_G(x_1,x_2,x_3,x_1)+u_1\left[\begin{array}{cc}
			x_2 & x_5\\
			x_3 & x_5\\
		\end{array}\right]\hat\theta_3+\left(x_2-x_3\right)\hat\theta_3^{\dagger}+u_2\left[\begin{array}{cc}
			x_2 & x_5\\
			x_3 & x_5\\
		\end{array}\right]\hat\theta_4+x_5\left(x_2-x_3\right)\hat\theta_4^{\dagger}\,.
	\end{aligned}
\end{equation}
This MF operator could be reduced further with respect to root $x_2=x_3$ according to Sec.~\ref{sec:stray}:
\begin{equation}
	\MF_{\MOY_{11}}\big|_{\substack{x_4=x_1\\
			x_6=x_5\\ x_3=x_2}}=\MF_++p_{n-1}(x_5,x_2)\hat\theta_4\,,
\end{equation}
where $p_{n-1}$ is a polynomial given by \eqref{ppoly}.

Now we consider the other root $x_4=x_5$, $x_6=x_1$ at the first step:
\begin{equation}
	\begin{aligned}
		& \MF_{\MOY_{11}}\big|_{\substack{x_4=x_5\\ x_6=x_1}}=\MF_G(x_1,x_2,x_3,x_5)+u_1\left[\begin{array}{cc}
			x_2 & x_1\\
			x_3 & x_5\\
		\end{array}\right]\hat\theta_3+s_1\left[\begin{array}{cc}
			x_2 & x_1\\
			x_3 & x_5\\
		\end{array}\right]\hat\theta_3^{\dagger}+u_2\left[\begin{array}{cc}
			x_2 & x_1\\
			x_3 & x_5\\
		\end{array}\right]\hat\theta_4+s_2\left[\begin{array}{cc}
			x_2 & x_1\\
			x_3 & x_5\\
		\end{array}\right]\hat\theta_4^{\dagger}\,.
	\end{aligned}
\end{equation}
This MF operator could be reduced further again by splitting the second 4-valent vertex, there are two roots:
\begin{equation}
	\MF_{\MOY_{11}}\big|_{\substack{x_4=x_5=x_6=x_1\\ x_3=x_2}}=\MF_+,\quad \MF_{\MOY_{11}}\big|_{\substack{x_4=x_5=x_2\\ x_6=x_3=x_1}}=\MF_-\,.
\end{equation}

Eventually, we could construct the following tree of wave function reductions:
\begin{equation}
	\begin{array}{c}
		\begin{tikzpicture}
			\node[draw,rounded corners =10] (A) at (0,0) {\scalebox{0.7}{$\Psi\left[\begin{array}{ccc|cc}
						x_1 & x_5 & x_3 & \theta_1 & \theta_3\\
						x_4 & x_6 & x_2 & \theta_2 & \theta_4
					\end{array}\right],\MF_{\MOY_{11}}$}};
			\node[draw,rounded corners =10] (B) at (6,1) {\scalebox{0.7}{$\Psi\left[\begin{array}{ccc|cc}
						x_1 & x_5 & x_3 & 0 & \theta_3\\
						x_1 & x_5 & x_2 & 0 & \theta_4
					\end{array}\right],\MF_{1}$}};
			\node[draw,rounded corners =10] (C) at (12,1) {\scalebox{0.7}{$\begin{array}{c}
						\Psi\left[\begin{array}{ccc|cc}
							x_1 & x_5 & x_2 & 0 & 0\\
							x_1 & x_5 & x_2 & 0 & \theta_4
						\end{array}\right],\\
						\MF_++p_{n-1}(x_5,x_2)\hat\theta_4
					\end{array}$}};
			\node[draw,rounded corners =10] (D) at (6,-1) {\scalebox{0.7}{$\Psi\left[\begin{array}{ccc|cc}
						x_1 & x_5 & x_3 & 0 & \theta_3\\
						x_5 & x_1 & x_2 & 0 & \theta_4
					\end{array}\right],\MF_{2}$}};
			\node[draw,rounded corners =10] (E) at (12,-0.5) {\scalebox{0.7}{$\Psi\left[\begin{array}{ccc|cc}
						x_1 & x_1 & x_2 & 0 & 0\\
						x_1 & x_1 & x_2 & 0 & 0
					\end{array}\right],\MF_+$}};
			\node[draw,rounded corners =10] (F) at (12,-1.5) {\scalebox{0.7}{$\Psi\left[\begin{array}{ccc|cc}
						x_1 & x_2 & x_1 & 0 & 0\\
						x_2 & x_1 & x_2 & 0 & 0
					\end{array}\right],\MF_-$}};
			\path (A) edge[->] node[above] {\scalebox{0.7}{$\begin{array}{c}
						x_4=x_1\\
						x_6=x_5
					\end{array}$}} (B);
			\path (B) edge[->] node[above] {\scalebox{0.7}{$\begin{array}{c}
						x_3=x_2
					\end{array}$}} (C);
			\path (A) edge[->] node[below] {\scalebox{0.7}{$\begin{array}{c}
						x_4=x_5\\
						x_6=x_1
					\end{array}$}} (D);
			\path (D) edge[->] node[above] {\scalebox{0.7}{$\begin{array}{c}
						x_5=x_1\\
						x_3=x_2
					\end{array}$}} (E);
			\path (D) edge[->] node[below] {\scalebox{0.7}{$\begin{array}{c}
						x_5=x_2\\
						x_3=x_1
					\end{array}$}} (F);
		\end{tikzpicture}
	\end{array}
\end{equation}

According to Sec.~\ref{sec:stray} wave function for MF operator $\MF_1$ takes the following form:
\begin{equation}\label{Psi1}
	\Psi_1=U_1\left(x_5^k\theta_4\psi_+(x_1,x_2)\right)+\MF_1\Theta_1,\quad 0\leq k\leq n-2\,,
\end{equation}
where $U_1$ is an appropriate dressing operator and $\psi_+\in H^*(\MF_+)$.

For MF operator $\MF_2$ we find following Sec.~\ref{sec:4v_split}:
\begin{equation}\label{Psi2}
	\Psi_2=U_2\left(\frac{1}{2}\left(1+\frac{x_5-x_3}{x_1-x_2}\right)\xi_+(x_1,x_2)+\frac{1}{2}\left(1+\frac{x_3-x_5}{x_1-x_2}\right)\xi_-(x_1,x_2)\right)+\MF_2\Theta_2\,,
\end{equation}
where $U_2$ is also a dressing operator, and $\xi_{\pm}\in H^*(\MF_{\pm})$, $\xi_+(x,x)=\xi_-(x,x)$.
Now we should impose matching point constraint from the first 4-valent vertex splitting $\Psi_{1}\big|_{x_5=x_1}=\Psi_{2}\big|_{x_5=x_1}$.
Actually,
\begin{equation}
	\scalebox{0.85}{$\MF_3:=\MF_1\big|_{x_5=x_1}=\MF_2\big|_{x_5=x_1}=\MF_G(x_1,x_2,x_3,x_1)+u_1\left[\begin{array}{cc}
		x_2 & x_1\\
		x_3 & x_1\\
	\end{array}\right]\hat\theta_3+\left(x_2-x_3\right)\hat\theta_3^{\dagger}+u_2\left[\begin{array}{cc}
		x_2 & x_1\\
		x_3 & x_1\\
	\end{array}\right]\hat\theta_4+x_1\left(x_2-x_3\right)\hat\theta_4^{\dagger}\,.$}
\end{equation}
Its cohomology coincides with cohomology of $\MF_1$ with substituted $x_5=x_1$.
So we impose the following condition:
\begin{equation}\label{constr}
	\begin{aligned}
		&U_1'\left(x_1^k\theta_4\psi_+(x_1,x_2)\right)-U_2'\left(\xi_+(x_1,x_2)+\frac{x_2-x_3}{x_1-x_2}\cdot \frac{\xi_+(x_1,x_2)-\xi_-(x_1,x_2)}{2}\right)=\MF_3\left(\Theta_2'-\Theta_1'\right)\,,
	\end{aligned}
\end{equation}
where primes denote proper projected functions to $x_5=x_1$.
We should note that dressing operators $U_{1,2}$ do not alter the lowest fermionic degree, so for the functions $\psi$ appearing at the first degree and $\xi_{\pm}$ appearing at the zeroth degree the only option not to belong to ${\rm Im}\,\MF_3$ is to be proportional to root $x_1-x_5$.
Requiring this proportionality from wave functions \eqref{Psi1}, \eqref{Psi2} we will find that the generic cohomological solution reads:
\begin{equation}
	\begin{aligned}
		\Psi=\frac{1}{2}\left(1+\frac{x_6-x_4}{x_1-x_5}\right) U_1(x_5^{k-1}(x_5-x_1)\theta_4\psi_+(x_1,x_2))+&\frac{1}{2}\left(1-\frac{x_6-x_4}{x_1-x_5}\right)U_2\left((x_5-x_1)\xi_-(x_1,x_2)\right)\,,\\
		& 1\leq k\leq n-2,\quad  \psi_+\in H^*(\MF_+),\quad \xi_-\in H^*(\MF_-)\,.
	\end{aligned}
\end{equation}
component $\xi_+$ is a projection to $x_5=x_1$ so it is annihilated, also mode $x_5^0\theta_4\psi_+$ is annihilated by constraint \eqref{constr}.
This relation is a wave function manifestation of MOY move \eqref{MOYloc_V}.

Having constructed wave functions explicitly one could solve for the cohomologies of the respective operator $\Op_L$.
This equation can be solved partly by separating variables analogously to Sec.~\ref{sec:invariant}.
We will present a detailed derivation elsewhere.
One would find that the wave function splits in three parts:
\begin{equation}
	\Psi_{\rm bipt}=V_1\psi_1(x_1,x_2)+\epsilon_1V_2\psi_2(x_1,x_2)+\epsilon_1\epsilon_2V_3\psi_3(x_1,x_2)\,,
\end{equation}
where $\psi_{1,2}\in H^*(\MF_+)$ and $\psi_3\in H^*(\MF_-)$, and $V_i$ are invertible dressing factor operators we do not need to know an explicit form of at the moment.
The horizontal morphism would reduce to:
\begin{equation}
	\begin{aligned}
		&\fD=\fD_G+\hat\epsilon_1V_2(x_1-x_2)V_1^{-1}+\hat\epsilon_2\hat\epsilon_1\hat\epsilon_1^{\dagger}V_3\Sigma V_2^{-1}\,,\\
		&\Sigma=\frac{\MF_G(x_1,x_2,x_1,x_2)-\MF_G(x_1,x_2,x_2,x_1)}{x_1-x_2}\,.
	\end{aligned}
\end{equation}

This statement is equivalent to Krasner's complex \cite{Krasner} for a bipartite vertex:
\begin{equation}
	\begin{array}{c}
		\begin{tikzpicture}[scale=0.6]
			\draw[thick, postaction={decorate},decoration={markings, mark= at position 0.9 with {\arrow{stealth}}}] (0,0.4) to[out=0,in=120] (1,-0.7);
			\draw[thick, postaction={decorate},decoration={markings, mark= at position 0.9 with {\arrow{stealth}}}] (0,-0.4) to[out=180,in=300] (-1,0.7);
			\draw[white, line width = 1.2mm] (-1,-0.7) to[out=60,in=180] (0,0.4);
			\draw[thick, -stealth] (-1,-0.7) to[out=60,in=180] (0,0.4) -- (0.1,0.4);
			\draw[white, line width = 1.2mm] (1,0.7) to[out=240,in=0] (0,-0.4);
			\draw[thick, -stealth] (1,0.7) to[out=240,in=0] (0,-0.4) -- (-0.1,-0.4);
			\node[above] at (-1,0.7) {$\scriptstyle x_1$};
			\node[below] at (1,-0.7) {$\scriptstyle x_2$};
			\node[above] at (1,0.7) {$\scriptstyle x_3$};
			\node[below] at (-1,-0.7) {$\scriptstyle x_4$};
			\node[above] at (0,0.4) {$\scriptstyle x_5$};
			\node[below] at (0,-0.4) {$\scriptstyle x_6$};
		\end{tikzpicture}
	\end{array}=\begin{array}{c}
		\begin{tikzpicture}
			\node (A) at (0,0) {$\begin{array}{c}
					\begin{tikzpicture}[scale=0.6]
						\draw[thick, postaction={decorate},decoration={markings, mark= at position 0.9 with {\arrow{stealth}}}] (-1,-0.7) to[out=60,in=300] (-1,0.7);
						\draw[thick, postaction={decorate},decoration={markings, mark= at position 0.9 with {\arrow{stealth}}}] (1,0.7) to[out=240,in=120] (1,-0.7);
						\node[above] at (-1,0.7) {$\scriptstyle x_1$};
						\node[below] at (1,-0.7) {$\scriptstyle x_2$};
						\node[above] at (1,0.7) {$\scriptstyle x_2$};
						\node[below] at (-1,-0.7) {$\scriptstyle x_1$};
					\end{tikzpicture}
				\end{array}$};
			\node (B) at (4,0) {$\begin{array}{c}
					\begin{tikzpicture}[scale=0.6]
						\draw[thick, postaction={decorate},decoration={markings, mark= at position 0.9 with {\arrow{stealth}}}] (-1,-0.7) to[out=60,in=300] (-1,0.7);
						\draw[thick, postaction={decorate},decoration={markings, mark= at position 0.9 with {\arrow{stealth}}}] (1,0.7) to[out=240,in=120] (1,-0.7);
						\node[above] at (-1,0.7) {$\scriptstyle x_1$};
						\node[below] at (1,-0.7) {$\scriptstyle x_2$};
						\node[above] at (1,0.7) {$\scriptstyle x_2$};
						\node[below] at (-1,-0.7) {$\scriptstyle x_1$};
					\end{tikzpicture}
				\end{array}$};
			\node (C) at (8,0) {$\begin{array}{c}
					\begin{tikzpicture}[scale=0.6]
						\draw[thick, postaction={decorate},decoration={markings, mark= at position 0.9 with {\arrow{stealth}}}] (-1,-0.7) to[out=60,in=120] (1,-0.7);
						\draw[thick, postaction={decorate},decoration={markings, mark= at position 0.9 with {\arrow{stealth}}}] (1,0.7) to[out=240,in=300] (-1,0.7);
						\node[above] at (-1,0.7) {$\scriptstyle x_1$};
						\node[below] at (1,-0.7) {$\scriptstyle x_2$};
						\node[above] at (1,0.7) {$\scriptstyle x_1$};
						\node[below] at (-1,-0.7) {$\scriptstyle x_2$};
					\end{tikzpicture}
				\end{array}$};
			\path (A) edge[->] node[above]{$\scriptstyle (x_1-x_2)\cdot$} (B) (B) edge[->] node[above]{$\scriptstyle \Sigma$} (C);
		\end{tikzpicture}
	\end{array}
\end{equation}


\section{Conclusion}

The powerful functional integral technique allows one to calculate correlators of Wilson loops in Yang-Mills theory.
In the particular case of the $3d$ Chern-Simons theory, it reduces to the RT formalism \cite{Reshetikhin:1990pr,Reshetikhin:1991tc,mironov2012character} for the evaluation of
topologically invariant HOMFLY-like polynomials, which can also be considered as Euler characteristics
of certain complexes.
Usually, Euler characteristic has two levels of description -- as an alternated sum of dimensions of spaces
(of all forms with different gradings) and as that of their zero modes (holomorphic or harmonic).
Once zero modes show up, one can study them separately and not obligatorily only alternatively summed
--  this leads to a generalization of Euler characteristic to Poincaré polynomial and, perhaps, even further.
Instead the integrals of curvatures and their functional-integral generalizations no longer work --
and a new formalism is needed.
Moreover, the study of zero modes requires a kind of Hamiltonian, which distinguishes them from everything else --
and conceptually this adds one extra dimension to the problem.
This addition is not important for Euler characteristic itself, but becomes crucial for Poincaré generalization.
In the case of the $3d$ Chern-Simons theory, this leads from RT to KR calculus, which naturally possesses formulations
in $4d$ and even $5d$ and $6d$ terms \cite{Witten:2011xiq,Witten:2011zz}.

If we are interested in building an efficient calculational technique, these general concepts should be reduced
to more compact and handleable kinds of formalism.
In this paper, we substitute a  sophisticated machinery of abstract matrix factorization by a straightforward
analysis of odd differential operators, associated with link diagrams.
The calculus is essentially cohomological, still applicable to open tangles --
what is crucially important for knot theory, at least for calculational purposes.
As such, this formalism seems very promising for further analysis of KR \cite{KRI,KRII} and superpolynomials \cite{Gukov:2004hz,Dunfield:2005si},
which so far remains restricted to just a few examples -- either for very small knots or to knots with
special additional structures, like torus \cite{Aganagic:2012ne,dunin2013superpolynomials}, twist \cite{dunin2022evolution} or pretzel \cite{anokhina2019nimble} knots.

We list here just a few straightforward applications, which hopefully will not take too much time to complete.

\begin{itemize}
	\item Generalizations from fundamental to other representations and algebras -- straightforward in the RT approach --
	imply the change from $w(x)=x^{n+1}$ for the fundamental representation of $\fs\fu_n$
	to more sophisticated functions of more sophisticated variables.
	Does the cabling procedure \cite{Anokhina:2013ica,cooper2012categorification,2013arXiv1309.3796W} lifted to operators $\Op_L$ allow one to produce new superpotentials and vertices?

	\item Bipartite knots, made entirely from antiparallel lock tangles, possess Kauffman planar decomposition \cite{Kauffman}
	for any $n$, not just $n=2$.
	This makes the theory of fundamental bipartite HOMFLY polynomials nearly trivial \cite{Anokhina:2024lbn} --
	though some problems persist for higher representations, even symmetric \cite{Anokhina:2024uso}.
	Planar decomposition is immediately seen in our operator formalism, and building a systematic bipartite KR theory,
	originated in \cite{Krasner,Anokhina:2025nti}, seems now within reach.
	
	\item $A$- and $C$-polynomials \cite{Garoufalidis_2006,Cpols} can now be lifted to the level of KR polynomials.
	A mixture of approaches of \cite{galakhov2025geometric,galakhov2025geometric-2} and the present paper seems especially interesting.

\end{itemize}

Coming back to the main story of this paper, the main challenge is to make it fully algorithmic and computerized --
so that operators, their zero modes and Poincare polynomial for a given link diagram could be found automatically.
In fact, we are already close to this level of understanding.
What remains rather abstract and requires more work, are certain steps in computation of the operators $\Op_L$,
especially for open tangles.
As to the zero modes of the Hamiltonian \eqref{Hamiltonian} on the space of graded polynomials,
the procedure is nearly algorithmic.
All the graded subspaces of polynomials are finite and the range of gradings, contributing to the zero modes,
could be estimated from counting the resolutions (MOY dimensions).
Surely, there is a need for optimization to make the procedure practical, and better algorithms are still to be found.

\section*{Acknowledgments}
We would like to thank Alexandra Anokhina and Radomir Stepanov for illuminating discussions.
The work was partially funded within the state assignment of the Institute for Information Transmission Problems
of RAS.


\appendix

\section{MF from topological strings}\label{app:MF}

One of the most popular model for topological strings is a 2d $\CN=(2,2)$ supersymmetric Landau-Ginzburg model \cite{Kapustin:2002bi, D-book_1}.
For simplicity we present its action for a single chiral multiplet with top complex scalar component $\phi$ and fermionic partners $\psi_{\alpha}$ .
The potential for this model is defined by a holomorphic superpotential function $W(\phi)$.
The action of this model reads:
\begin{equation}
	\begin{aligned}
	\CS_{\chi}(\phi,\psi_{\alpha},W)
	=\int dx^0dx^1\,\Bigg[&|\p_0\phi|^2-|\p_1\phi|^2+\I \bar\psi_-\left( \p_0+\p_1 \right)\psi_-+\I \bar\psi_+\left( \p_0-\p_1 \right)\psi_+-\\
	&-\frac{1}{4}|W'|^2-\frac{1}{2}W''\psi_+\psi_-+\frac{1}{2}\overline{W}''\bar\psi_+\bar\psi_-\Bigg]\,.
	\end{aligned}
\end{equation}
It is invariant (if no boundary of the worldsheet is present) with respect to the following SUSY transforms:
\begin{equation}
	\begin{aligned}
		&\delta \phi=\epsilon_+\psi_--\epsilon_-\psi_+\,,\\
		&\delta \psi_+=\I\bar\epsilon_-(\p_0+\p_1)\phi-\frac{1}{2}\epsilon_+ \overline{W}'\,,\\
		&\delta \psi_-=-\I\bar\epsilon_+(\p_0-\p_1)\phi-\frac{1}{2}\epsilon_-  \overline{W}'\,.
	\end{aligned}
\end{equation}
Supercharges anti-commute to translation symmetries, and if some boundary of the worldsheet is present some part of the supersymmetry should be broken as translation perpendicular to the boundary is no more symmetry of the problem.
We choose preserved subsymmetry to be of B-type by setting without loss of generality $\epsilon_+=\epsilon_-=\epsilon$.

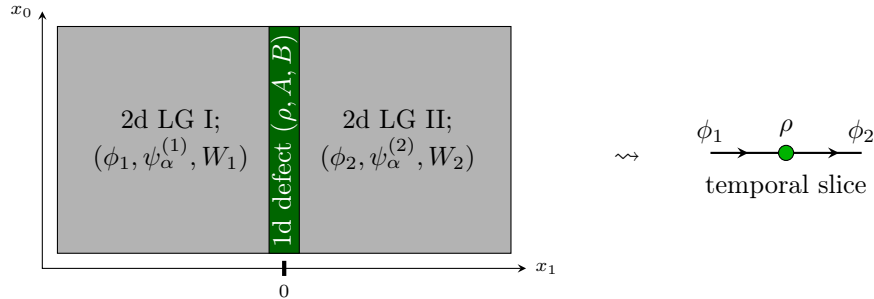
\begin{figure}[ht!]
	\centering
		$\begin{array}{c}
			\begin{tikzpicture}
			\draw[stealth-stealth] (-3.2,1.7) -- (-3.2,-1.7) -- (3.2,-1.7);
			\node[left] at (-3.2,1.7) {$\scriptstyle x_0$};
			\node[right] at (3.2,-1.7) {$\scriptstyle x_1$};
			\draw[ultra thick] (0,-1.8) -- (0,-1.6);
			\node[below] at (0,-1.8) {$\scriptstyle 0$};
			\draw[fill=white!40!gray] (-3,-1.5) -- (-3,1.5) -- (3,1.5) -- (3,-1.5) -- cycle;
			\node at (-1.5,0) {$\begin{array}{c}
					\mbox{2d LG I};\\
					(\phi_1,\psi_{\alpha}^{(1)},W_1)
				\end{array}$};
			\node at (1.5,0) {$\begin{array}{c}
					\mbox{2d LG II};\\
					(\phi_2,\psi_{\alpha}^{(2)},W_2)
				\end{array}$};
			\draw[fill=black!60!green] (-0.2,-1.5) -- (-0.2,1.5) -- (0.2,1.5) -- (0.2,-1.5) -- cycle;
			\node[white, rotate=90] at (0,0) {1d defect $(\rho,A,B)$};
		\end{tikzpicture}\end{array}\quad\rightsquigarrow\quad\begin{array}{c}
		\begin{tikzpicture}
			\draw[thick,postaction={decorate},decoration={markings, mark = at position 0.85 with {\arrow{stealth}}, mark = at position 0.25 with {\arrow{stealth}}}] (-1,0) -- (1,0);
			\draw[fill=\mygreen] (0,0) circle (0.1);
			\node[above] at (-1,0) {$\phi_1$};
			\node[above] at (1,0) {$\phi_2$};
			\node[above] at (0,0.1) {$\rho$};
		\end{tikzpicture}\\
		\mbox{temporal slice}
	\end{array}$
	\caption{A junction of two LG worldsheets.}\label{fig:junction}
\end{figure}

Consider a junction of two LG worldsheets located at spatial coordinate $x^1=0$ as it is depicted in Fig.~\ref{fig:junction}.
To describe the junction defect we put on it a 1d model of the Fermi multiplet \cite{Kapustin:2002bi,Galakhov:2020upa} with fermionic field $\rho$.
It is allowed to interact with the bulk chirals via superapotentials $A$ and $\bar B$ that are holomorphic and anti-holomorphic functions of boundary values $\bphi_i(x^0):=\phi_i(x^0,x^1=0)$.
From the fermionic part of the chiral multiplets it interacts only with combinations $\bpsi_i(x^0):=(\psi_+^{(i)}(x^0,x^1=0)-\psi_-^{(i)}(x^0,x^1=0))$:
\begin{equation}
	\CS_{\rm F}(\rho,A,B)=\int dx^0\,\left(\I \bar\rho\p_0\rho-\frac{1}{4}|A|^2-\frac{1}{4}|B|^2-\frac{1}{\sqrt{2}}{\rm Re}\,\rho\left(\sum\lm_i\bpsi^{(i)}\p_{\bphi_i}A-\sum\lm_i\overline{\bpsi^{(i)}\p_{\bphi_i}B}\right)\right)\,.
\end{equation}
SUSY transforms for field $\rho$ read:
\begin{equation}
	\delta\rho=\frac{1}{\sqrt{2}}\left(\epsilon\overline{A}+\bar\epsilon B\right)\,.
\end{equation}

For the total SUSY variation of combined actions we find:
\begin{equation}
	\begin{aligned}
	&\delta(\CS_\chi(\phi_1,\psi_{\alpha}^{(1)},W_1)+\CS_\chi(\phi_2,\psi_{\alpha}^{(2)},W_2)+\CS_{\rm F}(\rho, A,B))=\\
	&=\int dx^0\;{\rm Re}\left(\bar\epsilon\sum\lm_{i}\bpsi^{(i)}\p_{\bphi_i}\left(\frac{\I}{2}W_1-\frac{\I}{2}W_2-\frac{1}{4}AB\right)+(-1)^i\left(-\bar \epsilon\left(\bar\psi_+^{(i)}-\bar\psi_-^{(i)}\right)\p_1\phi+\bar \epsilon\left(\bar\psi_+^{(i)}+\bar\psi_-^{(i)}\right)\p_0\phi\right)\bigg|_{x^1=0}\right)\,.
	\end{aligned}
\end{equation}
For this system to remain supersymmetric we should impose constraints on a relation between bulk and defect superpotentials:
\begin{equation}\label{MFconstraint}
	AB=2\I(W_1-W_2)\,,
\end{equation}
as well for fields $\phi_i$ and $\bar\psi_+^{(i)}-\bar\psi_-^{(i)}$ we impose Neumann boundary conditions at the defect boundary, and for fields $\bar\psi_+^{(i)}+\bar\psi_-^{(i)}$ we impose Dirichlet boundary conditions correspondingly.

Now we consider a horizontal section of the worldsheet and Hamiltonian dynamics on it.
Wave functions spanning the Hilbert space of this system are functionals $\Psi[\phi_i(x),\bar \phi_i(x),\psi_{\alpha}^{(i)}(x),\rho]$ of spatial field distributions $\phi_i(x)$, $\psi_{\alpha}^{(i)}$ and a function of variable $\rho$.
The Hamiltonian of this supersymmetric system can be derived as an anti-commutator of supercharge operators $\CH=\frac{1}{2}\left\{\fQ,\fQ^{\dagger}\right\}$, where \cite{Galakhov:2020upa}:
\begin{equation}
	\fQ^{\dagger}_{\Lambda}=\sum\lm_i\int dx^1\left(-\I\left(\hat\psi_+^{(i)\dagger}-\hat\psi_-^{(i)\dagger}\right)\frac{\delta}{\delta\bar\phi_i}+e^{\Lambda}\left(\hat\psi_+^{(i)\dagger}+\hat\psi_-^{(i)\dagger}\right)\p_1\phi_i-e^{-\Lambda}\frac{\I}{2}\left(\hat\psi_+^{(i)}+\hat\psi_-^{(i)}\right)W_i\right)+\frac{1}{\sqrt{2}}\left(A\hat\rho+B\hat\rho^{\dagger}\right)\,,
\end{equation}
where we have already introduced rescaling $\Lambda$ according to localization prescription \cite{Witten:1982im}.
All the Hamiltonians obtained from $\fQ^{\dagger}_{\Lambda}$ for different $\Lambda$ correspond to different systems, nevertheless supersymmetric ground states for all of them correspond to cohomologies $H^*(\fQ_{\Lambda}^{\dagger})$ and are isomorphic among each other.
It is easy to calculate such cohomologies in the limit $\Lambda\to\infty$ since the dominant term localizes field configurations to constant modes $\p_1\phi_i=0$.
The constant modes coincide with values $\bphi_i$ at the boundary.
The wave function factorizes as
\begin{equation}
	\Psi[\phi_i(x),\bar \phi_i(x),\psi_{\alpha}^{(i)}(x),\rho]=\Psi_0(\bphi_i,\bar\bphi^{(i)},\bpsi^{(i)},\rho)\times\Psi\left[\mbox{heavy modes}\right]\,.
\end{equation}
where $\Psi_0$ is described as cohomology of the following effective supercharge:
\begin{equation}
	\fQ_0^{\dagger}=-\I\sum\lm_i\bpsi^{(i)\dagger}\frac{\p}{\p\bar\bphi_i}+\frac{1}{\sqrt{2}}\left(A\hat\rho+B\hat\rho^{\dagger}\right)\,.
\end{equation}
In this supercharge variables split, so we find its cohomologies in the following form:
\begin{equation}
	\Psi_0(\bphi_i,\bar\bphi^{(i)},\bpsi^{(i)},\rho)=\psi(\bphi_i,\rho)\times\prod\lm_i\bpsi^{(i)}\,,
\end{equation}
where $\psi$ are strictly holomorphic in all variables $\bphi_i$ and are cohomologies of the following effective supercharge:
\begin{equation}
	\fQ^{\dagger}=\frac{1}{\sqrt{2}}\left(A\hat\rho+B\hat\rho^{\dagger}\right)\,,
\end{equation}
that is \emph{analogous to MF operators} $\MF$ introduced in Sec.~\ref{sec:MF}.

It is easy to generalize this story to any number of 2d sheets and 1d junctions.
A temporal slice of the string worldsheet tube is encoded literally by a graph $\Gamma$ whose oriented edges correspond to intersections of 2d worldsheets with the slice, and vertices correspond to intersections with junction defect worldlines (see Fig.~\ref{fig:junction}).
Orientation on edges reflects worldsheet orientation.
The supersymmetric ground state wave function could be determined as cohomology of an effective supercharge:
\begin{equation}
	\fQ_{\Gamma}^{\dagger}=\sum\lm_{v\in{\rm vertices}}\fQ_{v}^{\dagger}\,,
\end{equation}
where local vertex supercharges $\fQ_{v}^{\dagger}$ correspond to matrix factorization.
We can easily generalize a junction defect to a vertex of any valency.
Due to \eqref{MFconstraint} the respective matrix factorization is constrained:
\begin{equation}
	\left(\fQ_{v}^{\dagger}\right)^2=2\I\sum\lm_{a: *\to v}W_a-2\I\sum\lm_{b: v\to *}W_b\,,
\end{equation}
where we split sums over edges $a$ flowing inward vertex $v$ and edges $b$ flowing outward vertex $v$.

\section{Quotients of polynomial rings}\label{app:factor}

Working with the framework of Khovanov-Rozansky double complexes one encounters often a necessity to describe a quotient ring.
For example, cohomologies of a simplistic nilpotent MF operator $\MF=p(x)\hat\theta$, where $p(x)$ is a polynomial, acting on a ring $\IQ[x]\oplus\IQ[x]\theta$ is isomorphic as a vector space to quotient ring $\IQ[x]/\langle p(x)\rangle:=\IQ[x]/(p(x)\IQ[x])$.
This fact has an analogous rephrasing as the short exact sequence of modules:
\begin{equation}
	\begin{array}{c}
		\begin{tikzpicture}
			\node(B) at (2.5,0) {$0$};
			\node(C) at (4,0) {$\IQ[x]$};
			\node(D) at (6,0) {$\IQ[x]$};
			\node(E) at (8,0) {$\IQ[x]/\langle p(x)\rangle$};
			\node(F) at (9.8,0) {$0$};
			\path (B) edge[->] (C) (C) edge[->] node[above] {$\scriptstyle p(x)\cdot$} (D) (D) edge[->] (E) (E) edge[->] (F);
		\end{tikzpicture}
	\end{array}\,.
\end{equation}

So here we gather a few handy quotient relations that might appear useful for the task of a quotient ring construction.

\subsection{Linear quotient}

Let us start with quotients modulo linear relations.
Consider a polynomial function $F$ in a ring of $N$ pairs of variables $z_i$ and $w_i$.
Apparently, one could work equally well with polynomials over fields of other unconstrained variables.
Let us emphasize the dependence of $F$ on $z_i$ only and omit the dependence on $w_i$.
Assume $z_i$ are variables we would like to exclude by relations $z_i=w_i$, $i=1,\ldots,N$.
In terms of rings we consider $F$ in the ring modulo ideals $z_i-w_i$.
So we decompose $F$ over a zeroth term with substituted $z_i=w_i$ and a sum of ideal members:
\begin{equation}\label{chain}
\begin{aligned}
	&F(z_1,\ldots,z_N)=F(w_1,\ldots,w_N)+\sum\lm_{i=1}^{N}(z_i-w_i)\times G_i\,,\\
	& G_i=\frac{F(w_1,\ldots,w_{i-1},z_i,z_{i+1},\ldots,z_N)-F(w_1,\ldots,w_{i-1},w_i,z_{i+1},\ldots,z_N)}{z_i-w_i}\,.
\end{aligned}
\end{equation}
Here all the ratios $G_i$ are polynomial integer quotients indeed as the denominators cancel factors in the numerators having respective roots at $z_i=w_i$.

Apparently, fractions $G_i$ are not defined uniquely, rather they are subjected to the gauge transforms parameterized by arbitrary polynomials $H$:
\begin{equation}
	G_i\to G_i+(z_j-w_j)H,\quad G_j\to G_j-(z_i-w_i)H\,.
\end{equation}

\subsection{Quadratic quotient}\label{app:Quad_quot}

Here we would like to discuss a specific quotient ring $\IQ[x_,x_2,x_3,x_4]/\langle s_1,s_2\rangle$, where
\begin{equation}
	s_1=x_1+x_2-x_3-x_4,\quad s_2=x_1x_2-x_3x_4\,.
\end{equation}
Here again an attempt to factor out a function by $s_i$ would lead to localization of the function to solutions $s_1=0$, $s_2=0$.
Let us assume that variables $x_3$ and $x_4$ are to be excluded, following Vi\`ete's theorem we find two possible roots:
\begin{equation}\label{root}
	(x_3=x_1,x_4=x_2)\mbox{ and }(x_3=x_2,x_4=x_1)\,.
\end{equation}

To construct the quotient of an arbitrary function $f(x_1,x_2,x_3,x_4)$ first we extract dependence on combinations $x_3+x_4$ and $x_3x_4$ by decomposing $f$ over symmetric and anti-symmetric part:
\begin{equation}
	f(x_1,x_2,x_3,x_4)=\frac{1}{2}\underbrace{\left(f(x_1,x_2,x_3,x_4)+f(x_1,x_2,x_4,x_3)\right)}_{\ell_1(x_1,x_2,x_3+x_4,x_3x_4)}+\frac{1}{2}(x_3-x_4)\underbrace{\frac{f(x_1,x_2,x_3,x_4)-f(x_1,x_2,x_4,x_3)}{x_3-x_4}}_{\ell_2(x_1,x_2,x_3+x_4,x_3x_4)}\,.
\end{equation}
Again for the ratio the denominator is canceled with a root in the numerator, so both functions $\ell_{1,2}$ are polynomials.
These polynomials are symmetric with respect to permutation $x_3\leftrightarrow x_4$, therefore they could be decomposed over elementary symmetric polynomials $x_3+x_4$ and $x_3x_4$.
Further we construct the linear quotient \eqref{chain} of $\ell_{1,2}$ for variables $z_1=x_3+x_4$, $z_2=x_3x_4$, $w_1=x_1+x_2$, $w_2=x_1x_2$.
Eventually we would arrive to the following decomposition of function $f$:
\begin{equation}\label{v12}
	f(x_1,x_2,x_3,x_4)=\frac{1}{2}(1+h)\times f(x_1,x_2,x_1,x_2)+\frac{1}{2}(1-h)\times f(x_1,x_2,x_2,x_1)+s_1v_1(f)+s_2v_2(f)\,.
\end{equation}
where $h:=\frac{x_3-x_4}{x_1-x_2}$, and linear maps $v_{1,2}$ represent integer quotients of $f$ for $s_1$ and $s_2$ respectively according to \eqref{chain}.
They are defined modulo the gauge transform:
\begin{equation}
	v_1(f)\to v_1(f)+s_2g,\quad v_2(f)\to v_2(f)-s_1g,\quad \forall g\,.
\end{equation}
Let us note that:
\begin{equation}
	h^2=\left(\frac{x_3-x_4}{x_1-x_2}\right)^2=1-s_1\frac{x_1+x_2+x_3+x_4}{(x_1-x_2)^2}+s_2\frac{4}{(x_1-x_2)^2}=1\;{\rm mod}\;\langle s_1,s_2\rangle\,.
\end{equation}
Thus we see that $P_{\pm}=\frac{1}{2}(1\pm h)$ are projectors to disjoint roots \eqref{root}:
\begin{equation}
	P_+^2=P_+\;{\rm mod}\;\langle s_1,s_2\rangle,\quad P_-^2=P_-\;{\rm mod}\;\langle s_1,s_2\rangle,\quad P_+P_-=P_-P_+=0\;{\rm mod}\;\langle s_1,s_2\rangle\,.
\end{equation}


\bibliographystyle{utphys}
\bibliography{biblio}

\end{document}